\documentclass{jfm}

\usepackage{pgfplots}
\pgfplotsset{compat=newest} 
\newif\iflocalcompile
\localcompilefalse   

\usepgfplotslibrary{external}

\iflocalcompile
\else
\newcommand{\includetikz}[1]{\includegraphics{tikzfigs/#1.pdf}}
\fi

\usepackage{graphicx}
\usepackage[labelsep=period]{caption} 

\usepackage{adjustbox}
\usepackage{newtxtext}
\usepackage{newtxmath}
\usepackage{mathrsfs}

\usepackage{tikz}

\usepackage{placeins} 

\usepackage{multirow} 
\usepackage{booktabs} 

\usepackage{natbib}
\usepackage{hyperref}
\hypersetup{
    colorlinks = true,	
    linkcolor   = blue,	
    citecolor  = blue,	
    urlcolor   = blue	
}

\usepackage[english]{babel}      
\usepackage{caption}             

\usepackage{enumitem} 

\usepackage{amsmath} 
\usepackage{cleveref}

\newcommand{\RomanNumeralCaps}[1]
\linenumbers

\newcommand{\figroot}{./fig}

\newcommand{\figtwo}{\figroot/2problem/}
\newcommand{\figthree}{\figroot/3baseflow/}
\newcommand{\figfourone}{\figroot/4.0wake/}
\newcommand{\figfourtwo}{\figroot/4.1vsDNS/}
\newcommand{\figfourthree}{\figroot/4.2sta/}
\newcommand{\figfourfour}{\figroot/4.3osc/}
\newcommand{\figfourfive}{\figroot/4.4fixedLB/}
\newcommand{\figfoursix}{\figroot/4.5differentchi/}

\shorttitle{Stability analysis of an oblate bubble pair}
\shortauthor{W. Q. Liu, J. M. Jiang, and J. Zhang.}

\title{A pair of oblate bubbles rising in-line: a linear stability analysis}
 
\author{Wei-Qiang Liu\aff{1},
	Jian-Ming Jiang\aff{1}
	\and Jie Zhang\aff{1}
	\corresp{\email{j\_zhang@xjtu.edu.cn}}}

\affiliation{\aff{1}State Key Laboratory for Strength and Vibration of Mechanical Structures, School of Aerospace, Xi’an Jiaotong University, Xi’an, China}

\begin{document}
\maketitle

\begin{abstract}
The stability of two bubbles rising initially in-line through a viscous liquid is revisited using a global linear stability analysis formulated within an Arbitrary Lagrangian–Eulerian framework, complemented by fully resolved Embedded Boundary Method simulations.  
Whereas previous studies attributed the promoted in-line stability of oblate bubbles to a deformation-enhanced wake entrainment, the present analysis demonstrates that the dominant stabilizing mechanism arises instead from an inclination-induced rotational feedback generated as the trailing bubble experiences the asymmetric shear of the leading bubble’s wake.  
This inclination–shear coupling, rather than deformation itself, governs the recovery of stability with increasing aspect ratio.  
Furthermore, the results reveal that the unstable drafting–kissing–tumbling mode originates from short-range, two-way coupling between the bubbles, whereas the asymmetric side-escape mode corresponds to a long-range, one-way interaction dominated by the trailing bubble response.  
In addition, a previously unreported oscillatory global mode emerges from the unsteady recirculation linking the two bubbles, acting as a hydrodynamic spring whose effective stiffness and damping govern the oscillation frequency and growth rate.  
Together, these findings identify inclination-induced lift as the primary mechanism controlling the stability of rising bubble pairs and provide a unified framework for interpreting their stationary and oscillatory transitions across a broad range of inertial and deformable regimes.
\end{abstract}



\section{Introduction}\label{sec:headings}

Buoyancy-driven bubbly flows are ubiquitous in both natural and industrial environments, ranging from breaking waves and hydrothermal vents to bubble columns and metallurgical reactors.  
Understanding the rise dynamics of individual bubbles and their mutual interactions remains a central problem for elucidating the fundamental mechanisms governing momentum, mass, and heat transfer in freely rising suspensions \citep{maeda2021viscid,legendre2025gas}.  
Among the many configurations observed in bubble swarms, the rise of an in-line bubble pair provides the simplest yet most informative prototype of hydrodynamic interaction.  
It encapsulates the essential coupling mechanisms, e.g. wake entrainment, drag reduction, and lift generation, that govern the collective dynamics of larger bubble populations \citep{zenit2001measurements,figueroa2018lifespan,ma2023fate,hessenkemper2025lagrangian}.  
Whether the in-line configuration remains stable or evolves toward an oblique arrangement determines the emergence of vertical chains or horizontal clusters in bubble swarms, making its stability a cornerstone for understanding self-organization in dispersed bubbly flows.

The in-line configuration was first examined under the assumption of spherical bubbles.  
In weakly inertial regimes corresponding to low Reynolds numbers, \citet{katz1996wake} observed that the two bubbles invariably collide and coalesce, a result confirmed experimentally in silicone oils by \citet{watanabe2006line}.  
This behaviour stems from the viscous attraction produced by wake region behind the leading bubble (LB), which tends to stabilize the in-line arrangement.  
At moderate Reynolds numbers, however, \citet{hallez2011interaction} numerically demonstrated that the in-line configuration becomes unstable, in agreement with earlier theoretical predictions \citep{harper1970bubbles,yuan1994line}.  
Under these conditions, the trailing bubble (TB) drifts laterally away from the symmetry axis.  
Depending on the response of the LB to this lateral deviation, two distinct unstable evolutions may occur: the well-known \emph{drafting–kissing–tumbling} (DKT) sequence, in which both bubbles depart from their initial trajectory, and the \emph{asymmetric side-escape} (ASE) scenario, in which the LB remains nearly unperturbed while the TB migrates laterally.  
These two instability scenarios have markedly different consequences for the spatial dispersion of bubbles in a plume.
Both originate from the combined action of a repulsive potential interaction and a shear-induced lift acting on the TB as it is immersed in the asymmetric wake of the LB, but differ in the degree to which the LB dynamically responds to this perturbation.
As the bubbles become deformable and adopt oblate shapes, the interaction dynamics change markedly.  
Although both the DKT and ASE scenarios persist in experiments \citep{kusuno2015experimental,kusuno2019lift} and direct numerical simulations (DNS) \citep{bunner2003effect,gumulya2017interaction,zhang2021three},  
a reversal of the stability trend is observed once the bubbles become sufficiently oblate, resembling   
a restabilized in-line configuration that the bubbles may even approach head-on collision.  
Recent three-dimensional (3D) Lagrangian tracking experiments by \citet{hessenkemper2025lagrangian} further showed that smaller, nearly spherical bubbles in plumes exhibit random pair alignments, confirming that less deformed bubbles are more sensitive to flow perturbations.  

This brief review highlights that determining how the bubble aspect ratio influences the stability of the in-line configuration is of essentially importance.  
To date, the prevailing interpretation attributes the restabilization observed for oblate bubbles to the enhanced viscous entrainment in the LB’s wake, 
which draws the TB back toward the symmetry axis and offsets the destabilising lift.  
Since neither the potential nor the shear-induced components of the lift can change sign within the investigated parameter range, 
this explanation has remained quite qualitative.  
The persistence of this ambiguity provides the main motivation for the present study, 
which aims to address two fundamental questions: \textbf{\emph{(a)}} Is the observed stabilization of oblate bubble pairs genuinely caused by deformation-enhanced wake entrainment, or does it originate from a different physical mechanism? \textbf{\emph{(b)}} Do the DKT and ASE evolutions represent two manifestations of a single instability, or do they correspond to distinct global modes of the coupled bubble–flow system?

The present work revisits this problem through a global linear stability analysis (LSA) formulated within an Arbitrary Lagrangian–Eulerian (ALE) framework.  
This formulation is essential because conventional LSAs, defined on fixed computational domains, cannot properly account for the evolving geometry arising from the relative motion of two bubbles.  
Classical LSAs have proved powerful for elucidating the onset of wake and path instabilities in isolated bodies.  
For fixed spheroids or disks, global analyses have established that the first loss of axisymmetry originates within the recirculation zone of the wake 
\citep{meliga2009unsteadiness,tchoufag2013linearwake,cano2016global}.  
When the body is allowed to move freely under buoyant forcing, the coupling between hydrodynamic and kinematic degrees of freedom gives rise to oscillatory modes responsible for the zigzag or spiral trajectories of rising bubbles \citep{tchoufag2014linearbubble,cano2016global} and disks 
\citep{assemat2012onset,tchoufag2014globaldisk}.  
In the present ALE–LSA approach, the time-dependent configuration is mapped onto a steady reference domain, allowing the linearised Navier–Stokes equations to be solved for the global eigenmodes of the coupled bubble–flow system.  
This methodology, previously applied to deformable single bodies \citep{bonnefis2024path,pfister2020fluid}, captures both the translational and rotational degrees of freedom of each bubble, 
thereby resolving the full structure of the two-way hydrodynamic coupling.

The results reveal that the stabilization of oblate bubble pairs does not originate from a deformation-dependent reversal of wake entrainment, as previously assumed, but instead from an inclination-induced rotational feedback.  
A small lateral disturbance generates a shear-induced torque on the TB that reorients its minor axis toward the outward side of the wake.  
As the rotation proceeds, an opposing inertial torque develops, and the two balance at a finite inclination angle where the bubble’s broadside faces the local shear.  
This orientation modifies the surface-pressure distribution, producing a restoring lift directed toward the symmetry axis and thereby counteracting the destabilizing potential and shear-induced components of the lift.  
The present analysis also clarifies the dynamical distinction between the DKT and ASE evolutions, which are shown to correspond to different global response modes of the coupled bubble–flow system.  
In the DKT regime, the hydrodynamic interaction between the two bubbles is strongly reciprocal:  
the self-rotation of each bubble plays a significant role, and their motions remain dynamically coupled throughout the instability growth.  
By contrast, the ASE regime behaves more like a one-way coupling problem, where the LB primarily imposes a shear environment, and the stability characteristics are governed almost entirely by the response of the TB.  
In addition, the analysis uncovers a previously unreported oscillatory global mode, which does not result from periodic vortex shedding as in isolated bubbles, but from the unsteady recirculation connecting the two bubbles.  
This confined vortical structure acts as a hydrodynamic spring that links their motions and mediates a phase-lagged exchange of momentum and torque.  
By combining the ALE–LSA framework with fully resolved Embedded Boundary Method (EBM) simulations, we isolate and quantify these mechanisms over a broad range of Reynolds numbers and aspect ratios.  
The resulting physical picture revises the classical interpretation of lift reversal and establishes a unified framework for predicting stability transitions in interacting bubble pairs—and, by extension, in more complex bubbly flows.

The paper is organized as follows.  
\hyperref[sec:Problem statement]{Section~\ref{sec:Problem statement}} introduces the problem formulation, the governing dimensionless parameters, and the numerical implementation of the global ALE–LSA framework.  
\hyperref[sec:BaseFlow-section]{Section~\ref{sec:BaseFlow-section}} describes the steady axisymmetric base flows and examines their dependence on bubble shape, separation distance, and Reynolds number, with particular emphasis on the equilibrium distance and the formation of the standing eddy connecting the two bubbles.  
\hyperref[sec:LSA_stationary]{Section~\ref{sec:LSA_stationary}} presents the stability characteristics of the stationary mode obtained from the ALE–LSA.  
Special attention is devoted to three aspects governing this mode:  the influence of bubble rotation, the role of mutual hydrodynamic coupling in the DKT and ASE regimes, and the effect of a realistic deformation contrast between the leading and trailing bubbles.  
The Embedded Boundary Method \citep{zhang2025lift,WEI2026114632} is further employed to compute the lift and torque acting on fixed in-line bubbles, 
providing quantitative support for the key role of bubble inclination in controlling the stability of the in-line configuration.  
\hyperref[sec:LSA_oscillatory]{Section~\ref{sec:LSA_oscillatory}} then examines the oscillatory global mode, which originates from the unsteady hydrodynamic coupling mediated by the inter-bubble recirculation.  
Finally, the main findings and concluding remarks are summarized in \hyperref[sec:conclusion]{Section~\ref{sec:conclusion}}.

\section{Problem statement and linear stability strategy}
\label{sec:Problem statement}

\subsection{Problem description}
\begin{figure}
	\centering
	\setlength{\tabcolsep}{0pt}
	\iflocalcompile
	\begin{tabular}[c]{ccc}
		\resizebox{0.325\textwidth}{!}{\input{\figtwo/model1.tex}} &
		\resizebox{0.3\textwidth}{!}{\input{\figtwo/model2.tex}} &
		\resizebox{0.335\textwidth}{!}{\input{\figtwo/model3.tex}}
	\end{tabular}
	\else
	\begin{tabular}[c]{ccc}
		\resizebox{0.325\textwidth}{!}{\includetikz{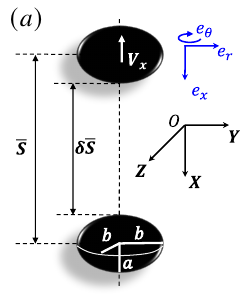}} &
		\resizebox{0.3\textwidth}{!}{\includetikz{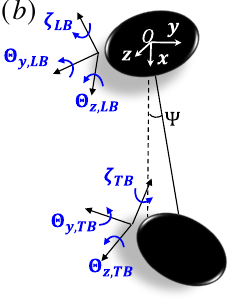}} &
		\resizebox{0.335\textwidth}{!}{\includetikz{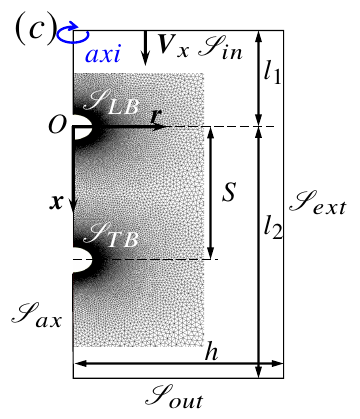}}
	\end{tabular}
	\fi
	\captionsetup{justification=justified, singlelinecheck=false, labelsep=period, width=\linewidth,font=small,format=plain}
	\caption{Sketch of the bubble-pair configuration(detailed descriptions are provided in the main text). 
$(a)$ Steady, axi-symmetric base flow and definition of the principal geometric parameters.
$(b)$ Typical unstable configurations following an infinitesimally imposed lateral perturbation.
$(c)$ Computational domain and spatial discretisation adopted for the ALE–LSA simulations.}
	\label{flow-configuration}
\end{figure}

The steady, axi-symmetric base configuration of the in-line bubble pair is sketched in 
\hyperref[flow-configuration]{figure~\ref{flow-configuration}$(a)$}, where the main geometric parameters are defined.  
The system is described in a stationary Cartesian frame $(X, Y, Z)$, or equivalently in a cylindrical coordinate system $(x, r, \theta)$. Two identical gas bubbles, hereafter referred to as the leading and trailing bubbles (LB and TB, respectively), rise in-line at a common velocity $-V_{x}\,\boldsymbol{e}_{x}$ through an incompressible Newtonian liquid of density $\rho$ and dynamic viscosity $\mu$. In the frame translating with the bubble pair, the surrounding liquid therefore moves toward the bubbles with velocity $V_{x}\,\boldsymbol{e}_{x}$.  
Each bubble is modelled as a non-deformable spheroid of fixed volume, characterised by its aspect ratio $\chi = b/a$, defined as the ratio of the major to minor semi-axes, and by the equivalent spherical radius $R=(b^{2}a)^{1/3}$. The Reynolds number, based on the rise velocity $V_{x}$ and the equivalent diameter $2R$, is $\Rey = 2\rho V_{x}R/\mu$. Since both bubbles rise at the same velocity in the base state, their centre-to-centre distance remains constant and is denoted by $\bar{S}$, while the dimensionless value normalised by $2R$ is $S$.  
The corresponding dimensionless surface-to-surface gap is  $\delta S = S - \chi^{-2/3}$, representing the minimum axial separation between the two interfaces.  
Bubble contact occurs at $\delta S=0$, that is, $S=\chi^{-2/3}$, which yields $S=1$ for spherical bubbles ($\chi=1$) and $S \approx 0.63$ for oblate bubbles with $\chi=2$. The numerical configuration corresponding to this steady, axi-symmetric base flow, constructed in the $xor$ meridional plane, is shown in  \hyperref[flow-configuration]{figure~\ref{flow-configuration}$(c)$} and described in detail in \hyperref[sec:LALE-numerical]{\S~\ref{sec:LALE-numerical}}.

Considering that the density and viscosity of the gas enclosed in the bubble are negligibly small compared with those of the surrounding liquid, the base flow is thus fully characterized by the dimensionless parameters $(\Rey,\,\chi,\,S)$, or equivalently $(\Rey,\,\chi,\,\delta S)$.  In real gas–liquid systems these parameters are not strictly independent, as they jointly depend on the fluid properties, bubble size, and release intervals.  Varying them independently, as done here, should therefore be viewed as a parametric continuation that isolates the respective influences of inertia, shape, and bubble–bubble interaction, even though such decoupling cannot be achieved experimentally. Throughout this study, the aspect ratio is restricted to $\chi \le 1.9$, a range over which the wake of the bubbles pair remain steady and axisymmetric, and this restriction ensures that any departure from the in-line configuration results exclusively from hydrodynamic interaction between the two bubbles rather than from intrinsic wake unsteadiness. This assumption is explicitly confirmed in \hyperref[app:wake]{Appendix~\ref{app:wake}}, and will also be discussed in \hyperref[sec:LSA_stationary]{\S~\ref{sec:LSA_stationary}}.

\label{sec:Problem}

\hyperref[flow-configuration]{Figure~\ref{flow-configuration}$(b)$} 
illustrates the reference configuration adopted to describe small deviations from perfect in-line alignment.  
In the absolute (laboratory) frame, the instantaneous positions and velocities of the leading and trailing bubbles are denoted by  
$\boldsymbol{X}_{LB} = (X,\,Y,\,Z)_{LB}$, 
$\boldsymbol{X}_{TB} = (X,\,Y,\,Z)_{TB}$,  
and  
$\boldsymbol{V}_{LB} = (-V_x,\,V_y,\,V_z)_{LB}$, 
$\boldsymbol{V}_{TB} = (-V_x,\,V_y,\,V_z)_{TB}$,  
respectively.  
For the computations, we introduce a second Cartesian frame $(x, y, z)$ translating with the LB, 
whose $x$-axis remains aligned with the vertical direction while the $y$–$z$ plane defines the transverse plane.  
In this moving frame, the dimensionless positions of the bubble centers are  $\boldsymbol{\widetilde{X}}_{LB} = (0,\,0,\,0)$ and $\boldsymbol{\widetilde{X}}_{TB} = (S,\,Y_{TB}-Y_{LB},\,Z_{TB}-Z_{LB})$.  
This coordinate system conveniently accounts for the relative displacement between the two bubbles and forms the reference framework for the subsequent stability analysis.
A lateral displacement of the TB defines an instantaneous inclination angle $\Psi = \tan^{-1}\!\left(\sqrt{(Y_{TB} - Y_{LB})^2 + (Z_{TB} - Z_{LB})^2}/S\right)$, which quantifies the deviation of the line joining the bubble centres from the vertical. Unlike in studies of isolated bubbles \citep{tchoufag2014linearbubble,cano2016global}, the present coordinate system does not rotate with the LB, as the relative motion of the TB renders such a frame unsuitable for describing their coupled dynamics.  Instead, the orientation of each bubble is described separately by the vector of small angular displacements  $\boldsymbol{\varXi}_{(LB,TB)} = (\zeta,\,\Theta_{y},\,\Theta_{z})_{(LB,TB)}$, representing roll, pitch, and yaw angles about their $(x,y,z)$ axes. For infinitesimal rotations, the corresponding angular velocity is  $\boldsymbol{\varOmega}_{(LB,TB)} = d\boldsymbol{\varXi}_{(LB,TB)}/dt$.  Hence, $\Psi$ characterises the revolution of the TB about the LB, while $\boldsymbol{\varXi}$ (or $\Theta$) quantifies its intrinsic rotation.  
When infinitesimal perturbations are superimposed on the steady base state, growing value of $\Psi$ indicates an unstable in-line configuration, whereas the remaining zero value corresponds to a stable alignment.  
These kinematic quantities are used consistently hereafter to describe both the steady base flow and its linear perturbations.

Note that the present configuration departs from realistic bubble pairs in two respects.  
First, experiments \citep{sanada2006interaction,maeda2021viscid} and three-dimensional VOF-DNS studies \citep{gumulya2017interaction,zhang2021three}  
have shown that the TB generally exhibits a smaller aspect ratio than the LB, since the low-pressure region in the wake of the latter tends to make the TB more spherical.  For clarity and tractability, both bubbles are here prescribed identical spheroidal shapes ($\chi_{LB}=\chi_{TB}=\chi$),  allowing the fundamental mechanisms destabilizing the in-line configuration to be isolated.  The influence of unequal aspect ratios is examined separately in \hyperref[sec:aspect_ratio]{\S~\ref{sec:aspect_ratio}}. Second, under realistic conditions the TB typically accelerates slightly faster than the LB before reaching equilibrium separation \citep{sanada2006interaction,gumulya2017interaction,zhang2021three}, so that $V_{x,TB} > V_{x,LB}$ in practice. For simplicity, both bubbles are here assumed to rise at the same mean velocity, so that the analysis focuses on the intrinsic stability of the steady in-line configuration. This approximation, which has limited influence on the instability mechanism, is consistent with previous numerical studies of bubble–bubble interactions \citep{legendre2003hydrodynamic,hallez2011interaction}.

\subsection{Governing equations and ALE formulation}
\label{sec:governing equations}

Because the gas density and viscosity are negligible compared with those of the surrounding liquid,  the internal motion within each bubble may be ignored and the gas pressure treated as spatially uniform \citep{magnaudet2007wake,tchoufag2013linearwake,tchoufag2014linearbubble,bonnefis2024path}.  
Accordingly, the dynamics of the system are entirely governed by the liquid motion in the external domain $\Upsilon(t)$, coupled with the translational and rotational motions of the two bubbles.  
As introduced earlier, the absolute velocity field is projected onto a reference frame translating with the LB, and in this frame, the incompressible liquid flow satisfies the mass and momentum equations:
\begin{subequations} \label{NS-Eulereq}
  \begin{align}
    \nabla \!\cdot\! \boldsymbol{u} &= 0, && \text{in } \Upsilon(t), \label{NSincompressible}\\[3pt]
    \rho \!\left( \frac{\partial \boldsymbol{u}}{\partial t} + (\boldsymbol{u} - \boldsymbol{V}_{LB})\!\cdot\!\nabla\boldsymbol{u} \right)
      &= \nabla \!\cdot\! \boldsymbol{\Sigma} (\boldsymbol{u}, p), && \text{in } \Upsilon(t), \label{NSmomentum}
  \end{align}
\end{subequations}
where $\boldsymbol{u}$ and $p$ denote the liquid velocity and pressure, respectively.  
The Newtonian stress tensor is $\boldsymbol{\Sigma} (\boldsymbol{u}, p) = -p\,\mathbb{I} + \mu\,(\nabla\boldsymbol{u} + \nabla\boldsymbol{u}^{T})$,  
with $\mathbb{I}$ the identity tensor, and $\boldsymbol{V}_{LB}$ is the translational velocity of the LB in the Lagrangian framework.

The motion of each bubble obeys Newton’s laws, which express the balance of hydrodynamic forces and torques acting on its surface $\mathscr{S}_{(LB,TB)}(t)$:
\begin{subequations} \label{Bubble-dynamics}
  \begin{align}
    M d\boldsymbol{V}_{(LB,TB)}/dt &= (M - \rho \mathscr{V})\,\boldsymbol{g} 
      + \int_{\mathscr{S}(t)} \boldsymbol{\Sigma}\!\cdot\!\boldsymbol{n}\,dS, 
      \label{dynamiceq}\\[3pt]
    \mathbb{J}\!\cdot\!d\boldsymbol{\varOmega}_{(LB,TB)}/dt &= 
      \int_{\mathscr{S}(t)} \boldsymbol{r}\!\times\!(\boldsymbol{\Sigma}\!\cdot\!\boldsymbol{n})\,dS, 
      \label{dynamiceqR}\\[3pt]
    d\boldsymbol{X}_{(LB,TB)}/dt &= \boldsymbol{V}_{(LB,TB)}, 
    \label{kinematic-1}\\[3pt]
    d\boldsymbol{\varXi}_{(LB,TB)}/dt &= \boldsymbol{\varOmega}_{(LB,TB)}, 
      \label{kinematic-2}
  \end{align}
\end{subequations}
where subscripts for the LB and TB are omitted for brevity.  
Here, $\boldsymbol{X}$ and $\boldsymbol{\varXi}$ denote the translational displacement and the angular displacement of the bubble in the Lagrangian description, respectively. 
$\boldsymbol{V}$ and $\boldsymbol{\varOmega}$ are the corresponding translational and angular velocities, and $\boldsymbol{r}$ is the position vector relative to the centre.  
The unit normal to the interface $\mathscr{S}$ is denoted by $\boldsymbol{n}$.  
Each bubble has a volume $\mathscr{V} = \tfrac{4}{3}\pi a b^{2}$, mass $M = \rho_b\mathscr{V}$, and diagonal inertia tensor $\mathbb{J} = J_{1}\,\mathbf{xx} + J_{2}\,(\mathbf{yy}+\mathbf{zz})$, with $J_{1} = \tfrac{2}{5}Mb^{2}$ and $J_{2} = \tfrac{1}{5}M(a^{2}+b^{2})$.  
Because the bubbles do not translate in the $x$-direction, the $x-$ component equations of \eqref{dynamiceq} and \eqref{kinematic-1} thus vanish.  
On each bubble surface, the interface is non-penetrable and shear-free, leading to the boundary conditions:
\begin{equation}\label{BC}
  \boldsymbol{u}\!\cdot\!\boldsymbol{n} = (\boldsymbol{V} + \boldsymbol{\varOmega}\!\times\!\boldsymbol{r})\!\cdot\!\boldsymbol{n}, 
  \qquad 
  \boldsymbol{n}\!\times\!(\boldsymbol{\Sigma}\!\cdot\!\boldsymbol{n}) = \boldsymbol{0},
  \qquad \text{on } \mathscr{S}(t).
\end{equation}
At infinity ($|\boldsymbol{x}|\!\to\!\infty$), the velocity field recovers the uniform far-field velocity $\boldsymbol{u} = (0, 0, 0)$. Lengths, velocities, time, and stresses are non-dimensionalised with $2R$, $V_{x}$, $2R/V_{x}$, and $\rho V_{x}^{2}$, respectively. The bubble mass and moment of inertia are scaled with $8\rho R^{3}$ and $3\rho R^{5}$.

To analyse stability in a configuration where the fluid domain evolves with the bubble motion, the governing equations are recast within an Arbitrary Lagrangian–Eulerian (ALE) framework.  
This formulation maps the time-dependent domain $\Upsilon(t)$ onto a fixed reference configuration $\Upsilon_{0}$, allowing all variables to be defined on a stationary computational mesh.  
The mapping $\mathcal{A}\!:\!\Upsilon_{0}\!\times\!\mathbb{R}^{+}\!\to\!\Upsilon(t)\!\times\!\mathbb{R}^{+}$ is defined as
\begin{equation}
  \boldsymbol{x} = \boldsymbol{x}_{0} + \boldsymbol{\xi}_{0}(\boldsymbol{x}_{0},t),
  \label{eq:mapping}
\end{equation}
where $\boldsymbol{x}_{0}$ denotes the position in the reference domain and $\boldsymbol{\xi}_{0}$ the mesh–extension displacement field. This auxiliary field transfers the rigid-body motions of both bubble interfaces ($\mathscr{S}_{LB}$ and $\mathscr{S}_{TB}$) smoothly into the surrounding liquid, 
ensuring a divergence-free and minimally distorted grid. The displacement field $\boldsymbol{\xi}_{0}$ is obtained by solving an elliptic extension problem in $\Upsilon_{0}$, subject to rigid-body boundary conditions on $(\mathscr{S}_{LB}~\!\cup~\!\mathscr{S}_{TB})$ and homogeneous conditions at the outer boundaries.  
This approach, initially developed by \citet{bonnefis2019etude} and refined by \citet{sierra2022dynamics} and \citet{bonnefis2024path}, provides a geometrically consistent description of translating and rotating bodies within a fixed reference domain.  
Unlike previous isolated-bubble formulations, the present implementation accounts for the simultaneous rigid-body motions of two spheroids, introducing a non-trivial geometric coupling between the two interfaces.

From the mapping \eqref{eq:mapping}, the local deformation of the fluid domain is characterised by
\[
  \mathbb{F}_{0} = \mathbb{I} + \nabla_{0}\boldsymbol{\xi}_{0}, 
  \qquad 
  J_{0} = \det(\mathbb{F}_{0}), 
  \qquad
  \boldsymbol{\Phi}_{0} = J_{0}\mathbb{F}_{0}^{-1},
\]
where $\mathbb{F}_{0}$ is the deformation gradient, $J_{0}$ the Jacobian determinant,  
and $\boldsymbol{\Phi}_{0}$ the metric operator linking the instantaneous and reference configurations.  
In this framework, the velocity and pressure fields, $\boldsymbol{u}_{0}(\boldsymbol{x}_{0},t)$ and $p_{0}(\boldsymbol{x}_{0},t)$, are defined on the fixed reference domain $\Upsilon_{0}$. The continuous and momentum equations \eqref{NS-Eulereq} then become:
\begin{subequations} \label{NS-ALEeq}
  \begin{align}
    \nabla_{0}\!\cdot\!(\boldsymbol{\Phi}_{0}\boldsymbol{u}_{0}) &= 0,
    && \text{in } \Upsilon_{0}, \label{NSincompressible-ALE}\\[3pt]
    J_{0}\rho_0\frac{\partial\boldsymbol{u}_{0}}{\partial t}
      + \rho_0(\nabla_{0}\boldsymbol{u}_{0}\boldsymbol{\Phi}_{0})
        \left(\boldsymbol{u}_{0} - \frac{\partial\boldsymbol{\xi}_{0}}{\partial t} - \boldsymbol{V}_{LB}\right)
      - \nabla_{0}\!\cdot\!\mathbb{L}_{0}(\boldsymbol{u}_{0},p_{0},\boldsymbol{\xi}_{0})
      &= \boldsymbol{0},
      && \text{in } \Upsilon_{0}, \label{MSmomentum-ALE}
  \end{align}
\end{subequations}
where $\mathbb{L}_{0}$ denotes the first Piola–Kirchhoff stress tensor, ensuring exact momentum conservation under the domain transformation. The hydrodynamic forces and torques acting on each bubble are obtained by integrating $\mathbb{L}_{0}$ over their respective surfaces, while the bubble dynamics remain expressed in the Lagrangian form \eqref{dynamiceq}–\eqref{kinematic-2}.  This coupled ALE formulation provides a conservative and geometrically consistent framework for computing the fluid–bubble interaction within a fixed reference geometry, preserving the geometric conservation law and avoiding mesh distortion. Explicit expressions for the transformation operators, the ALE stress tensor, and the weak variational formulation employed in the numerical implementation are provided in \hyperref[app:ALE_details]{Appendix~\ref{app:ALE_details}}.  
This ALE framework forms the foundation for the linearisation procedure developed in the next section \hyperref[sec:LALE-numerical]{\S~\ref{sec:LALE-numerical}}.

\subsection{Linearised-ALE formulation and numerical method}
\label{sec:LALE-numerical}

The objective of this analysis is to assess the stability of the in-line configuration of a rising bubble pair.  
The reference, or base flow, corresponds to a steady, axisymmetric state in which the liquid moves past both bubbles at velocity $V_x\boldsymbol{e}_x$, maintaining a fixed separation distance $S$ and a common aspect ratio $\chi$  
(see \hyperref[flow-configuration]{figure~\ref{flow-configuration}$(c)$}).  
This steady solution, denoted $\boldsymbol{\mathscr{Q}}^f_{0} = [\boldsymbol{U}_{0},P_{0}]$, satisfies the incompressible Navier–Stokes equations \eqref{NS-Eulereq} together with the boundary conditions \eqref{BC}.  
Since no relative motion exists between the bubbles, the base state is stationary and the ALE mapping reduces to the identity transformation. The base flow is computed numerically as a steady solution of the axi-symmetric momentum equations.

To examine the stability of this configuration, infinitesimal three-dimensional perturbations are superimposed on the base state. Such perturbations may induce relative motions between the bubbles, rendering the fluid domain time-dependent.  
The equations are therefore recast in the ALE framework introduced in \hyperref[sec:governing equations]{\S~\ref{sec:governing equations}}, allowing the problem to be formulated on a fixed reference domain while accounting for instantaneous displacements of the bubble interfaces. The coupled liquid–bubble system is described by the state vector
\[
\mathscr{Q} = [\mathscr{Q}^{f}, \mathscr{Q}^{b}] 
           = [\boldsymbol{u},\,p,\,\boldsymbol{\xi},\,\boldsymbol{X}_{k},\,\boldsymbol{\varXi}_{k},\,\boldsymbol{V}_{k},\,\boldsymbol{\varOmega}_{k}],
           \qquad (k = \text{LB, TB}),
\]
where $(\boldsymbol{u},p,\boldsymbol{\xi})$ are the fluid variables defined on $\Upsilon_{0}$, and $(\boldsymbol{X}_{k},\boldsymbol{\varXi}_{k},\boldsymbol{V}_{k},\boldsymbol{\varOmega}_{k})$ represent the position, orientation, translational velocity, and angular velocity of each bubble.  

Each variable is decomposed into a steady base state and a small perturbation:
\[
\mathscr{Q} = \mathscr{Q}_{0} + \epsilon\,\boldsymbol{q}', 
\qquad
\boldsymbol{q}' = [\boldsymbol{u},\,p,\,\boldsymbol{\xi},\,\boldsymbol{X}_{k},\,\boldsymbol{\varXi}_{k},\,\boldsymbol{V}_{k},\,\boldsymbol{\varOmega}_{k}]',
\]
where $\epsilon \ll 1$ and $\boldsymbol{q}'$ denotes the infinitesimal perturbation superimposed on the base flow $\mathscr{Q}_{0}$. Substituting this expansion into the ALE governing equations and retaining only first-order terms yields the linearised-ALE system,  which couples the fluid motion to the bubble dynamics through the linearised hydrodynamic forces and torques acting on the bubble surfaces. The corresponding weak variational form and explicit expressions of the linearised operators are given in \hyperref[app:ALE_details]{Appendix~\ref{app:ALE_details}}, following the formulation of \citet{tchoufag2014linearbubble} and \citet{bonnefis2024path}.

Because the base flow is steady and axisymmetric, perturbations are expanded in azimuthal Fourier modes:
\[
\boldsymbol{q}'(r,\theta,x,t)
= \widetilde{\boldsymbol{q}}(r,x)\,e^{im\theta+\lambda t} + c.c.,
\]
where $\widetilde{\boldsymbol{q}}(r,x)$ is the complex amplitude, $m$ the azimuthal wavenumber, and $\lambda = \lambda_{r} + i\lambda_{i}$ the complex eigenvalue.  
The real part $\lambda_{r}$ represents the growth rate, while the imaginary part $\lambda_{i}$ gives the oscillation frequency.  
Instability occurs for $\lambda_{r}>0$, and the leading global mode corresponds to the eigenpair with the largest $\lambda_{r}$.  
Axi-symmetric perturbations correspond to $m=0$, whereas $|m|=1$ modes describe helical disturbances whose linear combinations yield planar zigzag trajectories \citep{tchoufag2014globaldisk,bonnefis2024path}. Higher-order modes ($|m|\ge2$) are dynamically decoupled from the bubble motion and are therefore neglected. The present analysis focuses on the non-axisymmetric family $|m|=1$, which represents the first departure from perfect alignment.  
For each azimuthal mode, the linearised-ALE equations take the generalised eigenvalue form:
\[
\mathbb{A}_{m}\,\widetilde{\boldsymbol{q}} = \lambda\,\mathscr{B}_{m}\,\widetilde{\boldsymbol{q}},
\]
where $\widetilde{\boldsymbol{q}} = [\widetilde{\boldsymbol{q}}^{f},\,\widetilde{\boldsymbol{q}}^{b}]$  
collects the fluid and bubble degrees of freedom.  
The real part $\lambda_{r}$ determines exponential growth or decay,  
while $\lambda_{i}$ distinguishes stationary ($\lambda_{i}=0$) from oscillatory ($\lambda_{i}\neq0$) modes.  
The dominant eigenpairs $(\lambda,\widetilde{\boldsymbol{q}})$ are computed using an Arnoldi-based iterative solver applied to the discretised variational form.  
Convergence and validation are discussed in \hyperref[app:ALE_validation]{Appendix~\ref{app:ALE_validation}}.

Both the base-flow and linear stability analyses are performed with the open-source finite-element solver \textsc{FreeFem++} \citep{FreeFem}, where the present ALE framework is implemented by ourselves, following the methodology of \citet{bonnefis2024path}. Spatial discretisation employs Taylor–Hood $(P_{2},P_{1})$ elements for the velocity–pressure pair and quadratic $P_{2}$ elements for the mesh-extension field. The steady base state is obtained via Newton iteration, with linear systems solved by the sparse direct solver UMFPACK. The generalized eigenvalue problem arising from the linearised-ALE formulation is solved using a Krylov–Shur projection method available in SLEPc. Since the base flow is axi-symmetric, the problem is discretised on the meridional $xor$ plane centred on the LB. An unstructured Delaunay–Voronoi mesh, shown in \hyperref[flow-configuration]{figure~\ref{flow-configuration}$(c)$}, ensures adequate resolution near the bubble interfaces and in the wake. Besides the axi boundary at $\mathscr{S}_{axi}~(r=0)$, other boundary conditions are prescribed as follows: a uniform inflow $V_{x}\boldsymbol{e}_{x}$ at $\mathscr{S}_{in}~(x=-l_{1})$, a stress-free outlet at $\mathscr{S}_{out}~(x=l_{2})$, and an inviscid side boundary at $\mathscr{S}_{side}~(r=h)$. On $\mathscr{S}_{LB}$ and $\mathscr{S}_{TB}$, the shear-free and non-penetration conditions \eqref{BC} are imposed.

Accurate computation of the base flow and eigenvalues requires fine resolution near the interfaces and within the wake. Using $2R$ as characteristic length, the smallest grid spacing is $h_{\min} = 10^{-3}$ in the radial direction, corresponding to approximately $N_r \approx 20$ nodes across a viscous layer of thickness $5\Rey^{-1/2}$.  
Along the arc of each bubble, the number of surface nodes per diameter satisfies $N_{\theta} \approx 10\,\Rey^{1/2}$, with at least $N_{\theta}=200$ for $\Rey\le400$. The computational domain extends to $l_{1} = 120$, $l_{2} = 120$, and $h = 60$,  
substantially larger than that in \citet{bonnefis2024path}, ensuring negligible confinement effects.  For these settings, the relative variation of the non-axisymmetric eigenvalues ($|m|=1$) upon mesh refinement remains below $0.1\%$.  
Eigenvalues are computed using an Arnoldi-type Krylov subspace method (Krylov–Schur algorithm) combined with a shift-and-invert spectral transformation.
A relative residual tolerance of $10^{-8}$ is imposed, and the Krylov subspace dimension is automatically adapted during the iterations.
Under these settings, the growth rates and frequencies are found to be converged within $10^{-4}$ in nondimensional units.

All computations were performed under conditions consistent with the three-dimensional 
VOF–DNS provided by \citet{zhang2021three}, namely the corresponding isolated bubble remains rising vertically without performing path instability.  
The present analysis focuses on cases close to the neutral curves characererizing the stability transition, 
within $\Rey < 200$ and $\chi \leq 1.9$, ensuring that the base flow remains steady and free from wake unsteadiness.  
A detailed mesh-independence and validation study is provided in 
\hyperref[app:ALE_validation]{Appendix~\ref{app:ALE_validation}}, 
demonstrating that the solver accurately reproduces the neutral stability curves of 
a freely falling circular disk \citep{tchoufag2014globaldisk}, a freely rising isolated bubble \citep{tchoufag2014linearbubble},
and tandem-cylinder configurations \citep{tirri2023linear}, 
while also in excellent quantitative agreement with the numerical data \citep{hallez2011interaction,zhang2021three} 
for spherical in-line bubbles.


\section{Steady axi-symmetric base flow}
\label{sec:BaseFlow-section}

Before analysing the stability of the in-line configuration, we first characterize the steady, axi-symmetric base flow $\boldsymbol{\mathscr{Q}}_{0} = [\boldsymbol{U}_{0}(r, x),\, P_{0}(r, x)]$, which serves as the reference state for the linearised ALE formulation.  
The structure of this base flow largely determines both the onset and the nature of the unstable stationary and oscillatory modes identified in \hyperref[sec:LSA_stationary]{\S~\ref{sec:LSA_stationary}} and \hyperref[sec:LSA_oscillatory]{\S~\ref{sec:LSA_oscillatory}}, respectively.  
Two aspects are of primary importance:  
\emph{(a)} the vortical topology within the inter-bubble gap, whose evolution with the control parameters governs the emergence of the instability discussed later; and 
\emph{(b)} the equilibrium separation distance, which quantifies the strength of hydrodynamic coupling between the two bubbles.

\subsection{Flow structure}
\label{sec:flow-structure}

The topology of the flow in the inter-bubble gap plays an important role in the stability of the in-line configuration.  
In particular, the formation of a steady recirculating eddy in this region profoundly modifies the hydrodynamic coupling between the bubbles and provides the background flow from which the coupling instability emerges. Before turning to the linear analysis, we therefore characterize the structure and parameter dependence of this steady gap flow, which defines the base state of the system. For an isolated spheroidal bubble, \citet{dandy1986boundary} and \citet{blanco1995structure} showed that, beyond a critical oblateness ($\chi \gtrsim 1.65$), a steady standing eddy forms in the wake, marking the appearance of a closed recirculation region. This feature results from a finite-Reynolds-number balance between vorticity production at the shear-free interface and its downstream advection \citep{magnaudet2007wake}.  In the present in-line configuration, the threshold aspect ratio and Reynolds number for recirculation differ from the isolated case, since the confined gap flow is controlled by both $\Rey$ and the centre-to-centre distance $S$ (or equivalently the gap $\delta S$). 

\begin{figure}
    \begin{center}
        \setlength{\tabcolsep}{0pt}
        \iflocalcompile
        \begin{tabular}{@{\hskip 0pt}c@{\hskip 0pt}c@{\hskip -5mm}c@{\hskip 0pt}c@{\hskip 0pt}c@{\hskip 0pt}c@{\hskip 0pt}} 
                \resizebox{!}{0.35\textwidth}{%
                    \input{\figthree/chi1.3/chi-1.3-log-conneted.tex}
                } &
                
                \resizebox{0.14\textwidth}{!}{\raisebox{5cm}{\input{\figthree/chi1.3/chi1.3deltaS0.14Re30.tex}}} &
                \resizebox{0.14\textwidth}{!}{\raisebox{5cm}{\input{\figthree/chi1.3/chi1.3deltaS0.14Re100.tex}}} &
                \resizebox{0.14\textwidth}{!}{\raisebox{5cm}{\input{\figthree/chi1.3/chi1.3S1.2Re100.tex}}} &
                \resizebox{0.14\textwidth}{!}{\raisebox{5cm}{\input{\figthree/chi1.3/chi1.3S1.25Re100.tex}}}
        \end{tabular}
        \else
        \begin{tabular}{@{\hskip -5mm}c@{\hskip 0pt}c@{\hskip 0mm}c@{\hskip 0pt}c@{\hskip 0pt}c@{\hskip 0pt}c@{\hskip 0pt}} 
                \resizebox{!}{0.35\textwidth}{%
                    \includetikz{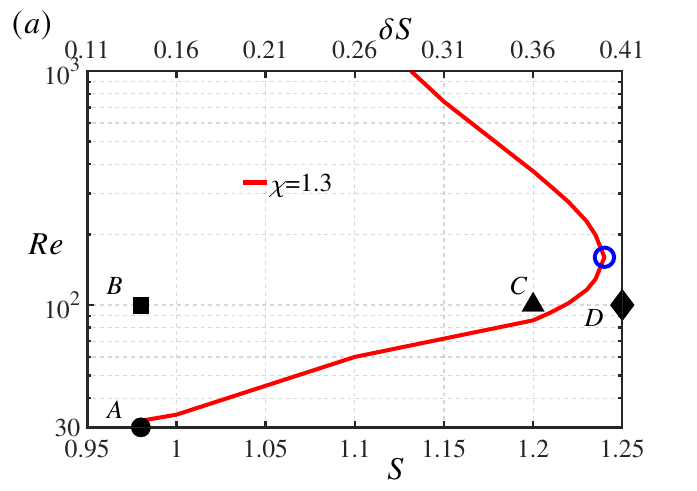}
                }&
                \raisebox{125pt}{\makebox[0pt][l]{\hspace{-10pt}{\fontsize{9pt}{11pt}\selectfont $(b)$ } }} & 
                \resizebox{0.14\textwidth}{!}{\raisebox{5cm}{\includetikz{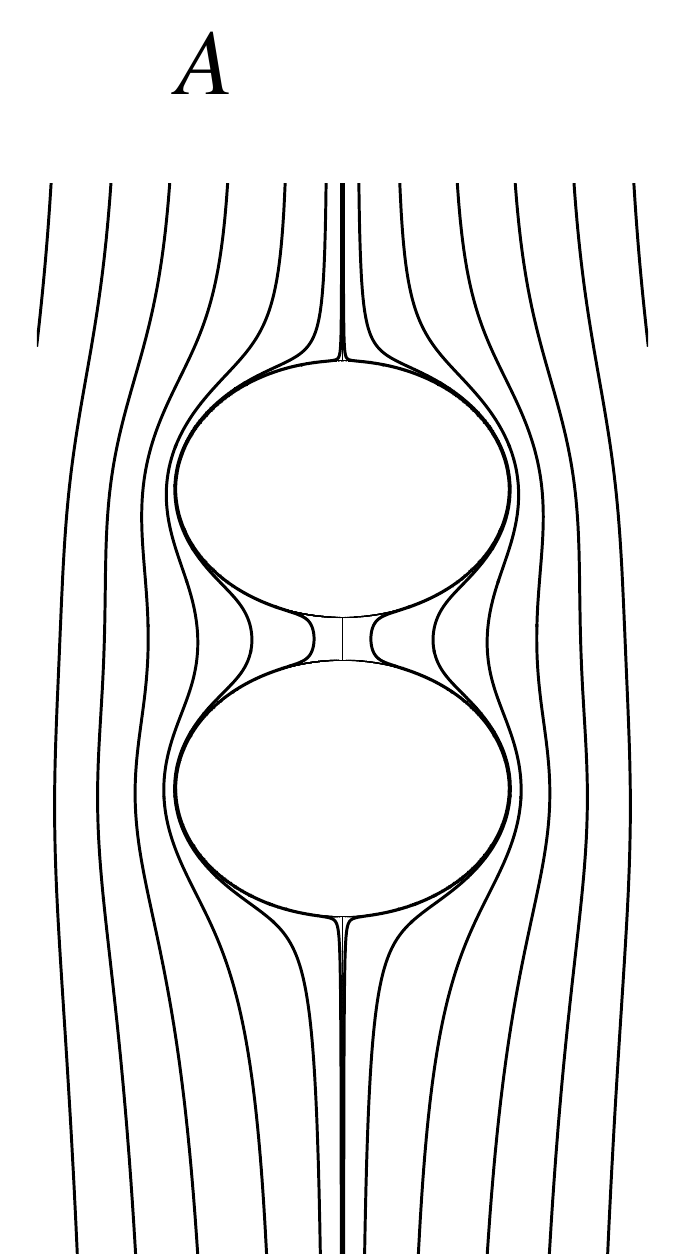}}} &
                \resizebox{0.14\textwidth}{!}{\raisebox{5cm}{\includetikz{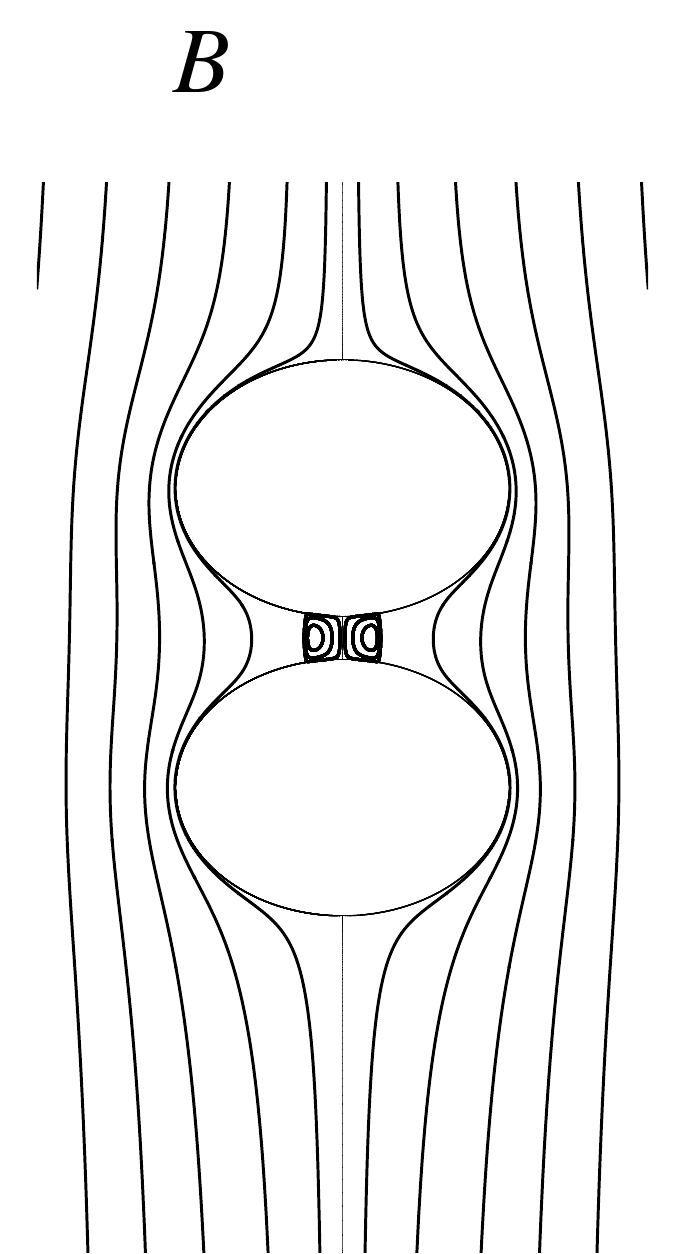}}} &
                \resizebox{0.14\textwidth}{!}{\raisebox{5cm}{\includetikz{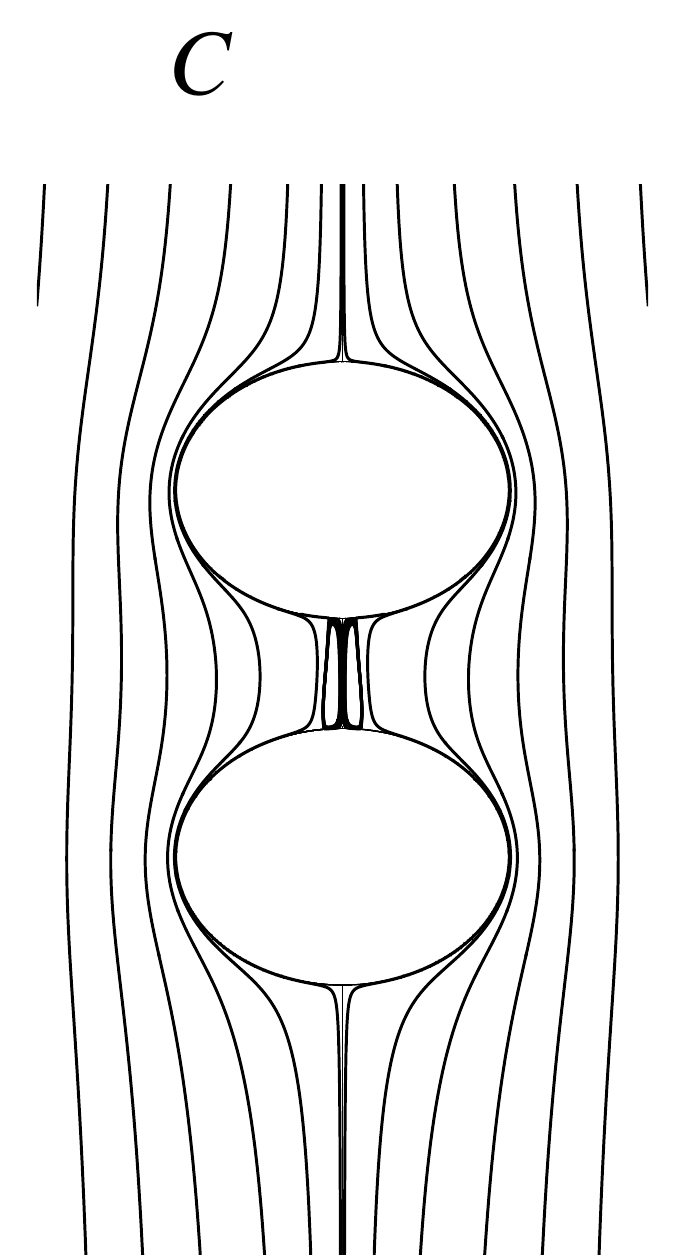}}} &
                \resizebox{0.14\textwidth}{!}{\raisebox{5cm}{\includetikz{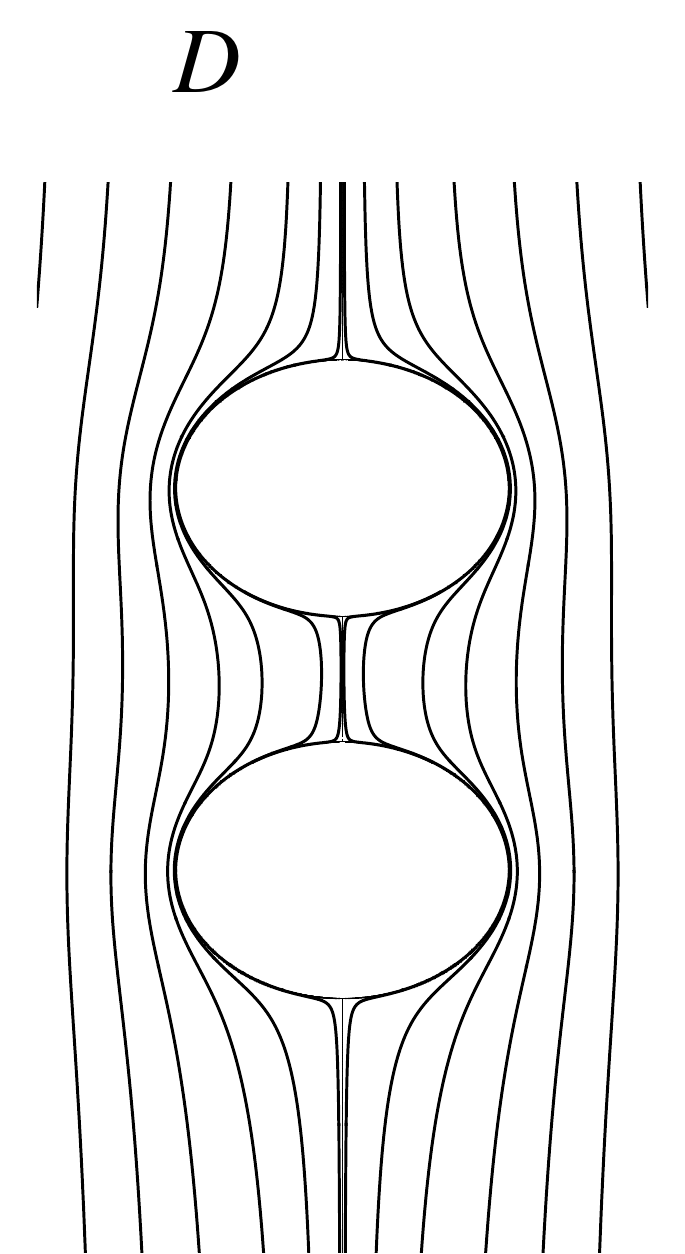}}}
        \end{tabular}
        \fi
    \end{center}
	\captionsetup{justification=justified, singlelinecheck=false, labelsep=period, width=\linewidth} 
	\caption{%
    Phase map associated with the formation of a standing eddy connecting the two bubbles for $\chi = 1.3$.
$(a)$ Critical curve in the $(S,Re)$ (or $(\delta S,Re)$) plane marking the onset of a connected recirculation between the two bubbles, open circle denotes the marginal configuration, corresponding to $Re \simeq 180$.
$(b)$ Streamlines illustrating representative flow structures for the cases indicated in $(a)$:
A $(\Rey, S) = (30, 0.98)$, B $(100, 0.98)$, C $(100, 1.20)$, and D $(100, 1.25)$. The connected eddy appears at points~B and~C, whereas points~A and~D correspond to the uni-direction flow structure.
}
\label{large_aspect_ratio-chi1.3}
\end{figure}

For moderately oblate bubbles ($\chi = 1.3$), \hyperref[large_aspect_ratio-chi1.3]{figure~\ref{large_aspect_ratio-chi1.3}$(a)$} maps the steady base flow topologies in the $(\Rey, S)$ plane, revealing two distinct regimes within the inter-bubble gap.  
Streamlines labelled \emph{A} and \emph{D} correspond to a monotonic, unidirectional flow without recirculation, whereas those labelled \emph{B} and \emph{C} display a continuous toroidal eddy connecting the wakes of the leading and trailing bubbles.  
This standing eddy arises solely from hydrodynamic interaction between the two bubbles, since an isolated bubble of the same aspect ratio does not exhibit wake recirculation.  
The transition between these regimes is identified by tracking the critical combinations of $(\Rey, S)$, or equivalently $(\Rey, \delta S)$, for which the connected recirculation first appears.  
For each fixed separation, the lower and upper branches of the transition curve, $\Rey_{\min}$ and $\Rey_{\max}$, are determined from the streamwise extent of the eddy.  
Below $\Rey_{\min}$, viscous diffusion prevents vorticity accumulation, while above $\Rey_{\max}$ the $\Rey^{-1/2}$ scaling of the interfacial vorticity flux reduces the circulation strength, suppressing the inter-bubble connection.  
The open symbol in the figure marks the marginal configuration, which occurs at $\Rey \approx 180$.

\begin{figure}
    \begin{center}
        \setlength{\tabcolsep}{0pt}
        \iflocalcompile
        \begin{tabular}{@{\hskip 0pt}c@{\hskip -5mm}c@{\hskip 0pt}c@{\hskip 0pt}c@{\hskip 0pt}c@{\hskip 0pt}} 
            \resizebox{!}{0.35\textwidth}{%
                \input{\figthree/chi1.9/chi-1.9-log-conneted.tex}
            } &
            \resizebox{0.14\textwidth}{!}{\raisebox{5cm}{\input{\figthree/chi1.9/chi1.9S2Re100.tex}}} &
            \resizebox{0.14\textwidth}{!}{\raisebox{5cm}{\input{\figthree/chi1.9/chi1.9S2.5Re100.tex}}} &
            \resizebox{0.14\textwidth}{!}{\raisebox{5cm}{\input{\figthree/chi1.9/chi1.9S2.5Re50.tex}}} &
            \resizebox{0.14\textwidth}{!}{\raisebox{5cm}{\input{\figthree/chi1.9/chi1.9S2.5Re200.tex}}}
        \end{tabular}
        \else
        \begin{tabular}{@{\hskip -5mm}c@{\hskip -0mm}c@{\hskip -0mm}c@{\hskip 0pt}c@{\hskip 0pt}c@{\hskip 0pt}c@{\hskip 0pt}} 
            \resizebox{!}{0.35\textwidth}{%
                \includetikz{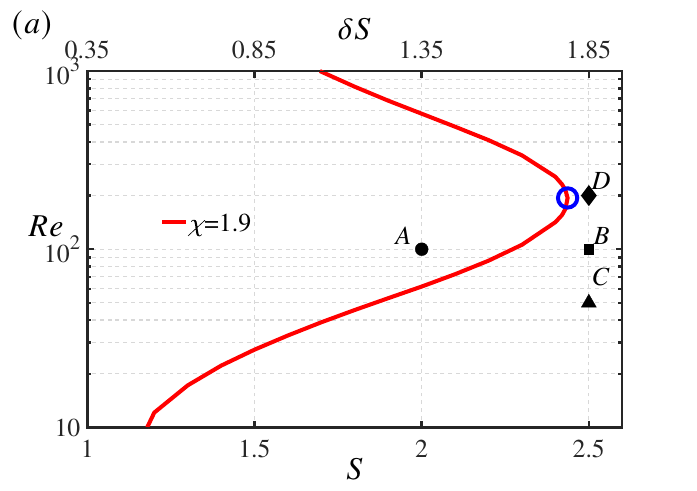}
            }&
            \raisebox{125pt}{\makebox[0pt][l]{\hspace{-10pt}{\fontsize{9pt}{11pt}\selectfont $(b)$ } }} & 
            \resizebox{0.14\textwidth}{!}{\raisebox{5cm}{\includetikz{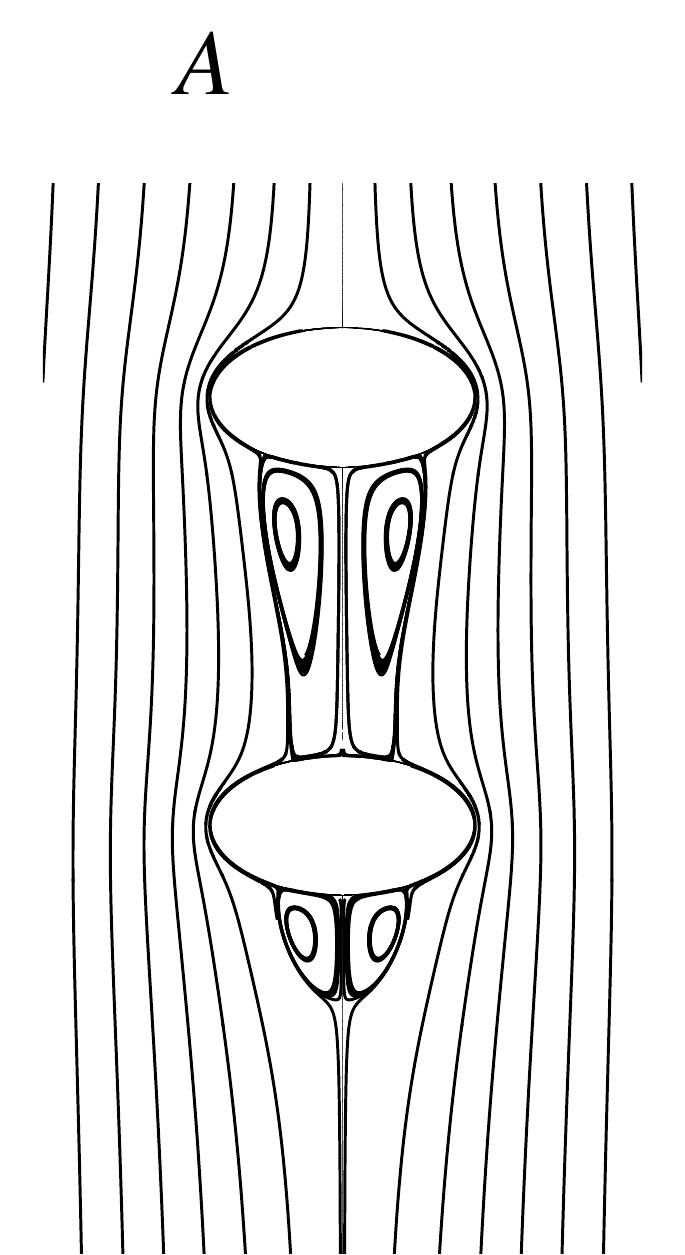}}} &
            \resizebox{0.14\textwidth}{!}{\raisebox{5cm}{\includetikz{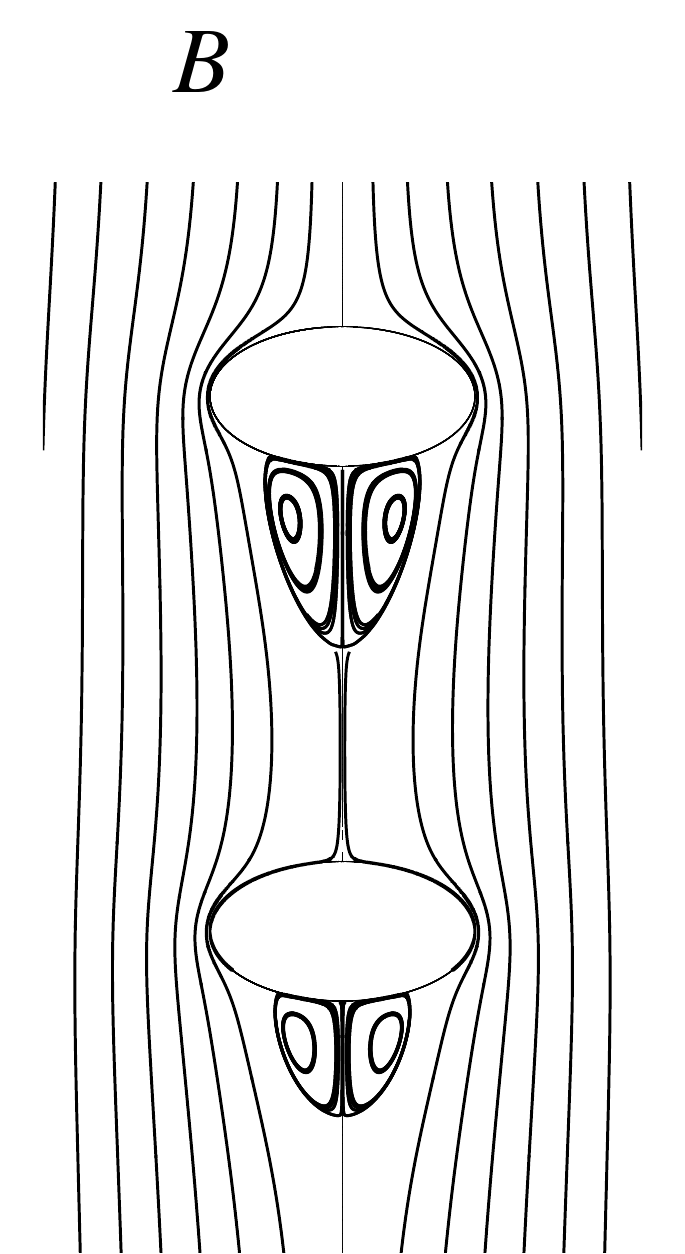}}} &
            \resizebox{0.14\textwidth}{!}{\raisebox{5cm}{\includetikz{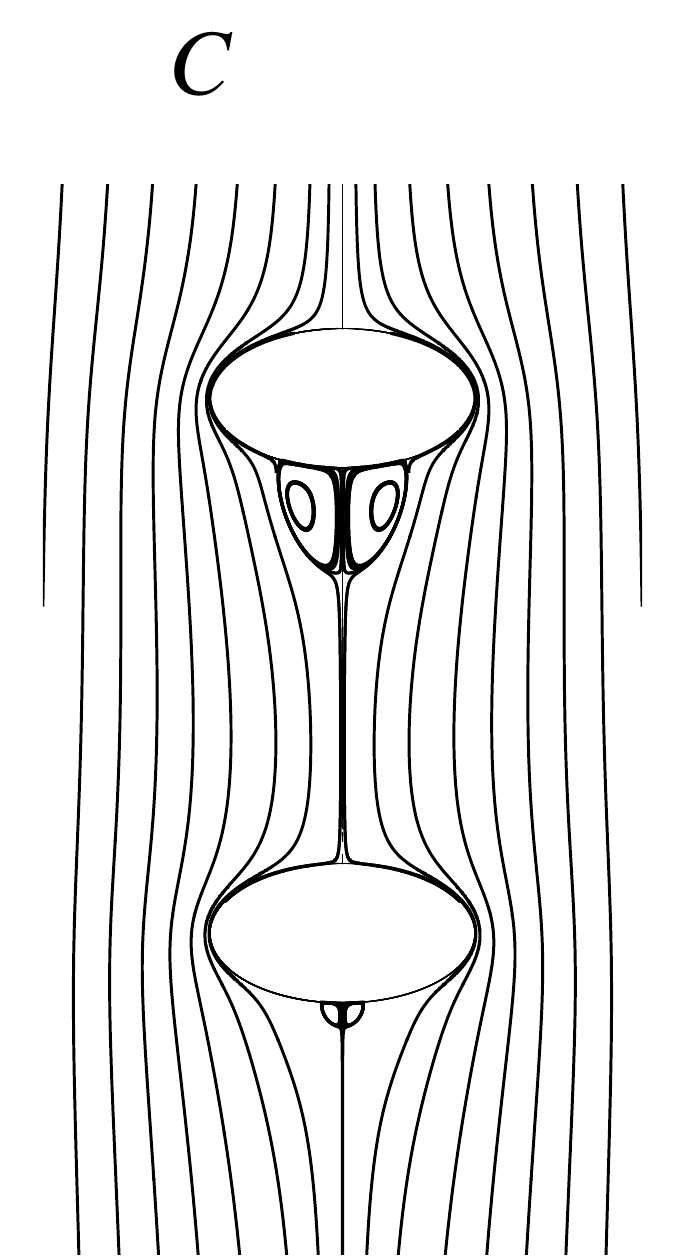}}} &
            \resizebox{0.14\textwidth}{!}{\raisebox{5cm}{\includetikz{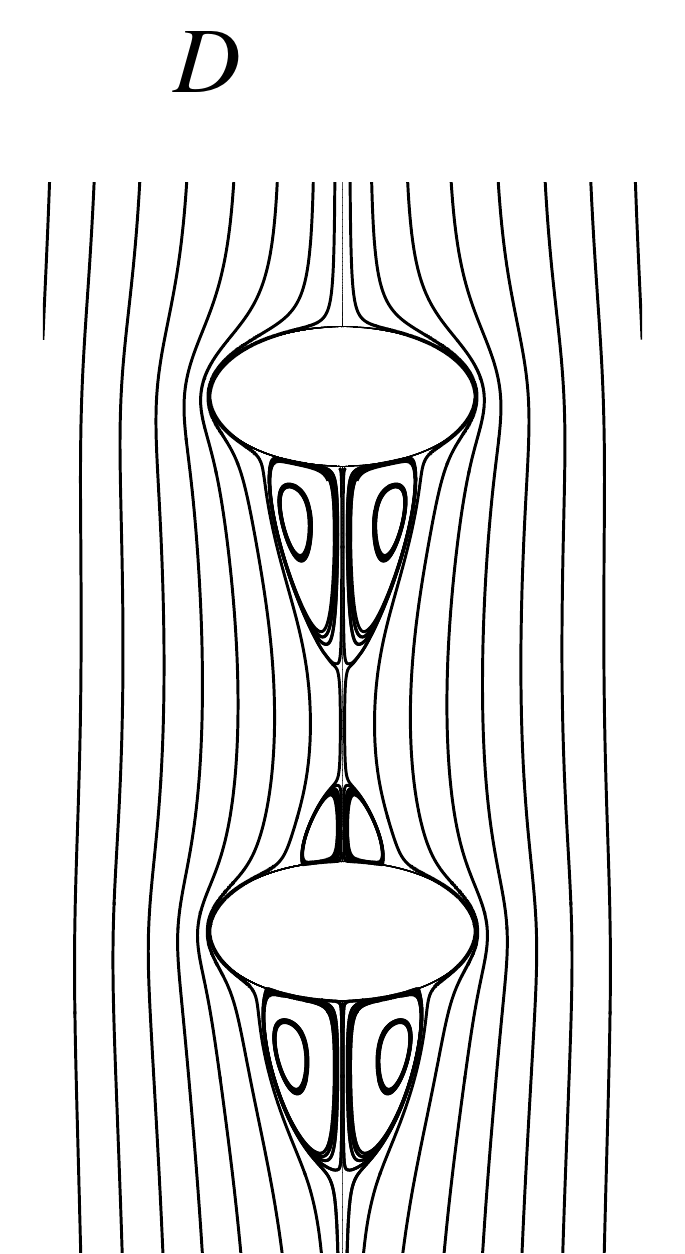}}}
        \end{tabular}
        \fi
    \end{center}
	\captionsetup{justification=justified, singlelinecheck=false, labelsep=period, width=\linewidth} 
	\caption{%
	Same as \hyperref[large_aspect_ratio-chi1.3]{figure~\ref{large_aspect_ratio-chi1.3}}, but for a more oblate shape with $\chi = 1.9$.
$(a)$ Critical curve in the $(S, Re)$ (or $(\delta S,Re)$) plane marking the onset of a connected recirculation between the two bubbles.
$(b)$ Streamlines illustrating the corresponding flow structures for the cases indicated:
A $(Re, S) = (100, 2.0)$, B $(100, 2.5)$, C $(50, 2.5)$, and D $(200, 2.5)$. The connected eddy appears at point~A, whereas points~B--D correspond to the detached eddy structure.
	}
	\label{large_aspect_ratio-chi1.9}
\end{figure}

For more oblate bubbles ($\chi = 1.9$), 
\hyperref[large_aspect_ratio-chi1.9]{figure~\ref{large_aspect_ratio-chi1.9}$(a)$} reveals a third flow topology in which a detached recirculating region forms between the two bubbles, as indicated by streamlines \emph{B}, \emph{C}, and \emph{D}.  
In addition, the parameter range supporting a connected recirculation widens markedly.  
The critical separation extends to $S \!\simeq\! 2.4$ (or $\delta S \!\simeq\! 1.75$), and the upper limit $\Rey_{\max}$ shifts to higher values.  
This trend reflects the $\chi^{7/2}\Rey^{-1/2}$ scaling of the surface vorticity flux: as $\chi$ increases, enhanced vorticity generation must be balanced by larger $\Rey$ to sustain comparable advection, thereby allowing the standing eddy to persist over a broader parameter range.  
The flow fields in \hyperref[large_aspect_ratio-chi1.9]{figure~\ref{large_aspect_ratio-chi1.9}$(b)$} confirm that, outside this finite $(\Rey,S)$ window bounded by the critical curves, the inter-bubble flow reverts to a detached-eddy topology rather than a unidirectional stream.  
A small upstream recirculation occasionally appears near the nose of the trailing bubble, resulting from vorticity convected from the leading-bubble shear layer and temporarily trapped ahead of the TB.

We do not attempt to present the complete set of critical curves for all aspect ratios, as the comparison between $\chi = 1.3$ and $\chi = 1.9$ already demonstrates that increasing oblateness enhances vorticity production and promotes the formation of a steady standing eddy within the inter-bubble gap.  
This connected recirculation, however, is not a passive wake feature:  it constitutes a closed conduit for momentum and vorticity exchange between the two bubbles, acting as a hydrodynamic spring that dynamically couples their motions.  
As will be shown in \hyperref[sec:LSA_stationary]{\S~\ref{sec:LSA_stationary}}, this coupling strongly influences the stability of the DKT-type stationary mode, while in \hyperref[sec:LSA_oscillatory]{\S~\ref{sec:LSA_oscillatory}} it underpins the oscillatory instability, where the periodic deformation of the gap flow sustains a self-excited lateral oscillation of the trailing bubble.

\subsection{Equilibrium distance}
\label{sec:equilibrium-distance}

As both bubbles rise at the same velocity $V_{x}$, and an equilibrium state is expected when the drag force acting on each bubble becomes identical.  
This defines an equilibrium configuration, characterised by the dimensionless centre-to-centre distance $S_{e}$, at which both bubbles ascend in the same velocity in tandem.  
In \hyperref[app:ALE_validation]{Appendix~\ref{app:ALE_validation}}, we examine how $S_{e}$ varies with the Reynolds number $\Rey$ for spherical bubble pair. It is shown there that the present results on spherical bubbles recover faithfully the DNS data of \citet{hallez2011interaction} and \citet{zhang2021three}, confirming the accuracy of the Newton iteration method and UMFPACK library in predicting the base flow of in-line configuration.

\begin{figure}
	\centering
	\setlength{\tabcolsep}{0pt} 
	\iflocalcompile
	\begin{tabular}{cc}
		\resizebox{0.5\textwidth}{!}{\input{\figthree/equilibrium_separation_distance_2D-Reconstant.tex}} &
		\resizebox{0.5\textwidth}{!}{\input{\figthree/equilibrium_separation_distance_2D-deltas.tex}}
	\end{tabular}
	\else
	\begin{tabular}{cc}
		\resizebox{0.5\textwidth}{!}{\includetikz{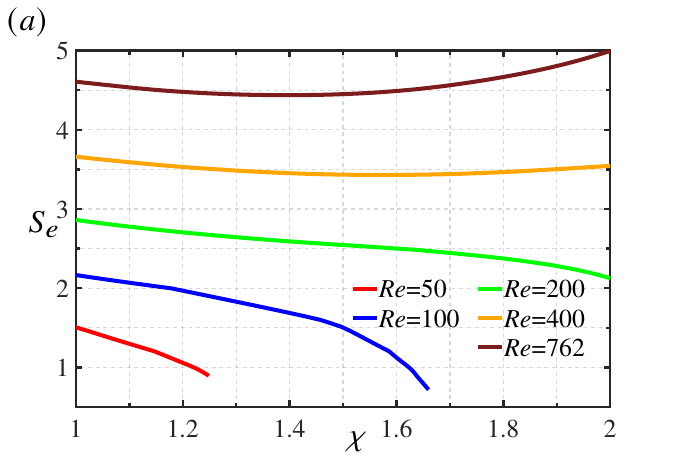}} &
		\resizebox{0.5\textwidth}{!}{\includetikz{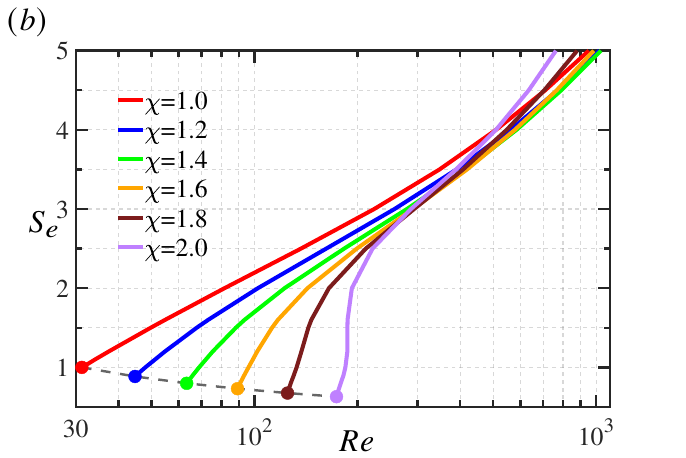}}
	\end{tabular}
	\fi	
	\captionsetup{justification=justified, singlelinecheck=false, labelsep=period, width=\linewidth} 
	\caption{Equilibrium centre-to-centre distance $S_e$ of the bubble pair in the steady, axi-symmetric base state, as a function of the bubble aspect ratio ($\chi$) and  Reynolds number ($\Rey$).
$(a)$ Variation of $S_e$ with $\chi$ for different Reynolds numbers.
$(b)$ Variation of $S_e$ with $\Rey$ for different aspect ratios.
Closed symbols in $(b)$ mark the minimum $S_e$ at which the two bubbles come into contact, corresponding to $S_e = S_{\min} = \chi^{-2/3}$ ($\delta S = 0$).}
	\label{equilibrium-distance}
\end{figure}

For non-spherical bubbles, \hyperref[equilibrium-distance]{figure~\ref{equilibrium-distance}$(a)$} presents the variation of the equilibrium separation $S_{e}$ with the aspect ratio $\chi$ for several Reynolds numbers in the range $50 \leq \Rey < 800$.  
At moderate Reynolds numbers ($50 \lesssim \Rey \lesssim 200$), $S_{e}$ decreases monotonically with $\chi$, in good agreement with the trends reported by \citet{zhang2021three}. This behaviour reflects the intensification of the surface vorticity flux as the bubbles become more oblate, 
since the flux entering the liquid scales as $\chi^{7/2}\Rey^{-1/2}$ \citep{magnaudet2007wake}.  
The resulting enhancement of vorticity within the inter-bubble gap amplifies the wake entrainment effect and lowers the local pressure, thereby increasing the suction acting on the TB and reducing the equilibrium distance $S_{e}$.

At higher Reynolds numbers ($400 \lesssim \Rey < 800$), however, $S_{e}$ exhibits a non-monotonic dependence on $\chi$: 
it first decreases, then rises again beyond $\chi \approx 1.5$.  
This reversal, not captured in previous VOF–DNS studies limited to $\Rey < 300$ \citep{zhang2021three}, marks a qualitative change in the wake interaction.  
At moderate $\Rey$, the low-pressure region forming in the wake of the LB dominates, drawing the TB closer and decreasing $S_{e}$.  
As $\Rey$ increases, the wake elongates and develops a broad, quasi-steady recirculation region, as already discussed in \hyperref[sec:flow-structure]{\S~\ref{sec:flow-structure}},   
For sufficiently oblate bubbles, the resulting vortex establishes an upstream stagnation surface that impinges on the front of the TB, raising the local pressure and weakening the suction effect.  
Interaction between the TB and the outer boundary of the LB’s recirculation region thus introduces a partial flow blockage, leading to a recovery of $S_{e}$ at large $\chi$.  
Hence, the non-monotonic variation of $S_{e}$ originates from the competition between wake-induced attraction and vortex-induced repulsion, 
the balance of which depends on both $\Rey$ and bubble oblateness.  
At fixed $\chi$, $S_{e}$ increases monotonically with $\Rey$, indicating that potential-flow effects gradually dominate over viscous entrainment, 
thereby promoting hydrodynamic repulsion between the bubbles.  
These combined trends are summarised in \hyperref[equilibrium-distance]{figure~\ref{equilibrium-distance}$(b)$}, which maps $S_{e}$ as a function of $\Rey$ for $1.0 \leq \chi \leq 2.0$.  
For $\Rey \lesssim 300$, $S_{e}$ decreases with $\chi$, whereas the opposite trend occurs for $\Rey \gtrsim 300$.  
When $S_{e} < \chi^{-2/3}$, the two bubbles are in contact and the equilibrium state ceases to exist, as indicated by the closed symbols on each curve.  
For spherical bubbles ($\chi = 1$), this limiting condition occurs at $\Rey \approx 31$, whereas for $\chi = 2$ it is delayed to $\Rey \approx 174$, highlighting the strong stabilizing influence of bubble oblateness on vertical alignment.

In summary, the equilibrium distance $S_{e}$ defines the steady reference configuration of the system. As will be shown in \hyperref[sec:LSA_stationary]{\S~\ref{sec:LSA_stationary}}, instability consistently develops at separations larger than $S_{e}$,  confirming that this equilibrium state represents the lower bound of the stable in-line branch.

\section{Stationary mode}
\label{sec:LSA_stationary}

As introduced in \hyperref[sec:Problem]{\S~\ref{sec:Problem}}, the instability of an in-line bubble pair manifests through the amplification of an infinitesimal angular displacement $\Psi$ from the vertical axis.  
This deviation reflects the action of a positive (i.e.\ destabilizing) lateral lift that drives the trailing bubble away from the wake of the leading bubble.  
The lift force acting on the TB, denoted by $F_{L}^{TB}$, may arise either from an intrinsic flow instability developing in its near wake or from the hydrodynamic coupling with the LB.  
Since the present study aims to elucidate how inter-bubble interactions affect the stability of the configuration and how this coupling depends on shape deformation, the parameter space must be chosen carefully to exclude the onset of flow (or, wake) instability.

For an isolated bubble, once the bubble exceeds a critical oblateness $\chi_{c}^{iso} \simeq 2.21$ \citep{magnaudet2007wake,ern2012wake}, flow instability originates from the emergence of a pair of counter-rotating streamwise vortices in the near wake, forming the characteristic double-threaded structure identified by \citet{mougin2001path}.  
This instability provides the lift responsible for lateral motion of the bubble.  
When the TB rises within the wake of another bubble, however, the shear imposed by the LB modifies the vorticity flux at the TB surface and effectively lowers the critical aspect ratio for flow destabilization, i.e.\ $\chi_{c}^{pair} < \chi_{c}^{iso}$ \citep{takagi1994drag}.  
Auxiliary computations presented in \hyperref[app:wake]{Appendix~\ref{app:wake}} confirm that $\chi_{c}^{pair} > 1.9$, indicating that for $\chi \leq 1.9$ the wake of the bubbles in the pair remain linearly stable.  

Accordingly, the present analysis is restricted to $\chi \leq 1.9$, ensuring that the instability discussed here does not originate from a local wake mechanism but rather from a global hydrodynamic coupling between the two bubbles.  
Further details of this verification are provided in \hyperref[app:wake]{Appendix~\ref{app:wake}} and are not repeated here. 
We now proceed to examine how bubble oblateness modifies this coupling and, consequently, the stability of the in-line configuration.

\subsection{Comparison with DNS results}
\label{sec:comparison_dns}

Before discussing the full ALE–LSA results, it is useful to recall the current understanding of the rising stability of an in-line bubble pair, after excluding the influence of local wake instabilities.  
For spherical, shear-free bubbles, both experiments and simulations consistently show that the in-line configuration is unstable in the inertial regime 
\citep{kok1993dynamics,harper1970bubbles,cartellier2001bubble,figueroa2005clustering,yin2008lattice}, represented by the lateral escape of the TB from the symmetry axis.  
To identify the physical origin of this destabilizing behaviour, \citet{hallez2011interaction} decomposed the lateral lift coefficient of the TB, 
defined as $C_{L}^{TB} = 2F_{L}^{TB}/(\pi \rho V_{x}^{2} R^{2})$, into three distinct contributions:
\begin{equation}
C_{L}^{TB} = C_{L}^{pot} + C_{L}^{vis} + C_{L}^{shear},
\label{eq:lift_decomposition}
\end{equation}
where the terms respectively represent the potential, viscous, and shear-induced components.  
The discussion below considers their relative roles in the limit $\Psi \rightarrow 0^{+}$, corresponding to an infinitesimal lateral perturbation of the in-line equilibrium.  
The potential contribution $C_{L}^{pot}$ arises from irrotational coupling between the two bubbles and is always positive, tending to drive the TB away from the symmetry axis.  
The viscous term, $C_{L}^{vis}\! \sim\! \mathcal{O}(\Rey^{-1} S^{-4}) f(\Psi)$, corresponds to the wake-entrainment effect, which draws the TB toward the axis and therefore acts as a negative, stabilizing lift.  
The shear-induced term $C_{L}^{shear}$ becomes dominant when the TB is partially immersed in the velocity deficit of the LB wake, producing a destabilizing lift that pushes it outward.  
All three contributions vanish for perfect alignment ($\Psi = 0$) and compete at finite offsets, their sum determining the direction of the net lift.  
Within the parameter range $20 \le \Rey \le 500$ and $1.25 \le S \le 5$,  \citet{hallez2011interaction} always found $C_{L}^{TB} > 0$ as $\Psi \rightarrow 0^{+}$, confirming that a spherical TB invariably experiences a destabilizing lift in the inertial regime, a crucial benchmark for the present analysis.

Extending this framework to spheroidal bubbles introduces two additional effects.  
First, bubble anisotropy modifies both the generation and transport of vorticity in the surrounding liquid.  
The maximum surface vorticity scales as $\chi^{8/3}$, while the associated flux scales as $\chi^{7/2}$ \citep{magnaudet2007wake}, indicating that flatter bubbles produce stronger vorticity and broader wakes.  
This intensification alters all three components in (\ref{eq:lift_decomposition}), leading to a non-trivial dependence of the net lift on the aspect ratio.  
Second, and more importantly, a spheroidal bubble drifting laterally is subjected to a hydrodynamic torque arising from the asymmetric shear imposed by the LB wake.  
This torque induces a small inclination $\Theta$ relative to the oncoming flow (see \hyperref[flow-configuration]{figure~\ref{flow-configuration}$b$}), introducing an additional inclination-induced lift on both bubbles.  
This contribution, absent for spherical bubbles, constitutes an effective fourth term in the lift decomposition (\ref{eq:lift_decomposition}).  

Previous experimental \citep{sanada2006interaction,sanada2009motion,duineveld1997bouncing} and numerical investigations \citep{gumulya2017interaction,zhang2021three,zhang2022three,kusuno2019lift,kusuno2021wake} have consistently reported that increasing the bubble aspect ratio $\chi$ enhances the stability of the in-line configuration.  
This trend has generally been interpreted as evidence that deformation amplifies the wake-entrainment effect, thereby increasing the magnitude of $C_{L}^{vis}$ until it overcomes the destabilizing contributions and reverses the sign of the total lift $C_{L}^{TB}$.  
Although the shear-induced lift $C_{L}^{shear}$ is also known to vary, or even reverse the sign, with bubble deformation \citep{adoua2009reversal,zhang2025lift}, \citet{zhang2021three} demonstrated that this variation alone is insufficient to explain the observed lift reversal on the TB.  
One of the main objectives of the present analysis is therefore to clarify the physical origin of the stabilization observed for increasingly oblate bubble pairs.  
Specifically, we aim to determine whether the enhanced stability results from a genuine reversal of the deformation-augmented wake-entrainment mechanism, or whether it originates from a distinct hydrodynamic process, namely, an inclination-induced restoring lift arising from the coupling between bubble orientation and the wake shear.

\begin{figure}
	\centering
	\setlength{\tabcolsep}{0pt} 
	\iflocalcompile
    \begin{tabular}{@{\hspace{-10pt}}c@{\hspace{-2pt}}c@{}c@{}}
		\resizebox{0.23\textwidth}{!}{\raisebox{10pt}{\input{\figfourtwo/fig-translation.tex}}} &
		\resizebox{0.42\textwidth}{!}{\input{\figfourtwo/neutralline-norotation-new.tex}}&
        \resizebox{0.42\textwidth}{!}{\input{\figfourtwo/chi1.6S3DdifferentRe.tex}}
	\end{tabular}
	\else
	\begin{tabular}{@{\hspace{-10pt}}c@{\hspace{-2pt}}c@{}c@{}}
		\resizebox{0.23\textwidth}{!}{\raisebox{10pt}{\includetikz{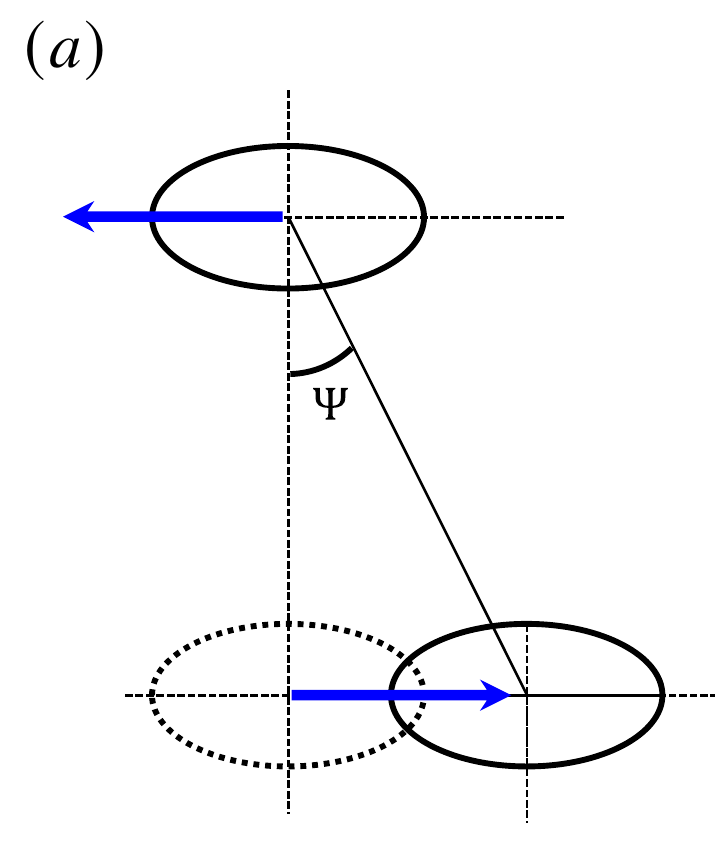}}} &
		\resizebox{0.42\textwidth}{!}{\includetikz{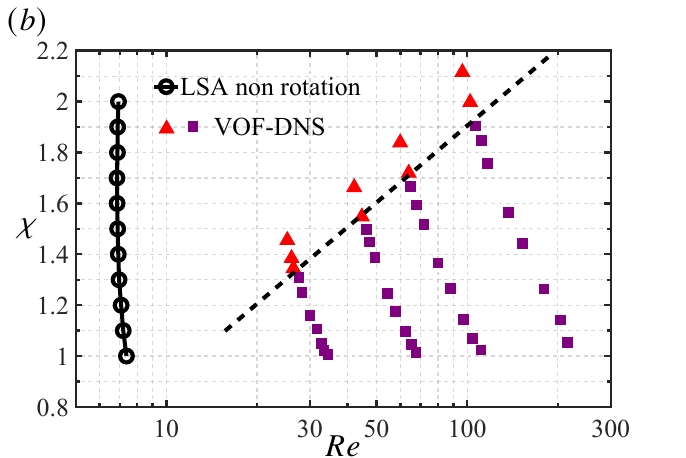}} &
        \resizebox{0.42\textwidth}{!}{\includetikz{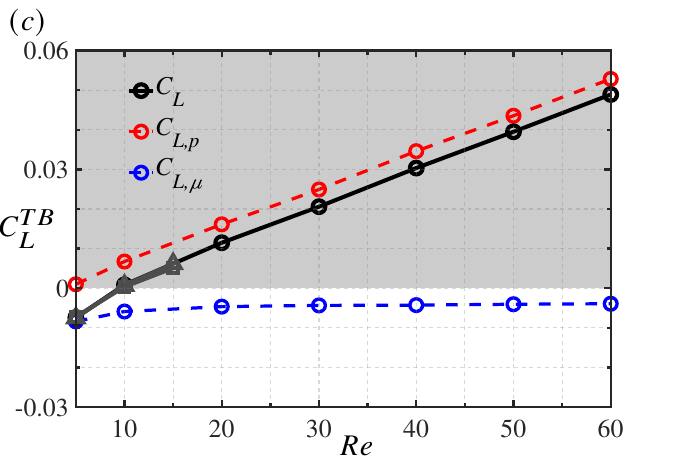}}
	\end{tabular}
	\fi	
	\captionsetup{justification=justified, singlelinecheck=false, labelsep=period, width=\linewidth} 
\caption{
Comparison between the present ALE–LSA results with rotational degrees of freedom suppressed, the VOF–DNS data of \citet{zhang2021three}, and the supporting results obtained from the EBM–DNS framework.  
$(a)$~Schematic of the ALE–LSA configuration. Blue arrows indicate that only lateral (transverse) translational motion is permitted, while rotations are constrained.  
$(b)$~Neutral curve of the stationary mode obtained from the non-rotating ALE–LSA (\protect\tikz\protect\draw[black, thick] (0,0) circle (3pt);) compared with the VOF–DNS results of \citet{zhang2021three} (their figure~21).  
The dashed line marks the transition boundary between the stable 
(\protect\tikz\protect\filldraw[red] (0,0) -- (6pt,0) -- (3pt,5pt) -- cycle;) and unstable 
(\protect\tikz\protect\filldraw[fill={rgb,255:red,128;green,0;blue,128}, draw={rgb,255:red,128;green,0;blue,128}] (0,0) -- (4pt,0) -- (4pt,4pt) -- (0,4pt) -- cycle;) 
regimes observed in the VOF–DNS.  
Symbols: 
\protect\tikz\protect\draw[black, thick] (0,0) circle (3pt);~neutral curve from the non-rotating ALE–LSA;  
\protect\tikz\protect\filldraw[red] (0,0) -- (6pt,0) -- (3pt,5pt) -- cycle;~stable bubble coalescence;  
\protect\tikz\protect\filldraw[fill={rgb,255:red,128;green,0;blue,128}, draw={rgb,255:red,128;green,0;blue,128}] (0,0) -- (4pt,0) -- (4pt,4pt) -- (0,4pt) -- cycle;~unstable lateral escape of the trailing bubble (TB).  
$(c)$~Lift force experienced by the TB computed using the EBM–DNS framework, showing the total ($C_L$), pressure ($C_{L, p}$), and viscous ($C_{L, \mu}$) lift coefficients as functions of $\Rey$ for $(\chi, S, S_r) = (1.6, 3, 0.1)$, corresponding to a horizontally aligned LB and TB ($\Theta_{LB} = \Theta_{TB} = 0^{\circ}$).  
Additional results for bubble pairs with $\chi = 1.0$ (open square) and $1.5$ (open triangle) are included for comparison.
}
\label{fig:LSA_comparison_no_rotation}
\end{figure}

To identify the dominant stabilizing mechanism associated with bubble oblateness, the present LSA employs a variable-separation strategy 
and systematically compares the results with the VOF–DNS data of \citet{zhang2021three}.  
In the first configuration, both bubbles are allowed to translate laterally while their rotational degrees of freedom are artificially constrained, therefore freezing $\Theta_{LB} = \Theta_{TB} = 0$.  
The corresponding results are shown in \hyperref[fig:LSA_comparison_no_rotation]{figure~\ref{fig:LSA_comparison_no_rotation}$(a)$--$(b)$}.  
Panel~$(a)$ sketches the configuration: two bubbles separated by a fixed distance $S = 4$ are free to move transversely but cannot rotate.  
Panel~$(b)$ displays the neutral curve (black line with open circles) of the most amplified azimuthal mode, $|m| = 1$, 
computed from the bubble state vector $\mathscr{Q}^{b} = [\boldsymbol{X}_{LB}, \boldsymbol{V}_{LB}, \boldsymbol{X}_{TB}, \boldsymbol{V}_{TB}]^{b}$.  
At these moderate Reynolds numbers, only a stationary mode ($\lambda_{i} = 0$) is detected, while no oscillatory branch ($\lambda_{i} \neq 0$) emerges.  
For comparison, the VOF–DNS data of \citet{zhang2021three}, marking the stable and unstable in-line regimes, are superimposed as red and purple symbols, respectively, and the dashed line interpolates the DNS-derived transition boundary.  
A pronounced discrepancy is observed.  
In the DNS, the critical Reynolds number increases sharply with the aspect ratio, from $Re_{c} \!\approx\! 30$ at $\chi \!=\! 1.3$ to $Re_{c} \!\approx\! 100$ at $\chi \!=\! 1.9$, confirming that increasing oblateness enhances the stability of the pair.  
In contrast, the non-rotating ALE–LSA predicts a nearly constant and much lower threshold ($Re_{c}\!\approx\! 7$), almost independent of $\chi$.  
This severe underestimation demonstrates that shape deformation alone does not stabilize the in-line configuration; instead, it reveals that a crucial mechanism is missing when the rotational degrees of freedom are suppressed.  

This conclusion is independently supported by dedicated simulations performed with the recently developed three-dimensional EBM solver \citep{zhang2025lift,WEI2026114632}.  
This solver directly resolves the surrounding flow field on Cartesian grids, allowing accurate evaluation of the hydrodynamic forces and torques acting on prescribed, fixed-shape bubbles.  
As outlined in \hyperref[app:EBM]{Appendix~\ref{app:EBM}}, the EBM formulation implicitly enforces shear-free interfacial conditions and allows the bubble shape and orientation to be specified independently.  
Here, two oblate bubbles of aspect ratio $\chi = 1.6$ are placed in-line at a centre-to-centre distance $S = 3$, with the TB slightly offset laterally by $S_{r} = 0.1$ (corresponding to an inclination angle $\Psi \simeq 2^{\circ}$).  
Both bubbles remain horizontal ($\Theta_{LB} = \Theta_{TB} = 0$), so that self-inclination is suppressed. \hyperref[fig:LSA_comparison_no_rotation]{Figure~\ref{fig:LSA_comparison_no_rotation}$(c)$} shows that the lift coefficient of the TB, $C_{L}^{TB}$, increases monotonically with $\Rey$ and becomes positive for $\Rey_{c} \!\approx\! 10$, corresponding to the transition from stabilizing to destabilizing lift.  
Additional computations for $\chi = 1.0$ and $1.5$ over the range $5 \le \Rey \le 15$ collapse onto the same critical value $\Rey_{c} \!\approx\! 10$, 
confirming that bubble oblateness alone does not modify the instability threshold.  
This finding fully supports the non-rotating ALE–LSA predictions shown in panel~$(b)$, thereby isolating the rotational coupling as the missing stabilizing ingredient.

\begin{figure}
	\centering
	\setlength{\tabcolsep}{0pt} 
	\iflocalcompile
	\begin{tabular}{@{\hspace{-10pt}}c@{\hspace{-2pt}}c@{}c@{}}
        \resizebox{0.23\textwidth}{!}{\raisebox{20pt}{\input{\figfourtwo/fig-rotation.tex}}} &
		\resizebox{0.42\textwidth}{!}{\input{\figfourtwo/neutralline-rotation-new.tex}} &
        \resizebox{0.42\textwidth}{!}{\input{\figfourtwo/chi1.6S3DSr0.1Re60differentdegree.tex}}
	\end{tabular}
	\else
	\begin{tabular}{@{\hspace{-10pt}}c@{\hspace{-2pt}}c@{}c@{}}
		\resizebox{0.23\textwidth}{!}{\raisebox{20pt}{\includetikz{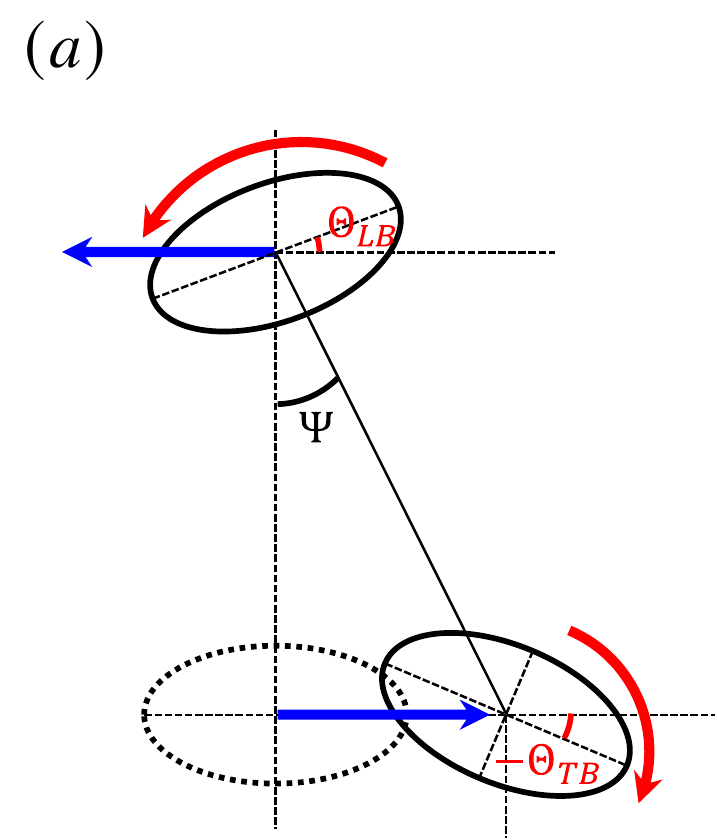}}} &
		\resizebox{0.42\textwidth}{!}{\includetikz{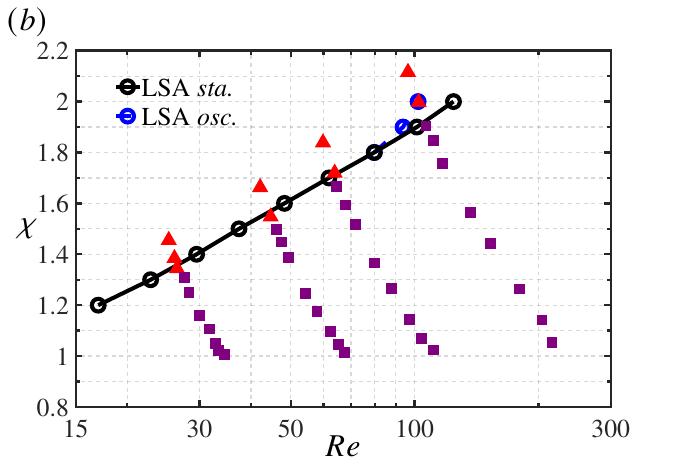}} &
        \resizebox{0.42\textwidth}{!}{\includetikz{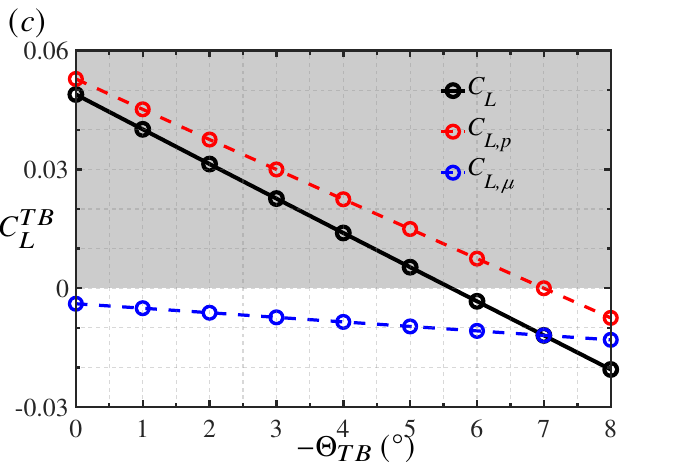}}
	\end{tabular}
	\fi	
	\captionsetup{justification=justified, singlelinecheck=false, labelsep=period, width=\linewidth} 
	\caption{Same as \hyperref[fig:LSA_comparison_no_rotation]{figure~\ref{fig:LSA_comparison_no_rotation}}, but with the ALE–LSA computations now including rotational degrees of freedom.
In panel~$(b)$, the additional blue symbols (\protect\tikz\protect\draw[blue, thick] (0,0) circle (3pt);) denote the oscillatory mode identified in the ALE–LSA spectrum. Panel $(c)$ is still obtained from the EBM solver, showing the variation of the lift components with the inclination angle of TB for $(\chi, S, S_r, \Rey) = (1.6, 3, 0.1, 60)$.
	}
	\label{fig:LSA_comparison_with_rotation}
\end{figure}

To approach a more realistic configuration, the second series of ALE–LSA computations allow each bubble to rotate freely about its own axis, 
as illustrated by the angular displacements $\Theta_{LB}$ and $-\Theta_{TB}$ in \hyperref[fig:LSA_comparison_with_rotation]{figure~\ref{fig:LSA_comparison_with_rotation}$(a)$}.  
The resulting neutral curve, obtained for the same azimuthal mode $|m| = 1$ but with the extended bubble state vector  $\mathscr{Q}^{b} = [\boldsymbol{X}_{LB}, \boldsymbol{V}_{LB}, \boldsymbol{\varXi}_{LB}, \boldsymbol{\Omega}_{LB}, \boldsymbol{X}_{TB}, \boldsymbol{V}_{TB}, \boldsymbol{\varXi}_{TB}, \boldsymbol{\Omega}_{TB}]$, is shown in 
\hyperref[fig:LSA_comparison_with_rotation]{figure~\ref{fig:LSA_comparison_with_rotation}$(b)$}.  
The solid and dashed lines correspond to the stationary ($\lambda_i = 0$) and oscillatory ($\lambda_i \neq 0$) branches, respectively, the latter being discussed in \hyperref[sec:LSA_oscillatory]{\S~\ref{sec:LSA_oscillatory}}.  
Allowing bubble rotation profoundly modifies the stability characteristics.  
The predicted critical Reynolds numbers, as well as the slope of the $Re_{c}$–$\chi$ relation, now exhibit excellent agreement with the VOF–DNS data.  
This improvement demonstrates unambiguously that the inclination response of the bubbles, arising from the asymmetric shear, provides a fundamental stabilizing feedback in the in-line configuration.  

This conclusion is further supported by dedicated simulations performed with the three-dimensional EBM solver.  
Here, two oblate bubbles of aspect ratio $\chi = 1.6$ are placed in-line at a distance $S = 3$, with the TB slightly offset laterally by $S_{r} = 0.1$.  
The LB remains horizontal ($\Theta_{LB} = 0$), while the TB is tilted by a prescribed angle $-\Theta_{TB}$.  
\hyperref[fig:LSA_comparison_with_rotation]{Figure~\ref{fig:LSA_comparison_with_rotation}$(c)$} shows that, at a fixed $\Rey = 60$, the lift coefficient $C_{L}^{TB}$ decreases sharply with increasing inclination and finally reverses sign for $5^{\circ} \!\lesssim\! -\Theta_{TB} \!\lesssim\! 6^{\circ}$.  
This reversal, dominated by the pressure contribution, demonstrates that the stabilizing effect originates from an inclination-induced redistribution of surface pressure rather than from deformation itself.  

The physical origin of this stabilization lies in the coupling between the TB’s inclination 
and the asymmetric vorticity field in the wake of the LB.  
When the TB drifts slightly away from the symmetry axis, 
the wake shear exerts a hydrodynamic torque that reorients its major axis by a small angle $-\Theta_{TB}$ 
until it is balanced by the opposing inertial torque.  
This inclination alters the surface-pressure distribution:  
the outer side, now more exposed to the incident flow, experiences a larger incidence angle 
and a higher pressure owing to the outward shift of the front stagnation point, 
whereas the inner side, lying within the slower near-wake region, 
is subjected to a lower pressure.  
The resulting pressure imbalance shifts the high-pressure zone outward from the wake centreline, 
producing a net negative lift that restores the TB toward alignment.  
In linear terms, this inclination-induced lift acts as a negative feedback 
approximately proportional to $(\cos\Theta_{TB}\sin\Theta_{TB})$ 
\citep{happel2012low,sanjeevi2017orientational}, 
thereby reducing the effective growth rate of lateral disturbances.  
This mechanism is analogous to the orientation-induced lift experienced by rigid spheroids 
in uniform flow \citep{ouchene2016new,sanjeevi2017orientational}, 
but opposite in sign to the classical Magnus lift on continuously rotating spheres 
\citep{bagchi2002effect}, which always drives outward motion.  
Furthermore, the torque acting on the TB increases with aspect ratio, 
such that more oblate bubbles experience stronger reorienting torques 
and thus larger inclination angles. This explains why more flattened bubble pair corresponds to a higher critical Reynolds number for transition to instability. The coupling between inclination and wake shear therefore counteracts the destabilizing potential and shear-induced lifts, 
recovering the DNS-observed trend that more oblate bubble pairs 
exhibit enhanced in-line stability.  

Together, the ALE–LSA predictions and the EBM–DNS results confirm 
that the stabilization of oblate in-line bubble pairs 
originates from an inclination-induced restoring lift, 
rather than from deformation-enhanced wake entrainment.  
This mechanism defines a distinct regime of lift generation 
and constitutes the dominant physical process governing 
the lateral stability of oblate bubble pairs in the inertial regime.

\subsection{Neutral curves}
\label{sec:neutral_curve}

After clarifying the mechanism by which bubble oblateness stabilizes the in-line configuration, 
the global linear stability of the stationary mode ($\lambda_{i} = 0$) was examined in detail 
with respect to non-axisymmetric perturbations of azimuthal order $|m| = 1$, 
which correspond to the most amplified lateral modes.  
For each aspect ratio $\chi \leq 1.9$, the Reynolds number and the centre-to-centre separation $S$ 
were varied until the real part of one eigenvalue, $\lambda_{r}$, changed sign, 
thereby defining the neutral stability boundary of the stationary branch.  

\begin{figure}
	\centering
	\setlength{\tabcolsep}{0pt}
	\iflocalcompile
	\begin{tabular}{@{}c@{\hspace{0em}}c@{}}
		\begin{minipage}[t]{0.5\textwidth}
			\vspace{-3.5pt} 
			\resizebox{\textwidth}{!}{%
				\input{fig/4.2sta/stan-neutralline-new.tex}
			}
		\end{minipage}
		&
		\begin{minipage}[t]{0.5\textwidth}
			\vspace{0pt} 
			\begin{tabular}{c}
				\resizebox{\textwidth}{!}{\input{\figfourthree/lambdar_chi1p6_differentS-2-nolabel-newscale.tex}} \\ [-1em]
				\resizebox{\textwidth}{!}{\input{\figfourthree/lambdar_S6_differentchi_nolabel-new}}
			\end{tabular}
		\end{minipage}
	\end{tabular}
	\else
	\begin{tabular}{@{}c@{\hspace{0em}}c@{}}
		\begin{minipage}[t]{0.5\textwidth}
			\vspace{-3.5pt} 
			\resizebox{\textwidth}{!}{%
				\includetikz{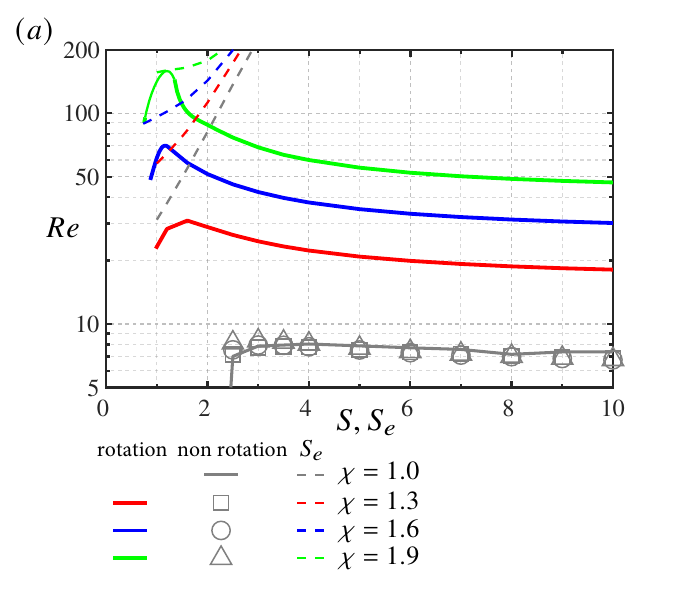}
			}
		\end{minipage}
		&
		\begin{minipage}[t]{0.5\textwidth}
			\vspace{0pt} 
			\begin{tabular}{c}
				\resizebox{\textwidth}{!}{\includetikz{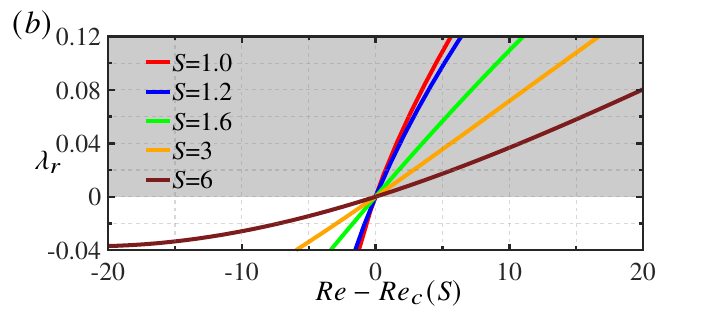}} \\ [-1em]
				\resizebox{\textwidth}{!}{\includetikz{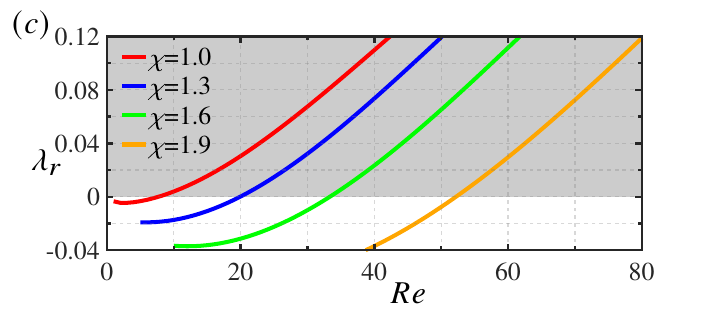}}
			\end{tabular}
		\end{minipage}
	\end{tabular}
	\fi
	\captionsetup{justification=justified, singlelinecheck=false, labelsep=period, width=\linewidth} 
	\caption{
	Stationary mode results from the ALE–LSA for bubble pairs as functions of Reynolds number $\Rey$, aspect ratio $\chi$, and separation distance $S$.
$(a)$ Neutral curves: coloured lines show results including rotational degrees of freedom for $\chi = 1.3$, $1.6$, and $1.9$, while grey lines and symbols indicate $\chi = 1.0$ and the corresponding spheroidal cases with rotation suppressed.
Dashed lines denote the equilibrium separation $\bar{S}_{e}$ for those $\chi$ obtained from the steady base flow. Note that the thin segment of $\chi = 1.9$ ($S \lesssim 1.6$) does not correspond to a real neutral curve but implies a transition from oscillatory mode to stationary mode, as described in \S~\ref{sec:LSA_oscillatory}.
$(b)$ Dependence of the growth rate $\lambda{r}$ on $(\Rey - \Rey_{c}(S))$ for various separations at fixed $\chi = 1.6$, where $\Rey_{c}(S)$ is the critical Reynolds number for each $S$.
$(c)$ Variation of $\lambda_{r}$ with $\Rey$ for different $\chi$ at a fixed $S = 6$.
In panels $(b)$ and $(c)$, shaded regions correspond to $\lambda_{r} > 0$, identifying unstable configurations.
	}
	\label{fig:neutral_curves}
\end{figure}

The resulting neutral curves for $\chi = 1$, $1.3$, $1.6$, and $1.9$ 
are displayed in \hyperref[fig:neutral_curves]{figure~\ref{fig:neutral_curves}$(a)$}.  
Open symbols and grey solid lines denote the results obtained for spherical and spheroidal bubbles 
with rotation artificially suppressed, 
while coloured solid lines correspond to the fully coupled solutions 
that include rotational degrees of freedom.  
The dashed lines of matching colours represent the equilibrium separations $S_{e}$ 
(see \hyperref[sec:equilibrium-distance]{\S~\ref{sec:equilibrium-distance}}) 
for the corresponding aspect ratios.  
In all cases, $S_{e}$ lies below the critical distance at which the in-line configuration loses stability, 
indicating that the pair becomes unstable before reaching its equilibrium spacing.

When bubble rotation is constrained (open symbols and grey lines), the critical Reynolds number $Re_{c}$ remains low and nearly insensitive to both $\chi$ and $S$ for $S \gtrsim 2.5$, collapsing to $Re_{c} \!\in\! [7,8]$.  
This value agrees closely with the EBM–DNS results shown in \hyperref[fig:LSA_comparison_no_rotation]{figure~\ref{fig:LSA_comparison_no_rotation}$(c)$}, 
further confirming that, in the absence of rotational freedom, the stabilizing influence of bubble oblateness is effectively suppressed.  
Consistently, \citet{hallez2011interaction} reported that spherical bubble pairs lose their in-line stability across the entire investigated range ($\Rey \geq 20$), in agreement with the present prediction.  
The weak dependence of $Re_{c}$ on $\chi$ in the non-rotating cases can be interpreted by examining the competing lift contributions in (\ref{eq:lift_decomposition}).  
As discussed by \citet{maeda2021viscid}, the shear-induced lift, $C_{L}^{shear}$, dominates the pairwise dynamics once a lateral perturbation develops, 
making its variation with bubble oblateness particularly relevant.  
The wake of the LB generates a rotational region of thickness $\delta^{wake}$ that increases with $\chi$; since $C_{L}^{shear} = C_{L}^{iso} - \alpha/(\delta^{wake} - \beta Re^{-1/2})$ grows with $\delta^{wake}$ at fixed $(\Rey, S)$ \citep{hallez2011interaction,adoua2009reversal}, this trend enhances the positive (destabilizing) lift acting on the TB.  
Here, $C_{L}^{iso}$ denotes the shear-induced lift on an isolated bubble, while $\alpha$ and $\beta$ are shape-dependent constants.  
At the same time, moderately increasing $\chi$ also intensifies vorticity generation at the bubble surface, amplifying $C_{L}^{iso}$ itself \citep{adoua2009reversal,zhang2025lift} and thus further reinforcing the destabilizing tendency.  
Conversely, the viscous correction $C_{L}^{vis}$, associated with wake entrainment behind the LB, strengthens with increasing $\chi$ and provides a stabilizing contribution.  
These two opposing tendencies, enhanced $C_{L}^{shear}$ and reinforced $C_{L}^{vis}$, largely compensate each other, leaving the critical threshold $Re_{c}$ almost independent of $\chi$ when bubble rotation is artificially suppressed.

When rotation is restored, the stability characteristics change fundamentally.  
As shown by the coloured lines in \hyperref[fig:neutral_curves]{figure~\ref{fig:neutral_curves}$(a)$}, the critical Reynolds number increases by more than an order of magnitude 
and exhibits a pronounced non-monotonic variation with the separation distance $S$.  
For each aspect ratio $\chi$, $Re_{c}$ first rises steeply for $0.8 \!\lesssim\! S \!\lesssim\! 1.2$, reaches a maximum near $S \!\approx\! 1.2$, 
and then decreases gradually toward a broad plateau at larger separations.  
This trend mirrors the spatial distribution of wake shear intensity behind the LB.  
At very small separations, the TB lies within the nearly irrotational core of the LB wake, where the transverse shear $\partial u_{x}/\partial r$ is weak and the destabilizing potential interaction thus dominates, leading to a low critical Reynolds number.  
As $S$ increases, the TB enters the annular shear layer where both $\partial u_{x}/\partial r$ and the associated hydrodynamic torque reach their maxima.  
This configuration produces the strongest inclination and, consequently, the most effective stabilizing lift.  
Beyond $S \!\approx\! 1.2$, the wake shear decays approximately as $\partial u_{x}/\partial r \!\sim\! (V_{x}r/X)\exp(-V_{x}r^{2}/X)$, so that the inclination–shear feedback weakens and $Re_{c}$ decreases. 
At sufficiently large separations ($S \!>\! 8$), $Re_{c}$ becomes nearly independent of $S$ and depends primarily on $\chi$. 
This behaviour indicates that, once the bubbles are far apart, the system’s stability is controlled by the intrinsic response of the TB immersed in a weak external shear, rather than by direct mutual-bubble coupling.  
Indeed, as the wake thickness $\delta^{wake} \!\propto\! S$ increases, the lift contribution $C_{L}^{wake} \!\rightarrow\! C_{L}^{iso}$, which depends solely on the TB aspect ratio and Reynolds number.  
Hence, in the asymptotic limit of large separation, the influence of the LB reduces to that of a steady far-field perturbation, and the stability characteristics revert to those of a single, freely rising oblate bubble.  

At a fixed separation, the critical Reynolds number increases monotonically with the aspect ratio, in quantitative agreement with the VOF–DNS results of 
\citet{zhang2021three} (see \hyperref[fig:LSA_comparison_with_rotation]{figure~\ref{fig:LSA_comparison_with_rotation}$(b)$}).  
As discussed previously, this trend reflects the strengthening of the shear-induced torque with increasing oblateness, which amplifies the inclination-driven stabilizing feedback.  
\hyperref[fig:neutral_curves]{Figure~\ref{fig:neutral_curves}$(b)$} further shows the variation of the growth rate $\lambda_{r}$ with $(\Rey - \Rey_{c})$ for several separations at $\chi = 1.6$, where $\Rey_{c}$ denotes the neutral threshold at which $\lambda_{r}$ changes sign.  
The curves for $S = 1.0$ and $1.2$ nearly coincide, while $\lambda_{r}$ decreases systematically with increasing $S$, indicating that weaker hydrodynamic coupling leads to slower amplification of perturbations.  
Conversely, at fixed separation ($S = 6$), \hyperref[fig:neutral_curves]{figure~\ref{fig:neutral_curves}$(c)$} shows that $Re_{c}$ increases monotonically with $\chi$, confirming that greater oblateness enhances the rotational response of the bubbles and thereby strengthens the stabilizing feedback mechanism.

\begin{figure}
	\centering
	\setlength{\tabcolsep}{0pt} 
	\iflocalcompile
	\begin{tabular}{cc}
		\resizebox{0.5\textwidth}{!}{\input{\figfourthree/chi1.6Re60Sr0.1differentS.tex}} &
		\resizebox{0.5\textwidth}{!}{\input{\figfourthree/S6DSr0.1Re60differentchi.tex}}
	\end{tabular}
	\else
	\begin{tabular}{cc}
		\resizebox{0.5\textwidth}{!}{\includetikz{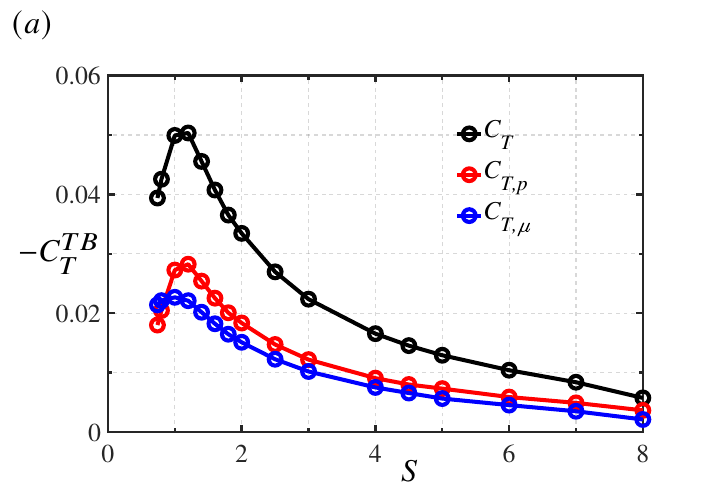}} &
		\resizebox{0.5\textwidth}{!}{\includetikz{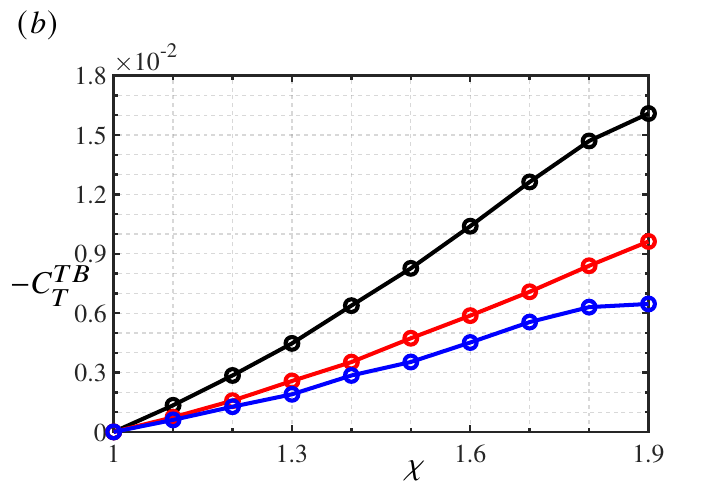}}
	\end{tabular}
	\fi	
	\captionsetup{justification=justified, singlelinecheck=false, labelsep=period, width=\linewidth} 
	\caption{
	Hydrodynamic torque $C_{T}^{TB}$ acting on the TB computed using the EBM–DNS framework \citep{WEI2026114632}, with the TB held horizontally, i.e. $\Theta_{TB} = 0^{\circ}$. $-C_{T}^{TB}$ is displayed for consistency with the sign convention adopted for the tilting angle $\Theta$.
$(a)$ Variation of the torque coefficients with separation, for fixed $(\Rey, \chi, S_{r}) = (60, 1.6, 0.1)$.
$(b)$ Variation of the torque coefficients with aspect ratio, for fixed $(\Rey, S, S_{r}) = (60, 6, 0.1)$.
In both panels, the total torque is decomposed into pressure and viscous contributions.
	}
	\label{fig:torque}
\end{figure}

Further evidence is provided by three-dimensional EBM–DNS simulations, which directly evaluate the hydrodynamic torque coefficient $C_{T}^{TB} = 2T^{TB}/(\pi R^{3}\rho V_{x}^{2})$ acting on the trailing bubble as a function of $(\Rey, S, \chi)$.  
For consistency with the sign convention adopted for the inclination angle $\Theta$, where clockwise rotation is defined as negative, the results are presented in terms of $-C_{T}^{TB}$.  
In these simulations, the TB was displaced laterally by $S_{r}=0.1$ while its orientation was kept horizontal ($\Theta_{TB}=0^{\circ}$), thereby isolating the torque generated solely by the shear imposed by the LB.  
For $\chi = 1.6$ and $\Rey = 60$, \hyperref[fig:torque]{figure~\ref{fig:torque}$(a)$} shows that $C_{T}^{TB}$ exhibits a pronounced non-monotonic dependence on $S$: 
it increases sharply, reaches a maximum near $S \!\approx\! 1.2$, and decreases thereafter, closely mirroring the shape of the neutral curve in \hyperref[fig:neutral_curves]{figure~\ref{fig:neutral_curves}$(a)$}.  
At fixed separation ($S = 6$) and $\Rey = 60$, \hyperref[fig:torque]{figure~\ref{fig:torque}$(b)$} shows that $C_{T}^{TB}$ increases monotonically with $\chi$, demonstrating that more oblate bubbles experience stronger hydrodynamic torques and hence larger inclination-induced stabilizing lifts.  
Together, these results confirm that the in-line stability of oblate bubble pairs is governed by the interplay between the wake-induced shear and the bubble geometry, which jointly determine the rotational feedback responsible for the inclination-driven lift.

\begin{figure}
	\centering
	\setlength{\tabcolsep}{0pt} 
	\begin{tabular}{cccc}
		\includegraphics[width=0.25\textwidth]{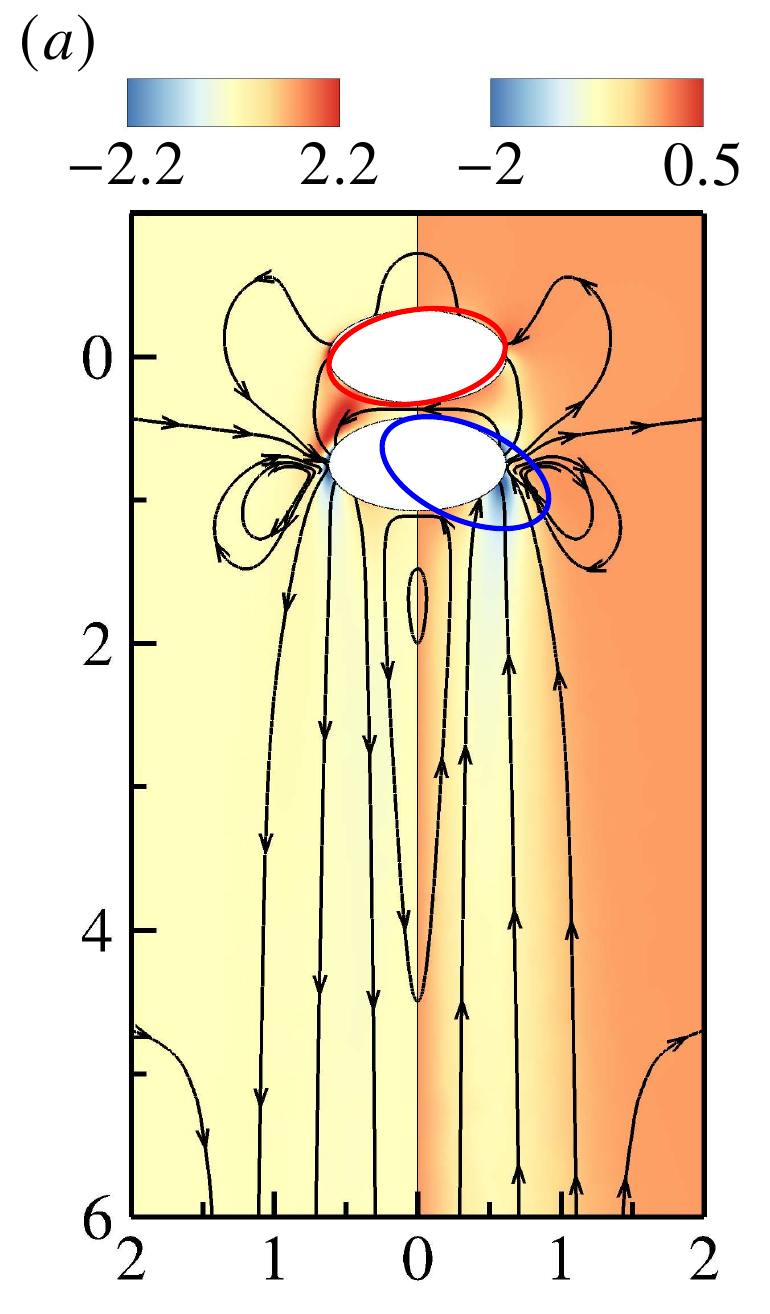} &
		\includegraphics[width=0.25\textwidth]{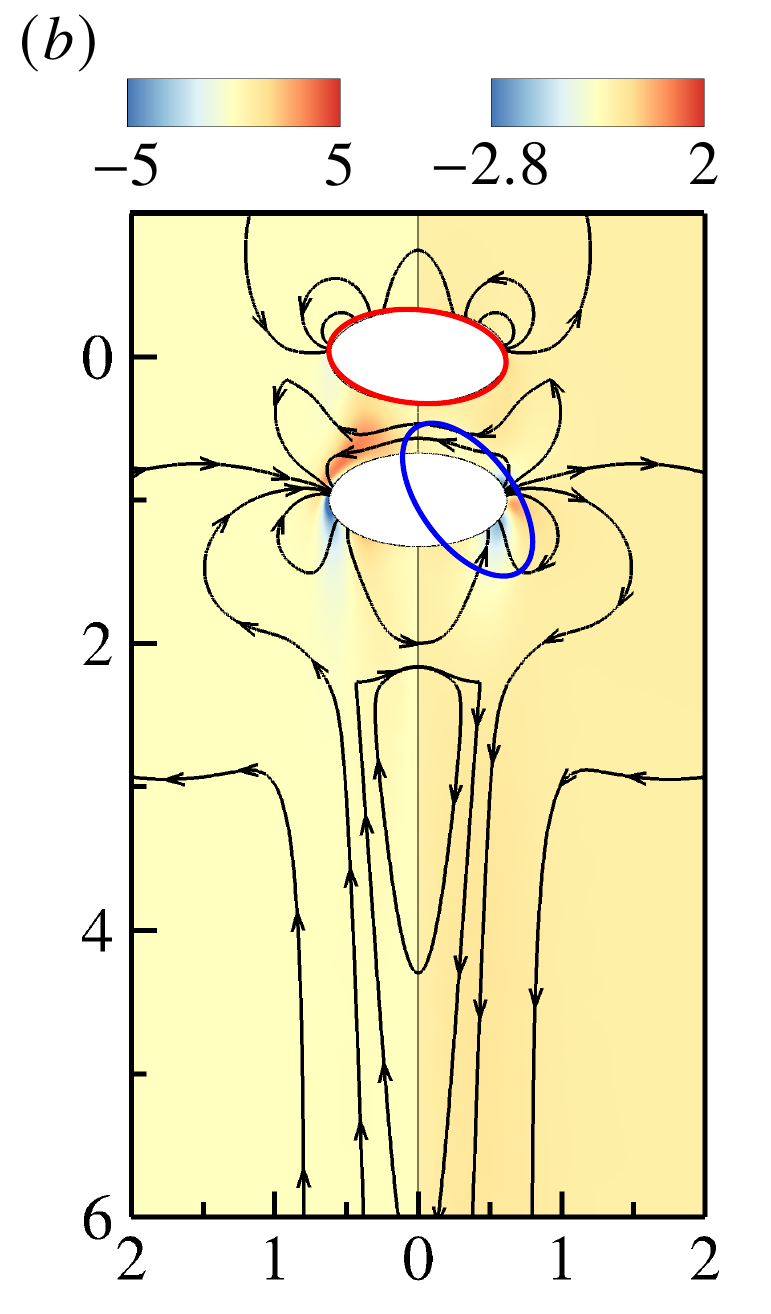} &
		\includegraphics[width=0.25\textwidth]{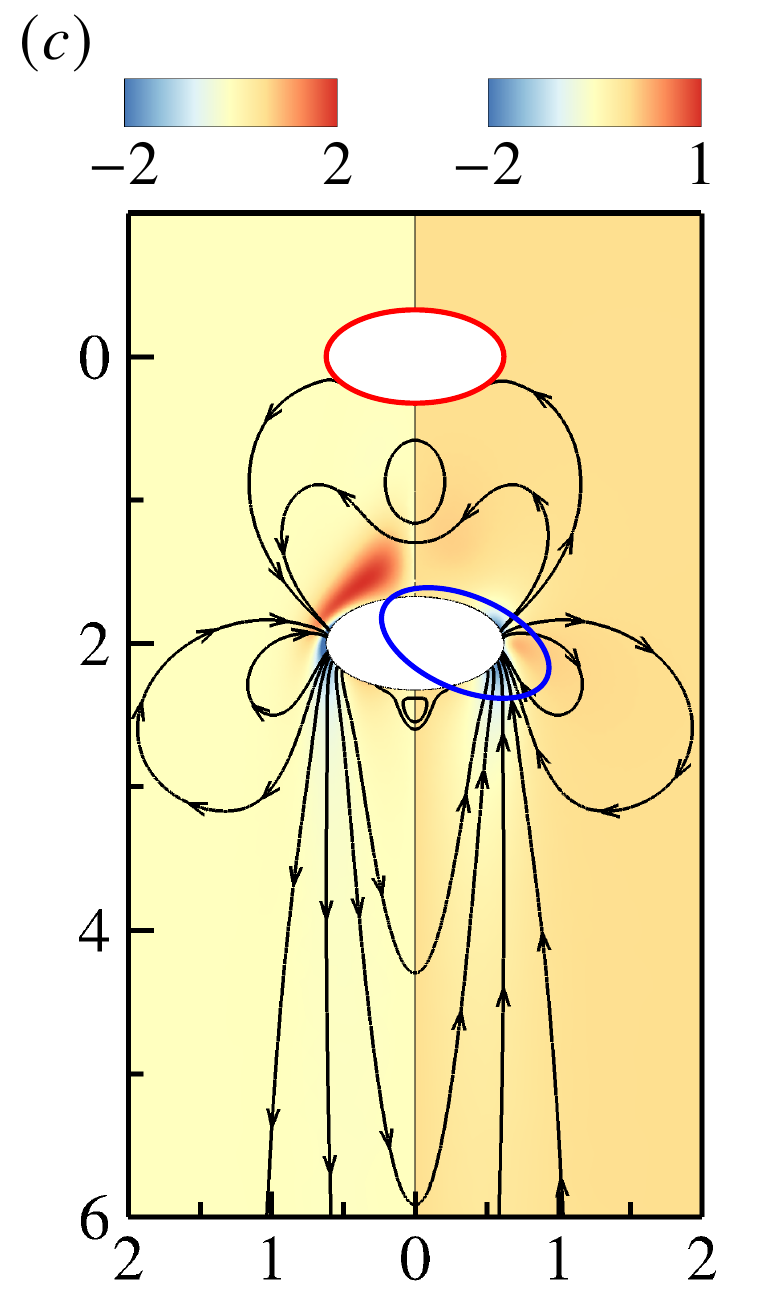} &
		\includegraphics[width=0.25\textwidth]{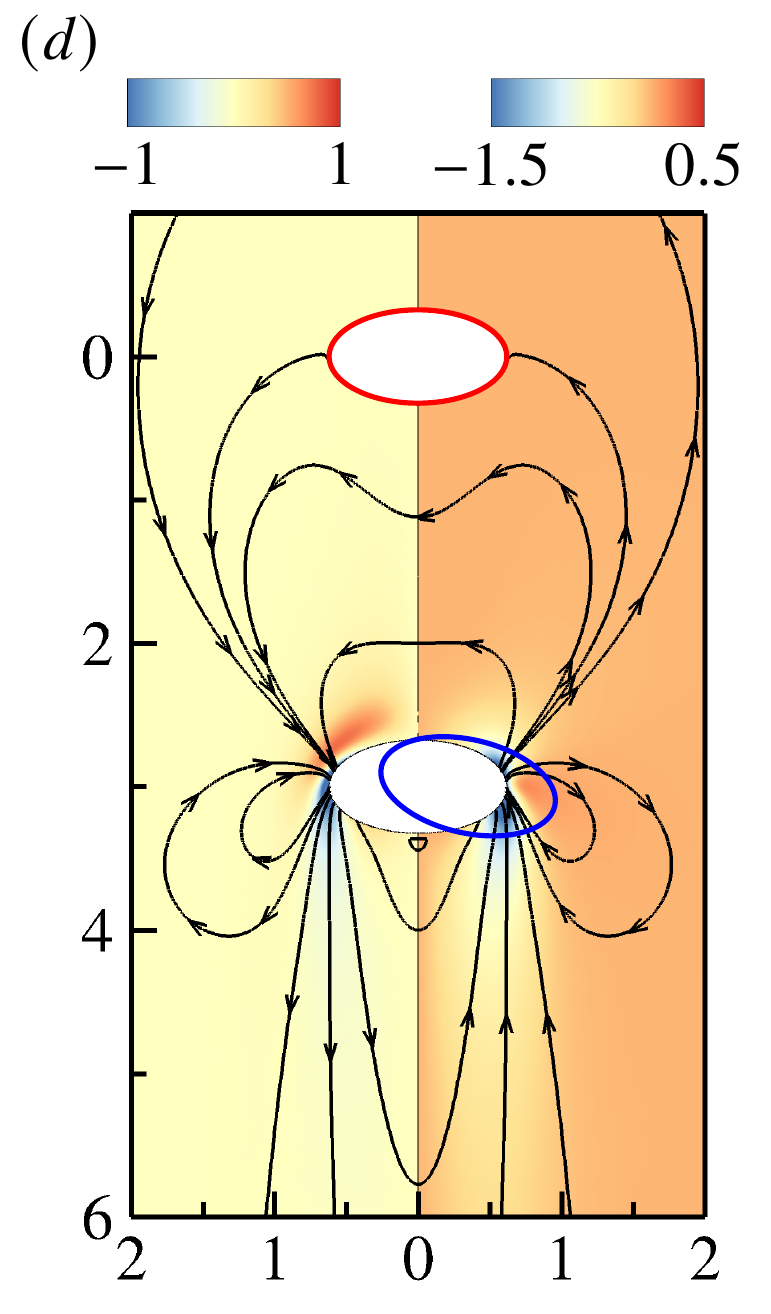}
	\end{tabular}
	\captionsetup{justification=justified, singlelinecheck=false, labelsep=period, width=\linewidth} 
	\caption{Structure of the stationary eigenmode for the bubble pair of $\chi = 1.9$ near the critical threshold, shown for different parameter combinations $(\Rey, S)$ from left to right:
$(a)$ $(90, 0.75)$, $(b)$ $(142, 1.0)$, $(c)$ $(90, 2.0)$, and $(d)$ $(73, 3.0)$.
In each panel, the left and right halves display iso-contours of the streamwise vorticity and velocity, respectively, with selected streamlines plotted in the reference frame of the base flow.
Red and blue contour lines mark the maximum angular displacements ($\hat{\eta}$) of the LB and TB, respectively.}
	\label{sta:mode:chi:1.9}
\end{figure}

The spatial structure of the stationary eigenmode for $\chi = 1.9$, taken slightly above the neutral threshold, 
is shown in \hyperref[sta:mode:chi:1.9]{figure~\ref{sta:mode:chi:1.9}}, 
providing further insight into the nature of the hydrodynamic coupling between the two bubbles.  
In the figure, the red and blue lines represent the real parts of the interfacial disturbances of the leading 
($\hat{\eta}_{LB}$) and trailing ($\hat{\eta}_{TB}$) bubbles, respectively, corresponding to the instant of maximal lateral excursion.  
The left and right half-panels display iso-contours of the streamwise vorticity ($\omega_x$) and velocity ($u_x$), 
while the superimposed streamlines indicate the perturbation flow field.  
In all cases, 
\hyperref[sta:mode:chi:1.9]{figure~\ref{sta:mode:chi:1.9}} 
reveals concentrated regions of streamwise vorticity within the inter-bubble gap and along the wake of the TB.  
Since the frozen-body LSA confirmed that the wakes remain linearly stable for $\chi \leq 1.9$ 
(see \hyperref[app:wake]{Appendix~\ref{app:wake}}), 
these vortical structures cannot originate from intrinsic wake instabilities.  
Instead, they are induced by the lateral motion of the bubbles and represent the fluid-mediated feedback that couples the two bubbles.  
The stabilization of the in-line configuration therefore arises from a genuinely global mechanism: 
the wake of the LB acts as a conduit for momentum and torque exchange, 
whose shear distribution and associated hydrodynamic torque govern the rotational response of the TB 
and, ultimately, the overall stability of the pair.

At small separations, $(\Rey, S) = (90, 0.75)$ ($\delta S \approx 0.1$), 
\hyperref[sta:mode:chi:1.9]{figure~\ref{sta:mode:chi:1.9}$(a)$} 
shows that both bubbles undergo pronounced angular displacements accompanied by intense local velocity perturbations.  
The TB rotates clockwise, whereas the LB tilts in the opposite direction, 
revealing a phase-opposed rotational response at short inter-bubble distances.  
This behaviour results from the strong hydrodynamic coupling within the narrow gap, as  
the clockwise rotation of the TB induces an inverted hydrodynamic torque on the LB, 
causing it to rotate anti-clockwise in reaction.  
At such small separations, the two bubbles interact strongly and separate laterally, corresponding to the DKT regime observed in experiments and DNS.    
When the separation increases to $(\Rey, S) = (142, 1.0)$ ($\delta S \approx 0.35$), 
panel~$(b)$ indicates that the disturbance amplitude around the LB decays rapidly, 
while that near the TB remains dominant.  
Both bubbles now rotate clockwise, and the TB displays a clear inclination whereas the LB exhibits only a mild tilt.  
This synchronous rotation arises because the inter-bubble recirculating eddy strengthens and broadens with increasing $S$.  
The clockwise rotation of the TB distorts this eddy, reversing the sense of the induced torque on the LB and aligning the rotation of the two bubbles.  
At still larger separations ($S = 2.0$ and $3.0$; panels~$c$ and $d$), 
the interaction becomes increasingly asymmetric and predominantly one-way:  
the TB continues to incline and drift laterally, while the LB remains nearly horizontal and only weakly perturbed. 
This configuration corresponds to the ASE regime, in which the TB departs laterally at long range without significantly altering the trajectory of the LB. In this regime, the LB wake merely acts as an external shear field that destabilizes the TB, while the feedback exerted by the TB on the leader becomes negligible.

A number of experimental and numerical studies have reported both the DKT and ASE scenarios within different regions of parameter space, each exhibiting markedly distinct dynamical features. Whether these two behaviours correspond to actually distinct unstable modes of the coupled bubble–flow system remains, however, an open question. The following section addresses this issue in more detail.

\subsection{Drafting-kissing-tumbling versus asymmetric side escape}
\label{sec:DKT_ASE}
\begin{figure}
    \centering
    \setlength{\tabcolsep}{0pt} 
    \iflocalcompile
    \begin{tabular}{cc}
        \resizebox{0.5\textwidth}{!}{\input{fig/4.2sta/LSA_displacement/chi1.9-LB-TB-LB-2.tex}} &
        \resizebox{0.5\textwidth}{!}{\input{fig/4.2sta/LSA_displacement/chi1.6-LB-TB-LB-2.tex}}
    \end{tabular}
    \else
    \begin{tabular}{cc}
        \resizebox{0.5\textwidth}{!}{\includetikz{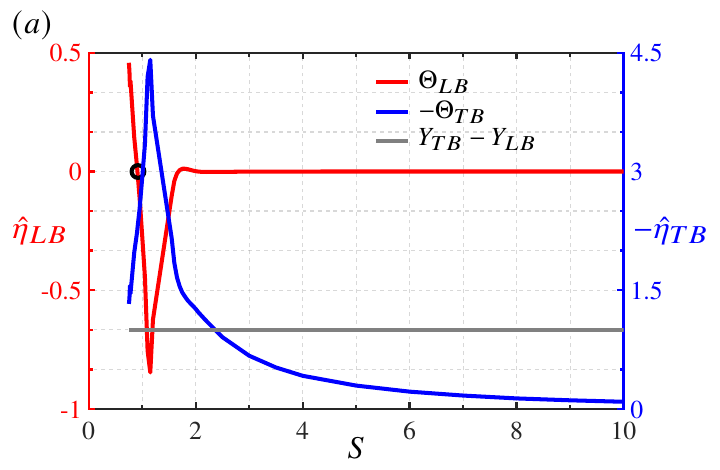}} &
        \resizebox{0.5\textwidth}{!}{\includetikz{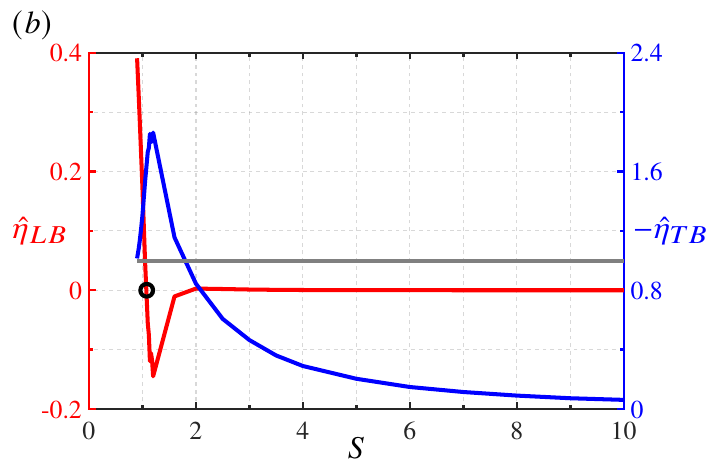}}
    \end{tabular}            
    \fi	
    \captionsetup{justification=justified, singlelinecheck=false, labelsep=period, width=\linewidth} 
    \caption{
Variation of the angular displacement of the bubbles in the stationary mode under different conditions.  
$(a)$~$\chi = 1.9$; $(b)$~$\chi = 1.6$.  
The open symbol (\protect\tikz\protect\draw[black, thick] (0,0) circle (3pt);) on the red curves marks the point $\hat{\eta}_{LB} = 0$, where the LB reverses its rotation direction from anti-clockwise to clockwise.  
The angular displacements of the leading and trailing bubbles are shown as functions of the centre-to-centre separation $S$ at the neutral threshold, normalized by their displacement difference, $(Y_{TB} - Y_{LB})$.  
The corresponding values of this displacement difference are indicated on the left axis.
}
    \label{sta:angular:chi:1.9}
\end{figure}

We first quantify the interfacial disturbances of both bubbles for $\chi = 1.9$.  
\hyperref[sta:angular:chi:1.9]{Figure~\ref{sta:angular:chi:1.9}$(a)$} displays the angular displacements $\hat{\eta}_{LB}$ (left axis) and $-\hat{\eta}_{TB}$ (right axis) as functions of the separation distance $S$, with Reynolds numbers corresponding to the respective critical thresholds for each $S$.  
Within the linear framework, the absolute amplitudes of these displacements are arbitrary, and only their relative magnitudes carry physical meaning.  
Accordingly, $\hat{\eta}_{LB}$ and $\hat{\eta}_{TB}$ are normalised by the lateral displacement difference $(Y_{TB} - Y_{LB})$, and the sign of $\hat{\eta}_{TB}$ is reversed for clarity.  
From the figure, the range $S \lesssim 1.7$ can be identified as the DKT regime, since both bubbles exhibit appreciable angular displacements at these short separations.  
For $S \gtrsim 1.7$, the response of the LB becomes negligible, decaying rapidly toward zero, whereas $-\hat{\eta}_{TB}$ remains finite, indicating the onset of the ASE regime.  
This progressive attenuation of the LB’s rotational response marks the transition from a two-way coupled DKT regime to a one-way coupled ASE regime dominated by the wake of the LB.
 
Still in \hyperref[sta:angular:chi:1.9]{figure~\ref{sta:angular:chi:1.9}$(a)$}, the TB consistently satisfies $-\hat{\eta}_{TB} > 0$, indicating a clockwise rotation across all separations, and its non-monotonic variation with $S$ closely follows the shape of the neutral curve.  
In contrast, the LB exhibits a distinct sign reversal: $\hat{\eta}_{LB} > 0$ (anti-clockwise rotation) for $S \le 0.92$, and $\hat{\eta}_{LB} < 0$ (clockwise rotation) for $S > 0.92$, with the transition marked by the open symbol on the curve.  
The range of clockwise rotation of the LB coincides precisely with the parameter domain where a connected recirculation forms within the inter-bubble gap. 
For instance, within the clockwise interval $0.92 < S < 1.7$ ($\hat{\eta}_{LB} < 0$), the critical Reynolds numbers identified in \hyperref[fig:neutral_curves]{figure~\ref{fig:neutral_curves}$(a)$} are $110 \lesssim \Rey_{c} \lesssim 160$, while the corresponding regime of connected recirculation in \hyperref[large_aspect_ratio-chi1.9]{figure~\ref{large_aspect_ratio-chi1.9}$(a)$} extends up to $S < 2.4$, fully encompassing this interval. 
This correlation confirms that the DKT interaction ($S \lesssim 1.7$) comprises two sub-modes: one in which the LB rotates anti-clockwise, and another in which it rotates clockwise once the connecting eddy dominates the inter-bubble dynamics.  
Across all separations, the inclination of the TB remains larger in magnitude than that of the LB, i.e.\ $-\hat{\eta}_{TB} > \hat{\eta}_{LB}$.  
A similar trend is observed for the bubble pair with $\chi = 1.6$, as shown in \hyperref[sta:angular:chi:1.9]{figure~\ref{sta:angular:chi:1.9}$(b)$}:  
while the TB always rotates clockwise, the LB again undergoes an anti-clockwise-to-clockwise transition, now occurring at a slightly larger separation $S \approx 1.08$, as marked on the curve.

\begin{figure}
	\centering
	\setlength{\tabcolsep}{0pt}
	\iflocalcompile
	\begin{tabular}{ccc}
		\resizebox{0.34\textwidth}{!}{\input{\figfourfive/chi1.9-LBfixed-sta.tex}} &
        \resizebox{0.34\textwidth}{!}{\input{\figfourfive/chi1.6-LBfixed-sta.tex}} &
        \resizebox{0.34\textwidth}{!}{\input{\figfourfive/chi1.3-LBfixed-sta.tex}} 
	\end{tabular}
	\else
	\begin{tabular}{ccc}
		\resizebox{0.34\textwidth}{!}{\includetikz{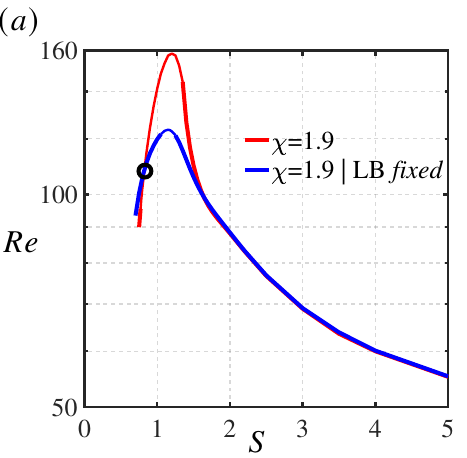}} &
        \resizebox{0.34\textwidth}{!}{\includetikz{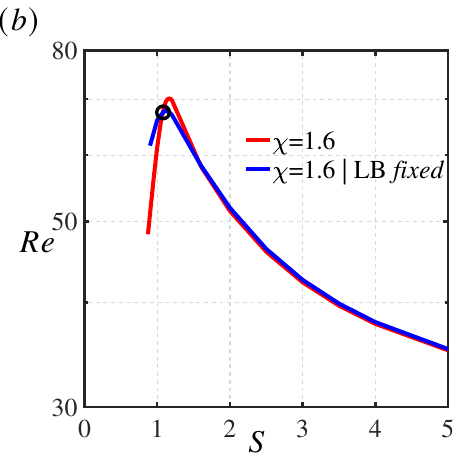}} &
        \resizebox{0.34\textwidth}{!}{\includetikz{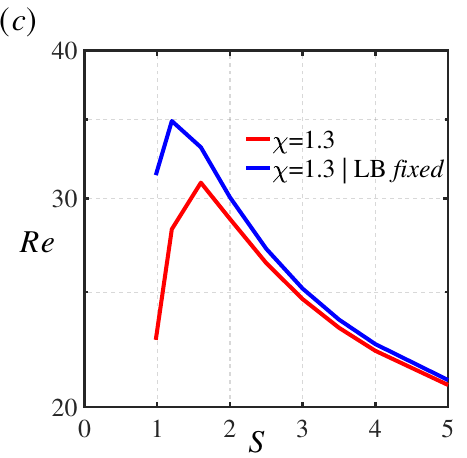}}
	\end{tabular}
	\fi
	\captionsetup{justification=justified, singlelinecheck=false, labelsep=period, width=\linewidth}
	\caption{ Neutral stability curves of the stationary mode in the $(\Rey, S)$ plane for $(a)$~$\chi = 1.9$, $(b)$~$\chi = 1.6$, and $(c)$~$\chi = 1.3$, comparing configurations where the LB is either fixed (blue line) or free to move (red line).  The open symbol (\protect\tikz\protect\draw[black, thick] (0,0) circle (3pt);) in panels~$(a)$ and $(b)$ marks the intersection point at which the freely moving case begins to exhibit a higher critical Reynolds number, $Re_c$, than the fixed-LB case.
}
	\label{LB:fixed:chi1.6and2.0:neutralcurve}
\end{figure}

The neutral stability curves of the stationary branch for $\chi = 1.9$ are shown in \hyperref[LB:fixed:chi1.6and2.0:neutralcurve]{figure~\ref{LB:fixed:chi1.6and2.0:neutralcurve}$(a)$}.  
Compared with the freely moving configuration (red line), fixing the leading bubble (blue line) strongly modifies the stability characteristics of the DKT branch at short separations ($S \lesssim 1.7$).  
In particular, the critical Reynolds number $Re_c$ is markedly reduced within $0.85 \leq S \leq 1.7$, where its maximum value drops from about $158.5$ to $123.6$, while a slight increase in $Re_c$ is observed at even smaller separations ($S < 0.85$, open symbol on the curve).  
This behaviour is directly related to the rotation characteristics in \hyperref[sta:angular:chi:1.9]{figure~\ref{sta:angular:chi:1.9}$(a)$}, which show that the freely moving LB reverses its rotation direction from clockwise to anti-clockwise at $S \approx 0.92$.  These results indicate that the rotational response of the LB plays a dual role in the DKT regime.  
When the LB rotates opposite to the TB, its counter-rotation induces an adverse torque that amplifies the lateral displacement of the pair, thereby destabilizing the in-line configuration.  
Conversely, when both bubbles rotate in the same direction, the LB contributes a cooperative torque that counteracts the lateral drift of the TB, producing a stabilizing feedback.  
Thus, the impact of the LB motion on stability within the DKT regime depends critically on its rotation direction, which in turn is governed by the presence and structure of the connected recirculation in the inter-bubble gap.  
At larger separations ($S \gtrsim 1.7$), approaching the ASE regime, the two-way hydrodynamic coupling between the bubbles becomes negligible.  
The stabilizing feedback associated with the LB’s motion vanishes, and the results for the one-way coupling collapse onto those of the freely moving configuration.  
This trend is fully consistent with the mode structures shown in \hyperref[sta:mode:chi:1.9]{figures~\ref{sta:mode:chi:1.9}} and \hyperref[sta:angular:chi:1.9]{\ref{sta:angular:chi:1.9}$(a)$}, confirming that the ASE regime represents a weakly coupled interaction in which the LB merely provides a shear background, while the stability of the pair is governed solely by the inclination and rotation of the TB.

Further comparison for $\chi = 1.6$ is presented in \hyperref[LB:fixed:chi1.6and2.0:neutralcurve]{figure~\ref{LB:fixed:chi1.6and2.0:neutralcurve}$(b)$}, 
which displays trends similar to those observed for $\chi = 1.9$.  
Fixing the LB again reduces the critical Reynolds number $Re_{c}$ over the range $1.08 \leq S < 1.6$, while producing a slight increase at smaller separations ($S < 1.08$).  
As shown in \hyperref[sta:angular:chi:1.9]{figure~\ref{sta:angular:chi:1.9}$(b)$}, the corresponding transition of the LB’s rotation from clockwise to anti-clockwise occurs precisely at $S \approx 1.08$.  
This coincidence reinforces the conclusion that, within the DKT regime, the influence of the LB’s motion on the system stability depends critically on its rotation direction, which is not incidental but is determined by the presence and strength of the connected recirculation within the inter-bubble gap.

As the aspect ratio decreases further, the inter-bubble recirculation disappears over the relevant parameter range, and the LB always rotates anti-clockwise, regardless of $S$.  
According to the above reasoning, this persistent counter-rotation should consistently destabilize the system.  
This prediction is confirmed in \hyperref[LB:fixed:chi1.6and2.0:neutralcurve]{figure~\ref{LB:fixed:chi1.6and2.0:neutralcurve}$(c)$}, which examines the bubble pair at $\chi = 1.3$.  
Here, fixing the LB increases $Re_{c}$ even at very small separations, in contrast to the behaviour observed for $\chi = 1.6$ and $1.9$.  
Since these small separations correspond to the DKT-type instability, the comparison across $\chi$ values demonstrates that the LB’s rotational response can either promote or suppress instability depending on its phase relationship with the TB’s motion.  

Taken together, these results demonstrate that the DKT and ASE regimes represent two distinct instability mechanisms of the in-line configuration.  
The DKT corresponds to a short-range, two-way coupling in which the motions of both bubbles strongly influence the system stability, while the ASE represents a long-range, essentially one-way interaction dominated by the TB’s response to the shear imposed by the LB wake.  
These two unstable scenarios are not sharply separated but can transform continuously: as reported by \citet{zhang2021three}, all DKT events disappear and transition to ASE when an initial angular displacement of $\Psi \!>\! 1^{\circ}$ is imposed on the in-line system.

\subsection{Effect of distinct aspect ratios of the two bubbles on stationary mode}
\label{sec:aspect_ratio}

As we stated in \hyperref[sec:Problem statement]{\S~\ref{sec:Problem statement}}, the leading and trailing bubbles in realistic in-line pair rarely possess identical shapes. Experiments and numerical simulations \citep{sanada2009motion,gumulya2017interaction,zhang2021three} consistently show that the TB is more spherical than the LB, as the low-pressure region in the LB wake tends to round the TB front and reduce its deformation. Configurations with $\chi_{TB} < \chi_{LB}$ are therefore typical, and it is reasonable to quantify how such shape asymmetry modifies the stability of the pair system.

\begin{figure}
	\centering
	\setlength{\tabcolsep}{0pt}
	\iflocalcompile
	\begin{tabular}{cc}
		\resizebox{0.5\textwidth}{!}{\input{\figfoursix/chi1.6-different-chi-new-2.tex}} &
		\resizebox{0.5\textwidth}{!}{\input{\figfoursix/chi1.9-different-chi-new-2.tex}}
	\end{tabular}
	\else
	\begin{tabular}{cc}
		\resizebox{0.5\textwidth}{!}{\includetikz{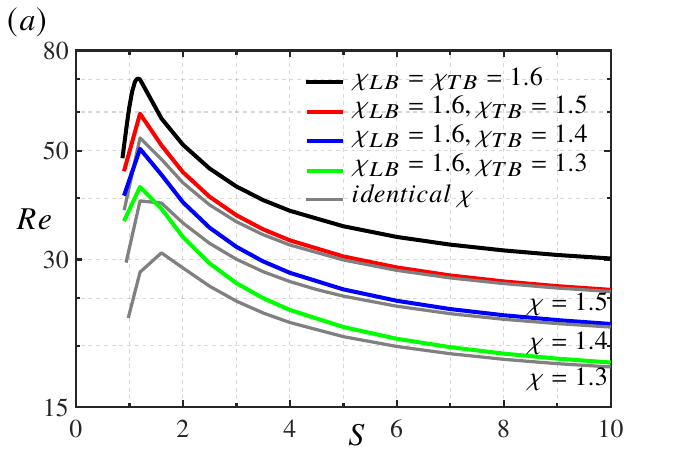}} &
		\resizebox{0.5\textwidth}{!}{\includetikz{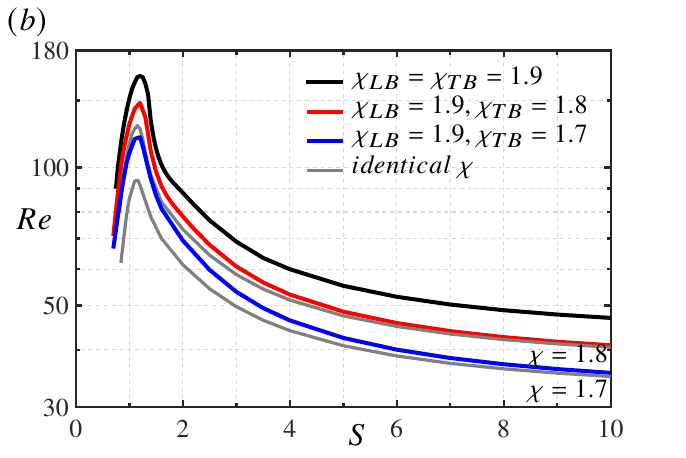}}
	\end{tabular}
	\fi
	\captionsetup{justification=justified, singlelinecheck=false, labelsep=period, width=\linewidth}
	\caption{Neutral stability curves of the stationary mode in the $(\Rey, S)$ plane for bubble pairs with unequal aspect ratios, i.e. a fixed LB aspect ratio $\chi_{LB}$ and a smaller TB aspect ratio $\chi_{TB} \leq \chi_{LB}$, representative of more realistic in-line configurations.
$(a)$ $\chi_{LB} = 1.6$, with $\chi_{TB} = 1.3$ (green), $1.4$ (blue), $1.5$ (red), and $1.6$ (black). Grey lines correspond to reference cases with identical aspect ratios, from bottom to top $\chi_{LB} = \chi_{TB} = 1.3$, $1.4$, and $1.5$.
$(b)$ $\chi_{LB} = 1.9$, with $\chi_{TB} = 1.7$ (blue), $1.8$ (red), and $1.9$ (black). Grey lines denote the reference cases with identical aspect ratios, from bottom to top $\chi_{LB} = \chi_{TB} = 1.7$ and $1.8$.}
	\label{chi1.6:different:chi}
\end{figure}

We first examine moderately oblate bubble pairs in which the deformation of the LB is fixed at $\chi_{LB} = 1.6$, while that of the TB is varied as $\chi_{TB} = 1.6$, $1.5$, $1.4$, and $1.3$. For reference, additional ALE–LSA results for pairs with identical but smaller aspect ratios,  namely $\chi_{LB} = \chi_{TB} = 1.5$, $1.4$, and $1.3$, are included to assess the influence of shape asymmetry between the two bubbles. The corresponding neutral curves are displayed in  \hyperref[chi1.6:different:chi]{figure~\ref{chi1.6:different:chi}$(a)$}. At large separations ($S > 4$), all unequal-$\chi$ configurations collapse onto the neutral curve  corresponding to the equal-$\chi$ case defined by $\chi_{TB}$. This convergence indicates that once the inter-bubble coupling weakens and the system enters the ASE regime,  the stability of the pair depends almost exclusively on the TB deformation. In this far-field limit, the wake of the LB merely provides an external shear environment,  while the inclination response and associated lift of the TB, governed by its own shape, dictate the overall stability. The LB’s contribution becomes passive, and direct two-way coupling plays only a secondary role.  

At smaller separations ($S < 4$), the neutral curves diverge and progressively separate from those of the symmetric reference cases ($\chi_{LB} = \chi_{TB}$). For instance, at $S = 1.2$, the neutral curve for  $(\chi_{LB}, \chi_{TB}) = (1.6, 1.5)$ lies between those corresponding to $\chi_{LB} = \chi_{TB} = 1.6$ and $\chi_{LB} = \chi_{TB} = 1.5$. In this near-field regime, approaching the DKT domain, the stability is governed by the  combined rotational responses of both bubbles.  
Each inclination generates a stabilizing torque opposing lateral drift, and reducing the deformation of either bubble weakens this feedback. Thus, at intermediate separations, the overall stability results from a balance between  the two coupled rotational feedbacks, whose relative strengths are set by the respective deformations of the LB and TB.  

For more oblate pairs with $\chi_{LB} = 1.9$ and $\chi_{TB} = 1.9$, $1.8$, and $1.7$, the stationary branches shown in \hyperref[chi1.6:different:chi]{figure~\ref{chi1.6:different:chi}$(b)$} exhibit similar trends to those described above and will not be discussed further.

\section{Oscillatory mode}
\label{sec:LSA_oscillatory}

As already shown in 
\hyperref[fig:LSA_comparison_with_rotation]{figure~\ref{fig:LSA_comparison_with_rotation}$(b)$}, the ALE–LSA results reveal that the neutral stability curve may exhibit an additional oscillatory branch depending on the parameters. This branch is characterized by a pair of complex-conjugate eigenvalues  $\lambda = \lambda_{r} \pm i\lambda_{i}$, corresponding to antisymmetric oscillations of the bubble pair about the vertical axis. The associated frequency $\lambda_{i}$ defines a reduced Strouhal number  $St = \lambda_{i}d/(2\pi V_{x})$. This finding is particularly remarkable because, for an isolated bubble,  the global oscillatory mode is well known to result from a Hopf bifurcation driven by periodic vortex shedding in the wake, the oscillation frequency being set by the shedding timescale \citep{tchoufag2014linearbubble,cano2016global}. In contrast, as demonstrated in  \hyperref[app:wake]{Appendix~\ref{app:wake}}, the flow past a frozen bubble pair remains entirely steady within the present parameter range. Hence, the oscillatory instability identified here differs fundamentally from the classical Hopf-type  path instability of an isolated bubble. Moreover, such an oscillatory mode has not been reported in any previous experiment or numerical simulation  for in-line bubble pairs below $\chi \approx 2.0$, highlighting the need for a dedicated investigation to uncover its physical origin.   

\subsection{Neutral curves}
\label{sec:oscillatory_curve}

\begin{figure}
	\centering
	\setlength{\tabcolsep}{0pt}
	\iflocalcompile
		\begin{tabular}{cc}
			\resizebox{0.5\textwidth}{!}{\input{\figfourfour/osci-neutralline.tex}} &
			\resizebox{0.5\textwidth}{!}{\input{\figfourfour/osci-neutralline-St.tex}}
		\end{tabular}
	\else
		\begin{tabular}{cc}
			\resizebox{0.5\textwidth}{!}{\includetikz{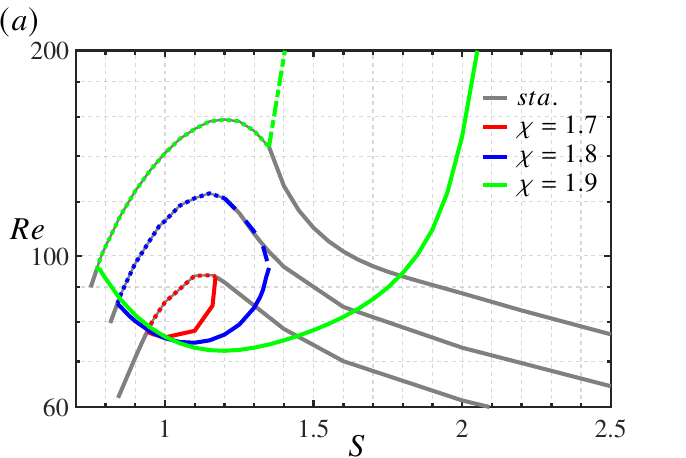}} &
			\resizebox{0.5\textwidth}{!}{\includetikz{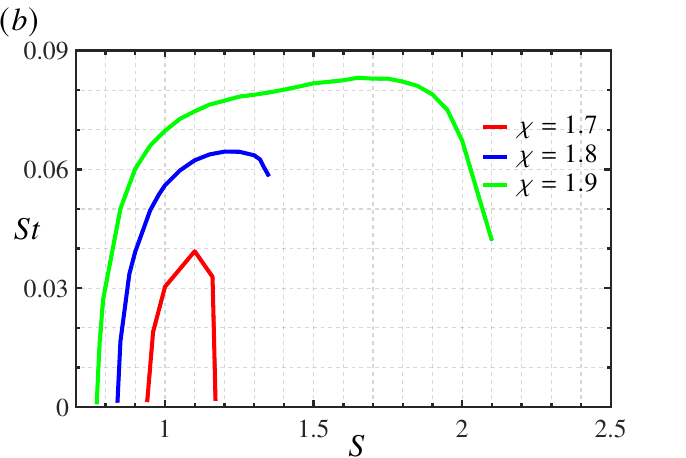}}
		\end{tabular}
	\fi
	\captionsetup{justification=justified, singlelinecheck=false, labelsep=period, width=\linewidth}
	\caption{
Neutral stability curves for the oscillatory and stationary modes at aspect ratios $\chi = 1.7$, $1.8$, and $1.9$.  
$(a)$ Neutral stability boundaries in the $(\Rey, S)$ plane.  
Coloured solid lines denote the thresholds of the oscillatory mode for different $\chi$, while grey lines represent the corresponding stationary branches.  
Dotted segments indicate \textbf{Case~(i)}, a \textit{mode conversion} where the oscillatory branch continuously transforms into a stationary one through a codimension-two ``exceptional" point.  
Dashed segments mark \textbf{Case~(ii)}, an \textit{alternating destabilization} regime in which the stabilization of the oscillatory branch coincides with the destabilization of the stationary one.  
The upwelling dash-dotted portion of the $\chi = 1.9$ curve corresponds to \textbf{Case~(iii)}, \textit{mode coexistence}, where both branches remain unstable over a finite range, each possessing its own threshold.  
$(b)$ Reduced frequency of the oscillatory mode, $St = \lambda_{i} d / (2\pi U_{0})$, as a function of separation distance $S$.
}
	\label{neutral:curve:osci}
\end{figure}

The neutral stability curves of the stationary and oscillatory branches are compared in \hyperref[neutral:curve:osci]{figure~\ref{neutral:curve:osci}$(a)$} for aspect ratios $\chi = 1.7$, $1.8$, and $1.9$, while the oscillatory mode disappears below $\chi = 1.6$. It arises only at small separations, and more specifically,  $S \lesssim 1.17$ for $\chi = 1.7$, $S \lesssim 1.35$ for $\chi = 1.8$, and $S \lesssim 2.1$ for $\chi = 1.9$, that is, precisely within the regime where a recirculation eddy connects the two bubbles (see \hyperref[sec:flow-structure]{\S~\ref{sec:flow-structure}}). The presence of this inter-bubble recirculation is essential,  as it provides the closed feedback loop required to sustain self-excited oscillations of the pair.  

This behaviour can be interpreted within a reduced-order hydrodynamic--spring analogy.
The recirculating flow developing in the inter-bubble region provides a lateral restoring response that may be characterized by an effective stiffness $K_{h}$, while the entrained surrounding liquid contributes an effective mass $M_{\mathrm{eff}}$.
Within this framework, the complex eigenvalue $\lambda = \lambda_{r} + i\lambda_{i}$ may be interpreted as that of a damped oscillator, with natural frequency $\omega_{0} = \sqrt{K_{h}/M_{\mathrm{eff}}}$ and damping ratio $\varsigma \sim -\lambda_{r}/\omega_{0}$.
Variations in the separation distance $S$ (or equivalently $\delta S$) and in the deformation aspect ratio $\chi$ modify the hydrodynamic coupling between the bubbles, thereby affecting both $K_{h}$ and $M_{\mathrm{eff}}$ and controlling the oscillation frequency and its amplification rate.
This analogy provides a useful framework for rationalizing the trends of $St$ reported in \hyperref[neutral:curve:osci]{figure~\ref{neutral:curve:osci}$(b)$}.
The dependence of the oscillation frequency on the separation distance $S$ is distinctly non-monotonic: $St$ increases with $S$ at small separations, reaches a maximum at an intermediate (but still small) distance, and then decreases as the bubbles move further apart.
At very small separations, the inter-bubble gap is strongly confined and supports a recirculating flow with pronounced pressure gradients.
A small lateral displacement of the trailing bubble then induces an asymmetric pressure distribution across the gap, resulting in a restoring force that increases rapidly as $\delta S$ decreases.
At the level of an order-of-magnitude estimate, this behaviour may be associated with a hydrodynamic stiffness scaling as $K_{h} \sim \Delta p/\delta S \sim \rho V_{\mathrm{rec}}^{2}/\delta S$, where $V_{\mathrm{rec}}$ denotes a characteristic velocity of the recirculating flow. At the same time, the effective inertia $M_{\mathrm{eff}}$ is expected to increase markedly as the gap narrows, owing to the strong co-acceleration of fluid trapped within the inter-bubble region. Previous studies on closely interacting bubbles and particles indicate that such confinement effects can lead to a rapid growth of added mass as the separation distance decreases \citep{takemura2003transverse}. Consistently, one may anticipate a scaling of the form $M_{\mathrm{eff}} \propto (1 + \delta S^{-n})$ with $n>2$, implying that, in the near-contact regime, the increase in effective inertia outweighs the growth of hydrodynamic stiffness. As a consequence, the ratio $K_{h}/M_{\mathrm{eff}}$, and hence the natural frequency $\omega_{0}$, remains limited and decreases as $\delta S \rightarrow 0$.
As the separation distance increases from this strongly confined regime, the effective inertia decreases more rapidly than the hydrodynamic stiffness, leading to an increase of $\omega_{0}$ and, consequently, of $St$.
The oscillation frequency reaches a maximum at an intermediate separation, corresponding to a configuration in which the balance between hydrodynamic stiffness and effective inertia is most favourable.
Beyond this distance, the hydrodynamic interaction between the bubbles weakens, the recirculating flow in the gap progressively decays, and the associated stiffness $K_{h}$ decreases, while $M_{\mathrm{eff}}$ remains close to its isolated-bubble value.
Consequently, both $\omega_{0}$ and $St$ decrease as the TB moves outside the region of strong interaction with the LB.
At sufficiently large separations, the diminishing hydrodynamic coupling suppresses the oscillatory mode, and the gap flow becomes effectively unidirectional.

Returning to \hyperref[neutral:curve:osci]{figure~\ref{neutral:curve:osci}$(a)$},  the neutral curves of the oscillatory branch form closed lobes in the $(S, \Rey)$ plane. For moderate aspect ratios ($\chi = 1.7$ and $1.8$), the instability remains confined to a narrow pocket  bounded by the stationary and oscillatory neutral lines. As the bubbles become more oblate ($\chi = 1.9$), this pocket widens and begins to overlap  with the stationary domain, reflecting the increased strength of the hydrodynamic coupling between the two bubbles.  
Within the hydrodynamic–spring framework, these lobes delineate parameter regions where the effective stiffness, $K_{h}$, and damping, $\varsigma$, change in sign or relative magnitude, driving the system from an underdamped (oscillatory) to an overdamped (stationary) response. Three generic transition scenarios can be distinguished:  
(i) \textit{Mode conversion}, in which the oscillatory mode continuously transforms into a stationary one, typically through a codimension-two “exceptional point” where a complex-conjugate pair of eigenvalues coalesces into two real roots; this behaviour corresponds to the dotted segments of the neutral curves. 
(ii) \textit{Alternating destabilization}, in which the instability domains of the oscillatory and stationary  branches remain distinct but alternate in dominance, such that the stabilization of one coincides with the destabilization of the other; this is observed along the dashed portions of the $\chi = 1.8$ curves. 
(iii) \textit{Mode coexistence}, where both branches remain unstable over a finite parameter range,  each possessing its own critical threshold, as indicated by the upwelling dash-dotted segment of the $\chi = 1.9$ curve.  
Among these three scenarios, the first two constitute the dominant regimes identified in the present analysis.  
They are discussed separately in the following sections, with emphasis on the underlying flow mechanisms  and the role of inter-bubble coupling in each case.

\subsection{Transitions of the oscillatory modes}
\label{sec:oscillatory_transition}

\begin{figure}
	\centering
	\setlength{\tabcolsep}{0pt}
	\iflocalcompile
		\begin{tabular}{cc}
			\resizebox{0.5\textwidth}{!}{\input{\figfourfour/casei/growth.tex}} &
			\resizebox{0.5\textwidth}{!}{\input{\figfourfour/casei/complexpath.tex}}
		\end{tabular}
	\else
		\begin{tabular}{cc}
			\resizebox{0.5\textwidth}{!}{\includetikz{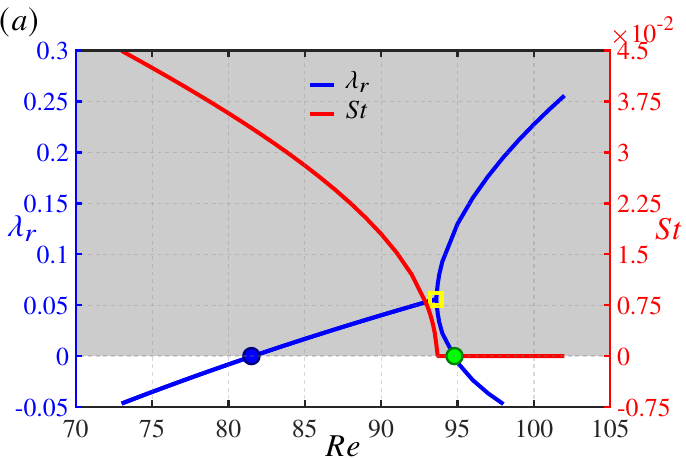}} &
			\resizebox{0.5\textwidth}{!}{\includetikz{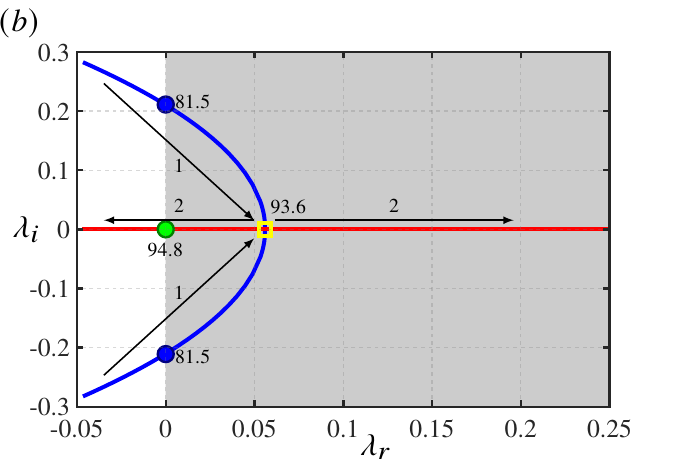}}
		\end{tabular}
	\fi
	\captionsetup{justification=justified, singlelinecheck=false, labelsep=period, width=\linewidth}
	\caption{Evolution of the leading eigenvalue with Reynolds number for $(\chi, S) = (1.8, 0.88)$, illustrating the transition scenario of Case~(i).  
$(a)$ Growth rate $\lambda_{r}$ (left axis) and reduced frequency $St$ (right axis) as functions of $\Rey$.  
$(b)$ Trajectory of the unstable eigenvalue in the complex plane, showing the merging and subsequent bifurcation of the conjugate pair.  
In both panels, the closed circles denote the transition points where $\lambda_{r} = 0$, and the open square marks the codimension-two ``exceptional’’ point at which the complex-conjugate pair associated with the oscillatory mode coalesces into two real eigenvalues, signalling the conversion from oscillatory to stationary instability. In panel~$(b)$, the numbered points correspond to the Reynolds numbers marking the transition thresholds identified in panel~$(a)$.
}
	\label{osci:sta1}
\end{figure}

\hyperref[osci:sta1]{Figure~\ref{osci:sta1}} illustrates the scenario of Case (i) for $\chi = 1.8$ and $S = 0.88$, where the instability switches from oscillatory to stationary as $\Rey$ increases.  
Panel~$(a)$ displays the growth rate $\lambda_{r}$ and the reduced frequency $St$ as functions of $\Rey$, while panel~$(b)$ shows the corresponding eigenvalue trajectories.  
At $\Rey \simeq 81$, a complex-conjugate pair crosses the imaginary axis, marking the onset of oscillatory instability.  
As $\Rey$ increases further, the imaginary part $\lambda_{i}$, and thus $St$, decreases continuously and vanishes near $\Rey \simeq 93.6$, where the two eigenvalues coalesce at a codimension-two exceptional point (open square) and split into two real branches.  
Beyond this point, one branch remains damped ($\lambda_{r}<0$) while the other becomes unstable ($\lambda_{r}>0$), producing a purely stationary instability that drives a steady lateral displacement of the TB.  
In the hydrodynamic–spring analogy, this exceptional point corresponds to the critical condition where the effective stiffness $K_{h}$ and damping $\varsigma$ exactly balance, converting the system from underdamped to overdamped behaviour.  
This sequence mirrors the oscillatory–stationary transition reported for isolated rising bubbles  
\citep{tchoufag2014linearbubble,bonnefis2024path}. In the present configuration, the ALE–LSA predicts an analogous conversion: an oscillatory mode arising at small separations that gradually evolves into a stationary one with increasing $\Rey$.

\begin{figure}
	\centering
	\setlength{\tabcolsep}{0pt}
	\iflocalcompile
	\begin{tabular}{cc}
		\resizebox{0.5\textwidth}{!}{\input{\figfourfour/caseii/growth.tex}} &
		\resizebox{0.5\textwidth}{!}{\input{\figfourfour/caseii/complexpath.tex}}
	\end{tabular}
	\else
	\begin{tabular}{cc}
		\resizebox{0.5\textwidth}{!}{\includetikz{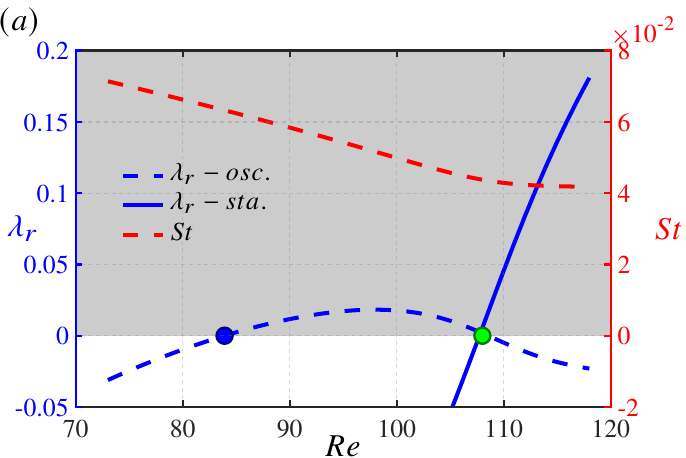}} &
		\resizebox{0.5\textwidth}{!}{\includetikz{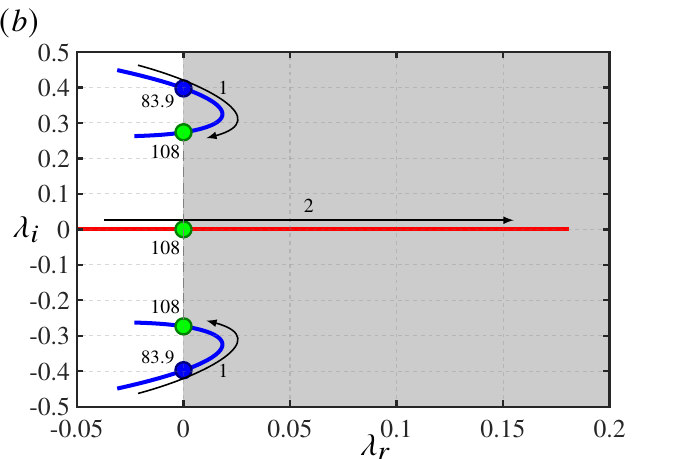}}
	\end{tabular}
	\fi
	\captionsetup{justification=justified, singlelinecheck=false, labelsep=period, width=\linewidth}
\caption{
Same as figure~\ref{osci:sta1}, but corresponding to the transition scenario of Case~(ii) at $(\chi, S) = (1.8, 1.3)$. $(a)$ Growth rate $\lambda_{r}$ and reduced frequency $St$ as functions of the Reynolds number $\Rey$.  
$(b)$ Trajectory of the unstable eigenvalue in the complex plane.
}
	\label{osci:sta2}
\end{figure}

For Case (ii), a distinct coupling mechanism is illustrated in \hyperref[osci:sta2]{figure~\ref{osci:sta2}} for $\chi = 1.8$ and $S = 1.3$, where the oscillatory branch stabilizes while the stationary branch becomes unstable, both transitions occurring near $\Rey \simeq 108$, as marked by the open symbol in the picture.  
As shown in \hyperref[osci:sta2]{figure~\ref{osci:sta2}$(a)$}, the two modes coexist but evolve nearly independently: the reduced frequency $St$ remains almost constant, indicating that the oscillatory branch retains its temporal nature up to its stabilisation threshold.  
With increasing $\Rey$, the complex-conjugate pair first grows unstable, then its amplification rate decreases and the pair re-enters the stable half-plane at $\Rey \simeq 108$.  
At the same Reynolds number, a distinct real eigenvalue crosses the imaginary axis, signalling the emergence of the stationary mode.  
The eigenvalue trajectories in \hyperref[osci:sta2]{figure~\ref{osci:sta2}$(b)$} highlight this clear exchange of stability.  
Such behaviour typifies global-mode competition: the amplification of one branch alters the base flow sufficiently to suppress the other.  
Within the hydrodynamic–spring picture, this exchange reflects a shift from reactive to quasi-static coupling, as the wake-induced interaction becomes dominated by steady pressure asymmetries rather than time-dependent recirculating motions.  
Physically, as $\Rey$ increases, the inter-bubble shear and deformation intensify, and the quasi-steady asymmetry in the gap ultimately outweighs the time-dependent recirculation, allowing the stationary mode to supersede the oscillatory one.

\begin{figure}
	\centering
	\setlength{\tabcolsep}{0pt}
	\iflocalcompile
	\begin{tabular}{cc}
			\resizebox{0.5\textwidth}{!}{\input{\figfourfour/caseiii/growth.tex}} &
			\resizebox{0.5\textwidth}{!}{\input{\figfourfour/caseiii/complexpath.tex}}
	\end{tabular}
	\else
	\begin{tabular}{cc}
		\resizebox{0.5\textwidth}{!}{\includetikz{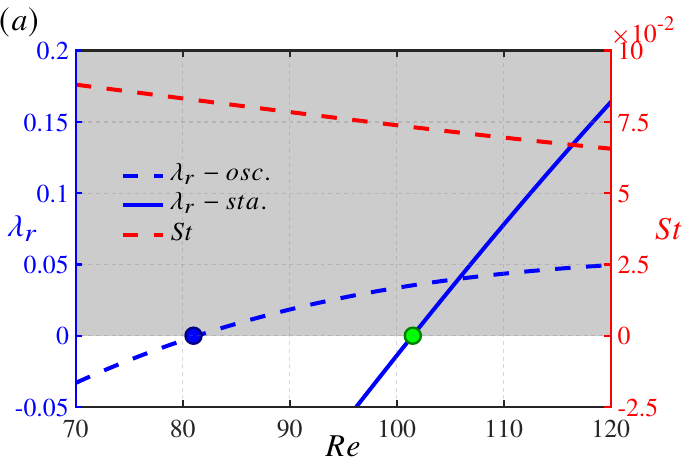}} &
		\resizebox{0.5\textwidth}{!}{\includetikz{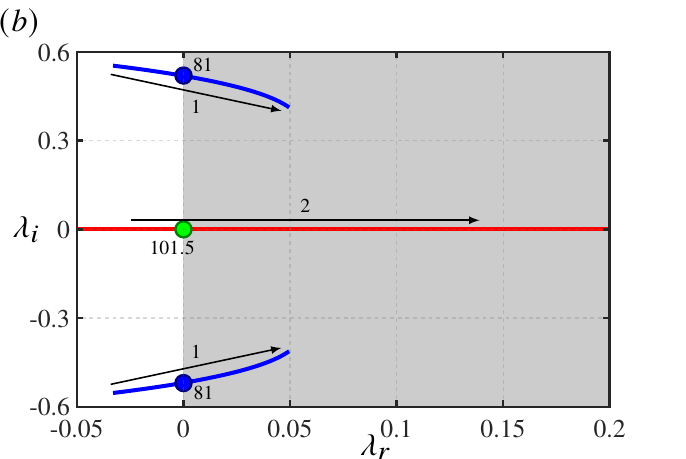}}
	\end{tabular}
	\fi
	\captionsetup{justification=justified, singlelinecheck=false, labelsep=period, width=\linewidth}
	\caption{Same as figure~\ref{osci:sta1}, but corresponding to the transition scenario of Case (iii) at $(\chi, S) = (1.9, 1.6)$.
$(a)$ Growth rate $\lambda_{r}$ and reduced frequency $St$ as functions of the Reynolds number $\Rey$.
$(b)$ Trajectory of the unstable eigenvalue in the complex plane, given numbers correspond to the threshold of $Re$ identified in panel $(a)$.}
	\label{osci:sta3}
\end{figure}

Case~(iii), representative of larger aspect ratios and wider separations  (e.g.\ $\chi = 1.9$, $S = 1.6$), corresponds to a regime in which the oscillatory and stationary branches coexist without notable spectral interaction, as illustrated in \hyperref[osci:sta3]{figure~\ref{osci:sta3}}, while beyond the explored range, the oscillatory branch is expected to restabilise, although a complete description would require extending the analysis to larger Reynolds numbers.
In this parameter region, the two modes remain well separated in frequency, implying that any coupling arises only through nonlinear effects once finite-amplitude motion develops. A weakly nonlinear analysis, similar to that of \citet{fabre2012steady} for buoyancy-driven bodies, would be necessary to quantify this interaction and the ensuing amplitude saturation. Within the hydrodynamic–spring framework, Case~(iii) corresponds to a configuration where the effective stiffness $K_{h}$ is moderate and the damping $\varsigma$ small but positive, so that the coupled system exhibits two natural responses: a low-frequency, quasi-static deformation (stationary mode) and a higher-frequency, weakly damped oscillation (oscillatory mode). At low to moderate $\Rey$, the oscillatory branch governs the TB motion, producing small-amplitude lateral oscillations. As $\Rey$ increases, viscous damping weakens and the quasi-steady asymmetry of the flow strengthens, causing the stationary branch to dominate and leading ultimately to a steady lateral escape of the TB.

\begin{figure}
	\centering
	\setlength{\tabcolsep}{0pt}
	\begin{tabular}{cccc}
		\includegraphics[width=0.25\textwidth]{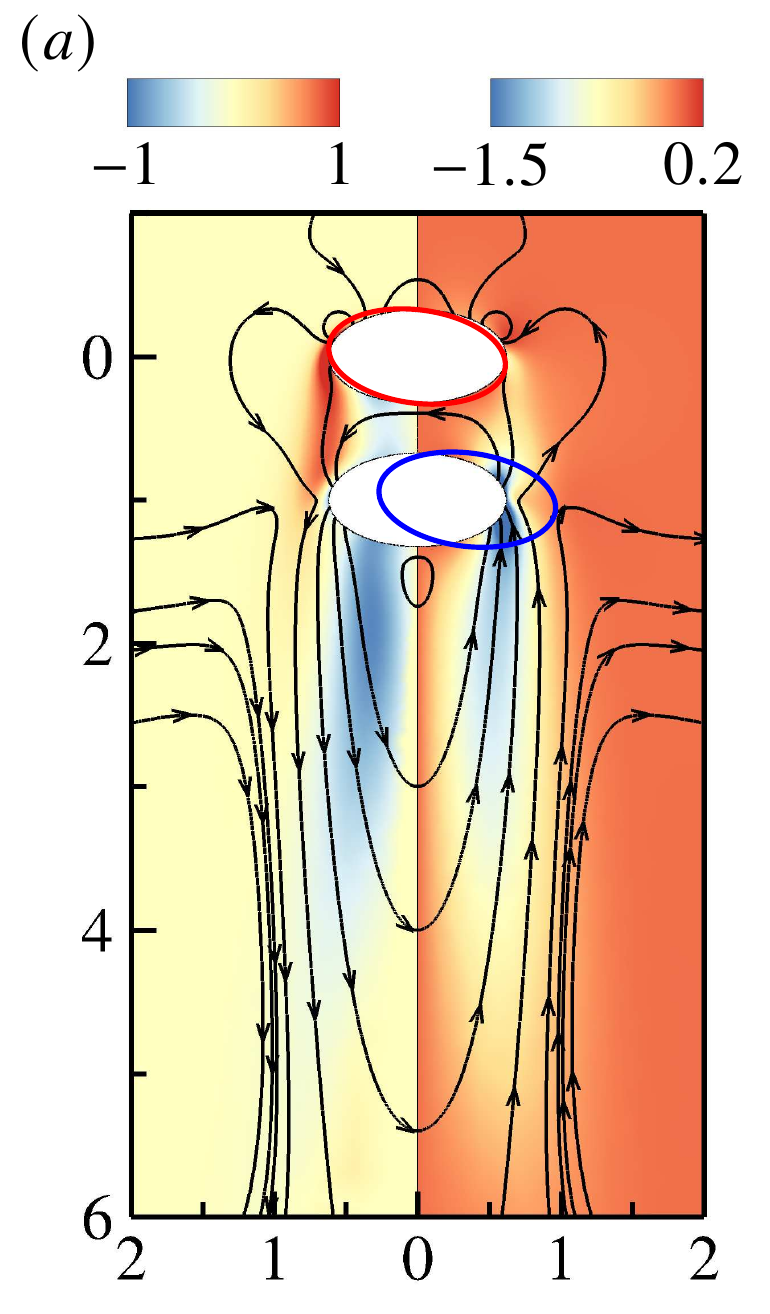} &
		\includegraphics[width=0.25\textwidth]{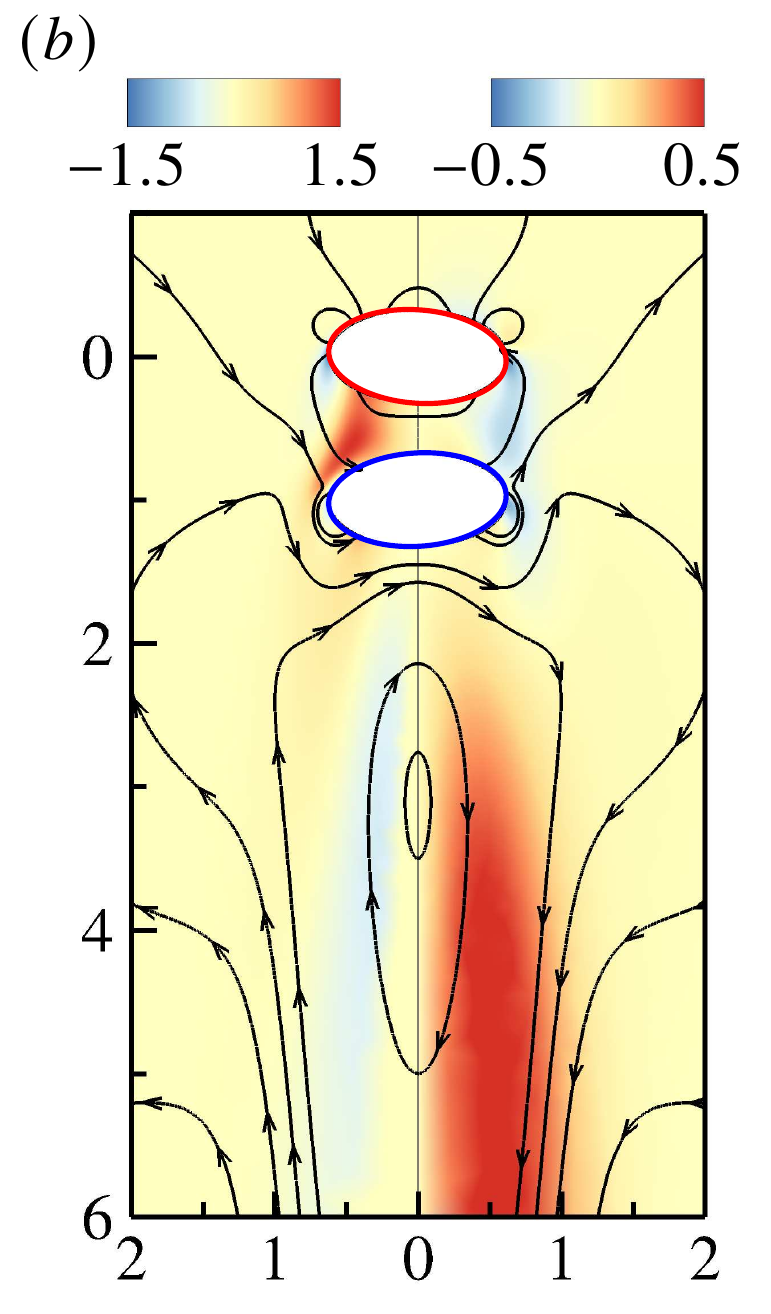} &
		\includegraphics[width=0.25\textwidth]{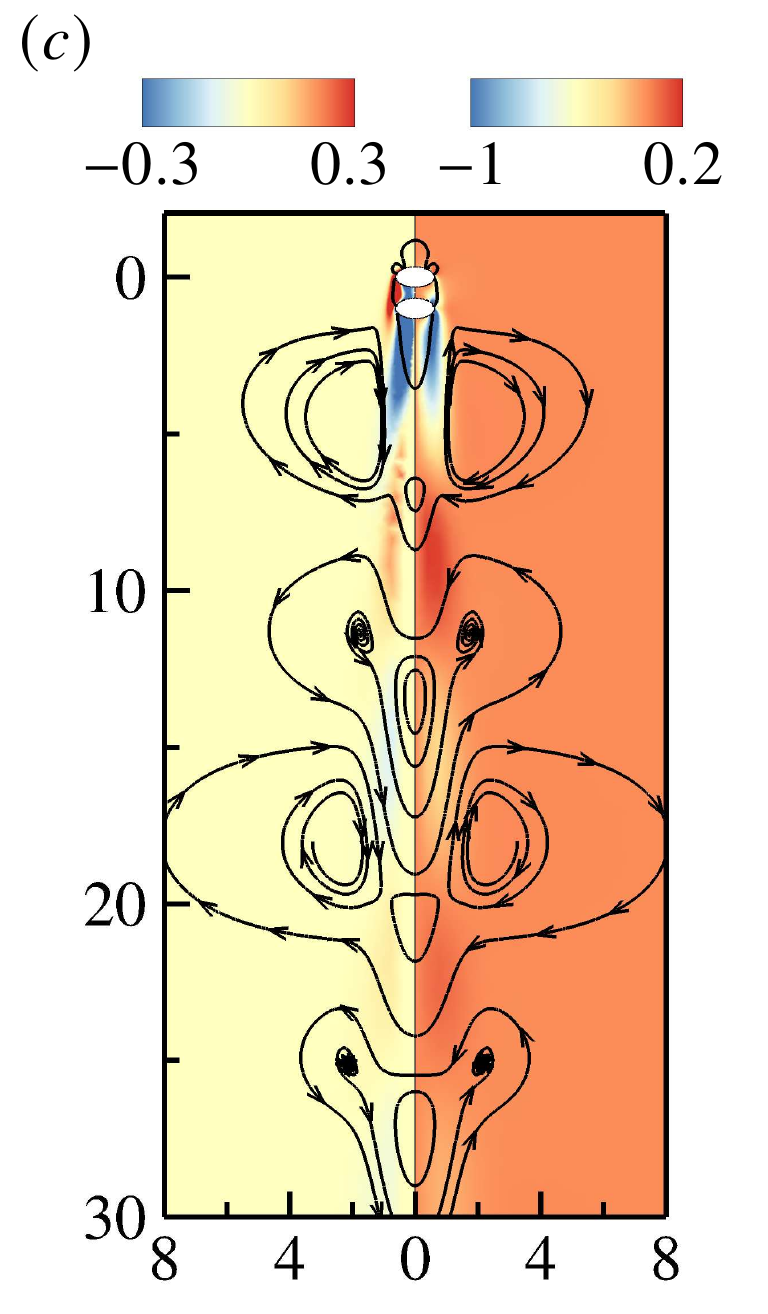} &
		\includegraphics[width=0.25\textwidth]{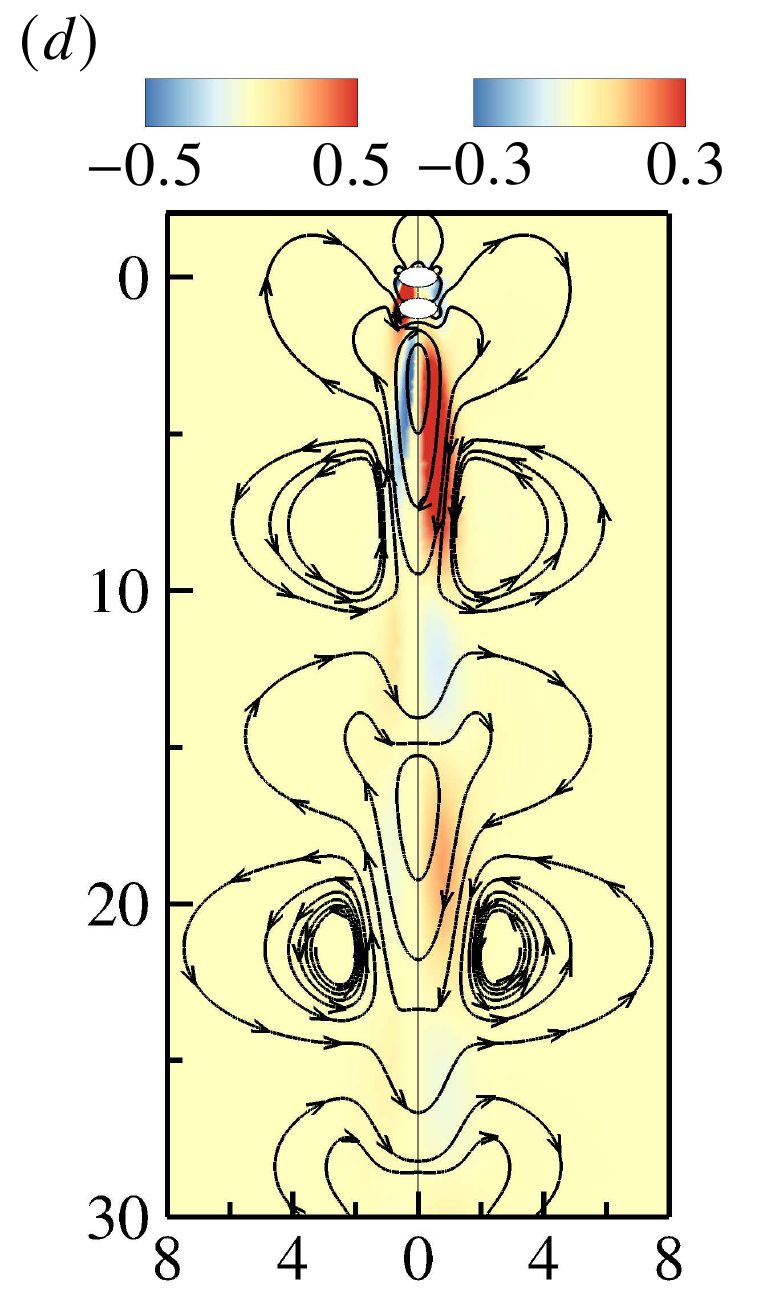}
	\end{tabular}

	\captionsetup{justification=justified, singlelinecheck=false, labelsep=period, width=\linewidth,format=plain}
	\caption{Spatial structure of the oscillatory mode near the onset of instability for $(\Rey, \chi, S) = (77, 1.9, 1.0)$.
In each panel, the left and right halves display iso-contours of streamwise vorticity and velocity, respectively, with selected streamlines shown in the reference frame of the base flow.
Red and blue contour lines denote the maximum angular displacements ($\hat{\eta}$) of the LB and TB, respectively.
$(a,c)$ Real part of the eigenmode, showing the flow disturbance near the bubble pair and further downstream in the wake.
$(b,d)$ Corresponding imaginary part of the eigenmode.}
	\label{osci:eigenmode}
\end{figure}

The spatial organisation of the oscillatory mode is illustrated in \hyperref[osci:eigenmode]{figure~\ref{osci:eigenmode}} for $\Rey = 77$, $\chi = 1.9$, and $S = 1.0$, i.e.\ close to the onset of instability.  
Panels~(\textit{a,c}) and~(\textit{b,d}) display, respectively, the real and imaginary parts of the eigenmode, which correspond to two phases of the oscillation separated by a quarter period, $\Delta t = \pi/(2\lambda_{i})$.  Their linear superposition reconstructs the time-periodic evolution of the disturbance over one full cycle.  This branch is associated with a pair of complex-conjugate eigenvalues and two mirror-symmetric azimuthal components ($|m|=1$).  
A purely planar combination of these components produces a zigzag trajectory, whereas a $\pi/2$ phase shift generates a helical path.  As shown in \hyperref[osci:eigenmode]{figure~\ref{osci:eigenmode}}, both bubbles undergo small but finite angular displacements, the trailing bubble exhibiting a larger amplitude owing to the stronger local shear in the wake of the leading one.  The perturbation field reveals intense streamwise vorticity concentrated within the inter-bubble gap and extending into the wake of the trailing bubble. Alternating regions of positive and negative vorticity appear both across the horizontal planes and along the streamwise direction, forming a longitudinal double-threaded structure reminiscent of the wake topology of an oscillating oblate bubble \citep{tchoufag2014globaldisk,bonnefis2024path}.  
The corresponding streamlines are antisymmetric with respect to the vertical mid-plane, confirming the counter-phase motion of the two bubbles.  This mode therefore represents a genuinely global instability, in which the rotation of the TB is synchronised with the unsteady vorticity sustained by the recirculation in the wake of the LB.  
The resulting coupling between the two bubbles through the standing eddy provides the physical backbone of the oscillatory instability.

To sum up, within the hydrodynamic–spring framework,  the oscillatory branch corresponds to an underdamped oscillation of a coupled fluid–spring system.
The inter-bubble recirculation acts as an elastic element that stores and releases energy,  the bubble inertia provides the effective mass, and viscous dissipation in the shear layer determines the damping. 
This oscillatory regime has not been reported in previous experimental or numerical studies.   Even in the recent VOF–DNS work of \citet{zhang2021three}, the bubble pair with $(\Rey, \chi) = (120, 2.0)$ falls within the oscillatory domain predicted in \hyperref[neutral:curve:osci]{figure~\ref{neutral:curve:osci}$(a)$}. However, upon re-examining their simulations,  no sustained oscillations were observed in the motion of either bubble prior to collision. This absence likely results from the narrowness of the parameter window  in which the oscillatory instability develops, typically restricted to short separations where the TB approaches the LB closely. In this limit, the lifetime of the oscillatory motion is short,  and nonlinear effects rapidly emerge, often obscuring the linear mechanism. A similar situation was reported by \citet{tchoufag2014linearbubble} and \citet{bonnefis2024path} for isolated bubbles, where the stationary oblique mode predicted by linear theory was rarely observed experimentally, owing to nonlinear saturation that quenches this branch before it dominates. Accordingly, dedicated DNS at $\chi \geq 1.7$  would be required in the future to delineate more precisely the extent of the oscillatory regime and to clarify its nonlinear evolution prior to bubble collision.

Finally, the hydrodynamic–spring analogy also suggests that,  as in the stationary mode, both the mutual interaction and the difference in bubble aspect ratios should significantly influence the oscillatory branches, particularly near the transitions between distinct modes. These effects are discussed in detail in \hyperref[app:oscillatory]{Appendix~\ref{app:oscillatory}} to avoid interrupting the main line of reasoning.

\section{Summary and conclusion}
\label{sec:conclusion}

The present study has elucidated the physical mechanisms governing the stability of two bubbles rising initially in-line through a viscous liquid.  
Using a global linear stability analysis formulated within an Arbitrary Lagrangian–Eulerian framework,  complemented by fully resolved three-dimensional simulations based on an Embedded Boundary Method, we have shown how hydrodynamic coupling, bubble rotation, and wake topology jointly control the stability of the pair.  
Three main contributions emerge from this work.  
First, the analysis reveals that the stabilization promoted by increasing bubble oblateness originates primarily from an inclination-induced rotational feedback,  rather than from a deformation-enhanced wake entrainment as previously assumed. Second, the results clarify the dynamical distinction between the  drafting–kissing–tumbling and asymmetric side-escape evolutions, which are shown to correspond to different global response modes of the coupled bubble–flow system. Finally, in addition to the classical stationary instability, a previously unreported oscillatory global mode has been identified and characterized.  

The results demonstrate that rotational freedom is an essential element of the coupled dynamics.   When rotation is artificially constrained, the analysis predicts unrealistically low instability thresholds, whereas allowing rotation restores quantitative agreement with three-dimensional direct numerical simulations. Rotation introduces a new class of lift, absent in spherical bubbles,  arising from the coupling between the bubble’s inclination and the asymmetric shear in the leader’s wake. A small lateral displacement of the trailing bubble generates a shear-induced torque that induces a compensating inclination,  and this reorientation modifies the surface-pressure field and produces a restoring lift directed toward the symmetry axis. The resulting inclination-induced feedback counteracts the destabilising potential and shear-induced lifts,  thereby explaining both the increase of the critical Reynolds number with bubble oblateness and its non-monotonic dependence on the inter-bubble spacing. The stationary mode thus corresponds to a steady lateral drift governed by the competition between destabilising hydrodynamic interactions and stabilising rotational torque,  consistent with transitions observed in experiments ad DNS.  

Two distinct stationary sub-modes of instability are further examined.   The DKT mode arises from a short-range, two-way coupling between the bubbles, in which the rotation direction of the leading bubble, governed by the connected recirculation in the inter-bubble gap, may either stabilise or destabilise the configuration. In contrast, the ASE mode corresponds to a long-range, one-way coupling, and  the LB merely imposes a shear environment that destabilises the TB, whose inclination and associated lift entirely control the instability.  

Furthermore, the oscillatory mode, not reported previously, arises from an unsteady coupling between the two bubbles through the standing recirculation formed in the inter-bubble gap.   This vortical structure behaves as a hydrodynamic spring: when the TB moves laterally, the induced cross-gap flow modifies the pressure and torque acting on the LB, which in turn alters the wake impinging on the TB. The resulting phase-lagged feedback sustains self-excited oscillations whose frequency varies non-monotonically with separation,  reflecting the interplay between the spring stiffness and the inertial response of the pair.  

The present analysis extends the classical framework of \citet{hallez2011interaction} by identifying a fourth, inclination-induced lift component that dominates for oblate bubbles. It quantitatively reproduces the transition between stable in-line and unstable escape configurations reported by \citet{zhang2021three}, but further demonstrates that the onset of instability in bubble pairs is not governed by the trailing-bubble wake. Instead, it results from a genuinely global feedback involving both bubbles and the vortical conduit linking them. 

Beyond the in-line configuration examined here, this mechanism provides a physical foundation for interpreting the collective dynamics of bubble swarms, where similar rotational and vortical couplings govern the emergence of vertical chains or lateral clustering. Future work will extend this framework to larger aspect ratios and to temporally deforming shapes,  in order to explore how wake instability and instant shape fluctuations interact to generate the complex, unsteady paths characteristic of realistic bubbly flows.

\section{Acknowledgement}

The authors gratefully acknowledge the support of the National Key R$\&$D Program of China under grants number 2023YFA1011000 and 2022YFE03130000, that of the NSFC under grants numbers W2511004 and 12472256. 
\vspace{2mm}\\
{\textbf{Declaration of Interests}}. The authors report no conflict of interest.\\
\vspace{1mm}\\
{\textbf{Author ORCIDs}}\\
{
Jie Zhang, ~https://orcid.org/0000-0002-2412-3617.
}

\appendix
\section{Details of the ALE–LSA method}
\label{app:ALE_details}

The ALE–LSA solver employed in the present study was developed in-house and closely follows the numerical framework introduced by \citet{bonnefis2024path}. Comprehensive implementation details are provided in the thesis of \citet{bonnefis2019etude}; here, we summarise only the essential formulation relevant to the bubble-pair configuration considered in this work. \hyperref[sec:A1:BF]{Section~\ref{sec:A1:BF}} introduces the base-flow problem and its governing equations. \hyperref[sec:A2:ALE]{Section~\ref{sec:A2:ALE}} presents the formulation of the perturbed system within the ALE framework and discusses the corrections introduced by the moving-grid description. Finally, \hyperref[Linearised:ALE:formulation]{Section~\ref{Linearised:ALE:formulation}} details the linearisation of the governing equations and derives the corresponding perturbation equations in ALE form.

\subsection{Base-flow equations}
\label{sec:A1:BF}

As described in \hyperref[sec:Problem]{\S~\ref{sec:Problem}}, the steady base flow, $\boldsymbol{\mathscr{Q}}^f_{0} = [\,\boldsymbol{U}_0(x,r), P_0(x,r)\,]$, is computed in a reference frame translating with the bubble pair. Following \citet{mougin2001path,mougin2002generalized,tchoufag2014globaldisk,tchoufag2014linearbubble}, the absolute velocity field $\boldsymbol{U}_0$ is projected onto this relative frame, which effectively fixes the bubbles and facilitates the stability analysis.

In this frame, the governing equations of the steady, axisymmetric base flow reduce to:
\begin{subequations} \label{base:flow:eq:abs}
\begin{align}
\nabla \!\cdot\! \boldsymbol{U}_0 &= 0, 
&& \text{in } \Upsilon_{0}, 
\label{BF:incompressibel:abs} \\[3pt]
\rho_0 \left( \boldsymbol{U}_0 + V_x \boldsymbol{e}_x \right) \!\cdot\! \nabla \boldsymbol{U}_0 
&= \nabla \!\cdot\! \boldsymbol{\Sigma}(\boldsymbol{U}_0, P_0), 
&& \text{in } \Upsilon_{0}, 
\label{BF:momentum:abs}
\end{align}
\end{subequations}
where $\boldsymbol{\Sigma}(\boldsymbol{U}_0, P_0) = -P_0\,\mathbb{I} + \mu(\nabla \boldsymbol{U}_0 + \nabla \boldsymbol{U}_0^{T})$  is the dimensional fluid stress tensor. On each bubble surface $\mathscr{S}_0$, the interface is non-penetrable and shear-free, yielding the boundary conditions:
\begin{equation}\label{BF:BC:abs}
\boldsymbol{U}_0 \!\cdot\! \boldsymbol{n} = -V_x\, \boldsymbol{e}_x \!\cdot\! \boldsymbol{n}, 
\qquad
\boldsymbol{n} \!\times\! \bigl(\boldsymbol{\Sigma}(\boldsymbol{U}_0, P_0)\!\cdot\! \boldsymbol{n}\bigr) = \boldsymbol{0},
\qquad 
\text{on } \mathscr{S}_0.
\end{equation}
At infinity ($|\boldsymbol{x}|\!\to\!\infty$), the velocity field decays to zero: $\boldsymbol{U}_0 \to \boldsymbol{0}$.
This formulation is consistent with the Eulerian-frame description of the perturbed bubble-pair system  introduced in \hyperref[sec:governing equations]{\S~\ref{sec:governing equations}}, governed by Eqs.~\eqref{NS-Eulereq}. For flow visualisation purposes (see \hyperref[sec:BaseFlow-section]{\S~\ref{sec:BaseFlow-section}}), it is often convenient to define the relative velocity:
\begin{equation}\label{trans:abstorel}
\boldsymbol{U}_0^{rel} = \boldsymbol{U}_0 + V_x \boldsymbol{e}_x,
\end{equation}
which represents the velocity in the frame moving with the bubbles.  
Both representations are equivalent:  
$\boldsymbol{U}_0$ is retained in the ALE and linear stability formulations for consistency with the governing equations, while $\boldsymbol{U}_0^{rel}$ is used primarily for illustrative purposes, as it more clearly reveals  the wake and inter-bubble flow structures.

\subsection{ALE formulation of the perturbed system}
\label{sec:A2:ALE}

For perturbations of azimuthal order $|m|=1$, the bubble pair deviates from its initial in-line configuration, producing relative motion between the bubbles and a time-dependent physical domain $\Upsilon(t)$.   To decompose the system into a steady base flow and a small perturbation, $\mathscr{Q} = \mathscr{Q}_0 + \epsilon\,\boldsymbol{q}'$, the time-varying domain is mapped onto a fixed reference domain,  naturally chosen as the fluid region of the base flow. This mapping defines the ALE framework $\boldsymbol{x} = \boldsymbol{x}_0 + \boldsymbol{\xi}_0(\boldsymbol{x}_0, t)$, where $\boldsymbol{x}_0$ denotes the position in the reference configuration and $\boldsymbol{\xi}_0$ is the local displacement of the mesh. The ALE formulation is applied only to the flow variables $[\boldsymbol{u},p]$, while the bubble dynamics, governed by \eqref{Bubble-dynamics}, remain fully Lagrangian.  
The mapping defined above provides the foundation for reformulating both the material derivative and the spatial differential operators in ALE form, as detailed below.

To clarify the effect of the ALE mapping, it is convenient to distinguish  three sets of coordinates: the \emph{Lagrangian} coordinates $\boldsymbol{x}_L$, the \emph{Eulerian} coordinates $\boldsymbol{x}$, and the \emph{ALE} coordinates $\boldsymbol{x}_0$ defined on the reference domain. Their relationships are expressed as:
\begin{equation} \label{Coordinate:relationship}
    \boldsymbol{x} = \phi(\boldsymbol{x}_L, t), 
    \qquad 
    \boldsymbol{x} = \varphi(\boldsymbol{x}_0, t),
    \qquad
    \boldsymbol{x}_0 = \psi(\boldsymbol{x}_L, t).
\end{equation}
A scalar field $f$ may thus be written equivalently as $f(\boldsymbol{x}_L,t)$, $f(\boldsymbol{x},t)$, or $f(\boldsymbol{x}_0,t)$, depending on the chosen frame of reference.
For the conventional Lagrangian and Eulerian descriptions, the material derivative is given by:
\begin{equation} \label{Lag:to:Euler}
    \frac{d f(\boldsymbol{x}_L,t)}{d t}
    = \frac{\partial f(\boldsymbol{x},t)}{\partial t}
    + \boldsymbol{u} \!\cdot\! \nabla_{\boldsymbol{x}} f,
\end{equation}
where $\boldsymbol{u} = \partial \boldsymbol{x}(\boldsymbol{x}_L,t) / \partial t = \partial \phi(\boldsymbol{x}_L,t) / \partial t$ is the local fluid velocity in the Eulerian frame.
In the ALE formulation, the time-dependent mapping between the physical and reference domains must be accounted for when evaluating time derivatives.  
Applying the chain rule to the mapping 
$\boldsymbol{x} = \varphi(\boldsymbol{x}_0,t) 
 = \boldsymbol{x}_0 + \boldsymbol{\xi}_0(\boldsymbol{x}_0,t)$  
yields the ALE form of the material derivative:
\begin{equation} \label{Lag:to:ALE:3}
    \frac{d f(\boldsymbol{x}_L,t)}{d t}
    = \frac{\partial f(\boldsymbol{x}_0,t)}{\partial t}
    + \left[\,\boldsymbol{u}(\boldsymbol{x},t)
    - \frac{\partial \boldsymbol{\xi}_0(\boldsymbol{x}_0,t)}{\partial t}\,\right]
    \!\cdot\! \nabla_{\boldsymbol{x}} f(\boldsymbol{x},t).
\end{equation}
The same expression applies to vector fields.  
The additional term, $\partial \boldsymbol{\xi}_0(\boldsymbol{x}_0,t)/\partial t$, represents the velocity of the moving reference mesh. It effectively modifies the convective velocity, accounting for the relative motion between the deforming physical domain  and the fixed reference configuration.

We now address the modification of the spatial differential operators induced by the ALE mapping.  
The Piola transformation is derived from the requirement that the integral forms of the conservation laws for mass and momentum remain invariant under the mapping between the physical and reference configurations.
Under the mapping $\boldsymbol{x} = \varphi(\boldsymbol{x}_0, t)$, the infinitesimal line, surface, and volume elements transform according to:
\begin{subequations}\label{transformations}
\begin{align}
    d\boldsymbol{x} &= \boldsymbol{F}_0\, d\boldsymbol{x}_0, 
    &\hfill
    d\boldsymbol{x}_0 &= \boldsymbol{F}_0^{-1}\, d\boldsymbol{x}, \\[3pt]
    d\boldsymbol{s} &= J_0\, \boldsymbol{F}_0^{-T}\, d\boldsymbol{s}_0, 
    &\hfill
    d\boldsymbol{s}_0 &= J_0^{-1}\, \boldsymbol{F}_0^{T}\, d\boldsymbol{s}, \\[3pt]
    d\Upsilon &= J_0\, d\Upsilon_0, 
    &\hfill
    d\Upsilon_0 &= J_0^{-1}\, d\Upsilon,
\end{align}
\end{subequations}
where $\boldsymbol{F}_0 = \partial \boldsymbol{x}/\partial \boldsymbol{x}_0$ 
is the deformation gradient tensor and 
$J_0 = \det(\boldsymbol{F}_0)$ its Jacobian determinant.
To preserve the integral form of the conservation equations within the ALE framework, the volume integral of any scalar field $f$ over the physical domain must be equivalent to that over the reference domain.  
This condition reads:
\begin{equation}\label{key}
    \int_{\Upsilon_0} f_0(\boldsymbol{x}_0, t)\, d\Upsilon_0
    = \int_{\Upsilon(t)} f(\boldsymbol{x}, t)\, d\Upsilon
    = \int_{\Upsilon_0} J_0\, f(\boldsymbol{x}(\boldsymbol{x}_0, t), t)\, d\Upsilon_0.
\end{equation}
Since $\Upsilon_0$ is arbitrary, this equivalence implies  $f_0 = J_0 f$ and, conversely, $f = J_0^{-1} f_0$.  
For vector and tensor fields, ensuring the equivalence of flux integrals under the ALE mapping uniquely determines the corresponding Piola transformations.  
Consequently, all spatial differential operators appearing in the governing equations must be consistently transformed onto the reference domain.

Because the full expressions for vector and tensor transformations are lengthy, only the relevant results are summarised here. Combining the corrected material derivative (derived in \hyperref[sec:A2:ALE]{\S~\ref{sec:A2:ALE}}) with the Piola-based transformation of the spatial operators, the Navier–Stokes equations in the Eulerian frame~\eqref{NS-Eulereq} can be rewritten in their ALE-consistent form as Eq.~\eqref{NS-ALEeq}. In this framework, the stress tensor appearing in the momentum equation is expressed in terms of the first Piola–Kirchhoff tensor $\mathbb{L}_0$, defined by:
\begin{equation}\label{first:Piola:kirchhoffL:stres:tensor}
    \mathbb{L}_0(\boldsymbol{u}_0, p_0, \boldsymbol{\xi}_0) = \boldsymbol{\sigma}_0(\boldsymbol{u}_0, p_0, \boldsymbol{\xi}_0)\,      \boldsymbol{\Phi}_0^T(\boldsymbol{\xi}_0),
\end{equation}
where $\boldsymbol{\sigma}_0$ denotes the Cauchy stress tensor expressed in the stress-free configuration,
\begin{equation}\label{Cauchy:stress:tensor}
    \boldsymbol{\sigma}_0(\boldsymbol{u}_0, p_0, \boldsymbol{\xi}_0) = -p_0 \, \mathbb{I} + \mu\, J_0^{-1} \left[\left(\nabla_0 \boldsymbol{u}_0\right)\boldsymbol{\Phi}_0(\boldsymbol{\xi}_0) + \boldsymbol{\Phi}_0^{T}(\boldsymbol{\xi}_0)\left(\nabla_0 \boldsymbol{u}_0\right)^{T}\right].
\end{equation}

In summary, the introduction of the mesh displacement $\boldsymbol{\xi}_0$ allows the time-dependent physical domain $\Upsilon(t)$ to be mapped onto a fixed reference configuration $\Upsilon_0$.  This transformation enables the Navier–Stokes equations, originally formulated in the Eulerian description, to be expressed consistently within the ALE framework, thereby facilitating the linear stability analysis of the deforming bubble-pair system.

\subsection{Linearised ALE formulation}
\label{Linearised:ALE:formulation}

Within the ALE framework, the reference domain $\Upsilon_{0}$ remains fixed in time, allowing the flow variables to be decomposed into a steady base flow and a time-dependent perturbation field. Equations~\eqref{base:flow:eq:abs}, \eqref{BF:BC:abs}, and \eqref{NS-ALEeq} describe, respectively,  
the governing equations for the base flow and the complete system expressed in the ALE framework,  with all quantities defined on the reference domain $\Upsilon_{0}$. Subtracting the base-flow equations from the full ALE system yields the governing equations for the perturbations:
\[
\boldsymbol{q}' = [\,\boldsymbol{u}',\,p',\,\boldsymbol{\xi}_0,\,\boldsymbol{X}'_{k},\,\boldsymbol{\varXi}'_{k},\,\boldsymbol{V}'_{k},\,\boldsymbol{\varOmega}'_{k}\,].
\]
Linearising these equations with respect to the perturbation amplitude $\epsilon$ then provides the linearised governing system for disturbances in the ALE framework:
\allowdisplaybreaks 
\begin{subequations}\label{ALE-LSA:eq} 
\begin{align}
\rho_0\frac{\partial\boldsymbol{u}'}{\partial t} &+ \rho_0\left( \left(\boldsymbol{U}_0 + V_x \boldsymbol{e}_x\right)\cdot\nabla_0\boldsymbol{u}' + \left(\boldsymbol{u}' - \boldsymbol{V}'_{LB} \right) \!\cdot\! \nabla_0\boldsymbol{U}_0 \right) \notag \\
 & -\rho_0\frac{\partial\boldsymbol{\xi}_0}{\partial t}\cdot\nabla_0\boldsymbol{U}_0 + \rho_0(\nabla_0\boldsymbol{U}_0)\boldsymbol{\Phi}^{\prime}(\boldsymbol{\xi}_0)\left(\boldsymbol{U}_0 + V_x \boldsymbol{e}_x\right) \notag \\ 
 &= \nabla_0\!\cdot\! \left[ \boldsymbol{\Sigma}^{\prime}(\boldsymbol{U}_0,P_0,\boldsymbol{\xi}_0) + \boldsymbol{\Sigma} (\boldsymbol{u}',p') \right] \quad &&\mbox{in\ } \Upsilon_{0} \label{LSA-momentum} \\ 
 0 &= \nabla_0\cdot\boldsymbol{u}' + \nabla_0\cdot\left( \boldsymbol{\Phi}^{\prime}(\boldsymbol{\xi}_0) \boldsymbol{U}_0 \right) \quad &&\mbox{in\ } \Upsilon_{0} \label{LSA-incompressible} \\ 
 \boldsymbol{0} &= \nabla_0 ^ 2 \boldsymbol{\xi}_0 \quad &&\mbox{in\ } \Omega_f \label{LSA:extension-displacement} \\ 
 \boldsymbol{0} &= \boldsymbol{u}' - \left( \boldsymbol{V}'_k + \boldsymbol{\Omega}'_k \times \boldsymbol{r} \right) \quad &&\mbox{on\ } \mathscr{S} \label{LSA:continuity-velocity} \\ 
 \boldsymbol{0} &= \boldsymbol{\xi}_0 - (\boldsymbol{X}'+\boldsymbol{\Xi}'\times\boldsymbol{r}-\boldsymbol{X}'_{LB}) \quad &&\mbox{on\ } \mathscr{S} \label{LSA:continuity-displacement}\\ 
 \boldsymbol{0} &= \int_{\mathscr{S}} \left( \boldsymbol{\Sigma}^{\prime}(\boldsymbol{U}_0,P_0,\boldsymbol{\xi}_0) +\boldsymbol{\Sigma} (\boldsymbol{u}',p') \right) \cdot \boldsymbol{n} \, dS \quad &&\mbox{on\ } \mathscr{S} \label{LSA:translation-dynamic} \\ 
 \boldsymbol{0} &= \int_{\mathscr{S}} \boldsymbol{r} \times \left( \left( \boldsymbol{\Sigma}^{\prime}(\boldsymbol{U}_0,P_0,\boldsymbol{\xi}_0) +\boldsymbol{\Sigma} (\boldsymbol{u}',p') \right) \cdot \boldsymbol{n} \right) \, dS \quad &&\mbox{on\ } \mathscr{S} \label{LSA:rotation-dynamic} \\ \frac{d\boldsymbol{X}'_k}{dt} &= \boldsymbol{V}'_k, \quad \frac{d\boldsymbol{\Xi}'_k}{dt} = \boldsymbol{\Omega}'_k \label{LSA:kinematics} 
\end{align} 
\end{subequations}
Here,
\begin{align}
\boldsymbol{\Phi}'(\boldsymbol{\xi}_0) &= [\,\boldsymbol{\Phi}_0(\boldsymbol{\xi}_0) - \mathbb{I}\,]_{\text{linear}} = (\nabla_0\!\cdot\!\boldsymbol{\xi}_0)\,\mathbb{I} - \nabla_0 \boldsymbol{\xi}_0, \\[4pt]
\boldsymbol{\Sigma}'(\boldsymbol{U}_0, P_0, \boldsymbol{\xi}_0) &= \boldsymbol{\Sigma}(\boldsymbol{U}_0, P_0)\,\boldsymbol{\Phi}'^{T}(\boldsymbol{\xi}_0) - \mu \left[(\nabla_0 \boldsymbol{U}_0)(\nabla_0 \boldsymbol{\xi}_0)+ (\nabla_0 \boldsymbol{\xi}_0)^T (\nabla_0 \boldsymbol{U}_0)^T \right].
\end{align}

Since the base flow is steady and axisymmetric, the perturbations can be expressed as separable time–azimuthal modes:
\begin{equation}\label{pre-eigenvalue-eq}
\boldsymbol{q}' =
\begin{pmatrix}
\boldsymbol{q}_{\mathrm{ALE}}^{\prime\,f}(x,r,\theta,t)\\[3pt]
\boldsymbol{q}^{\prime\,b}(x,y,z,t)
\end{pmatrix}
=
\begin{pmatrix}
\widetilde{\boldsymbol{q}}^{f}_A(r,x)\, e^{im\theta}\\[3pt]
\widetilde{\boldsymbol{q}}^{b}
\end{pmatrix}
e^{\lambda t} + c.c.,
\end{equation}
where $c.c.$ denotes the complex conjugate.
The fluid components $\widetilde{\boldsymbol{q}}^{f}_{\mathrm{ALE}}$ are expressed in cylindrical coordinates $(r,\theta,x)$, with the azimuthal dependence isolated through the wavenumber $m$, while the bubble components $\widetilde{\boldsymbol{q}}^{b}$ are represented in Cartesian coordinates $(x,y,z)$. Among the bubble variables, only a subset contributes to the global dynamics, depending on the azimuthal mode $m$ \citep{tchoufag2014globaldisk,tchoufag2014linearbubble}. These active components can be identified by analysing the symmetry of the force and torque terms appearing in Eqs.~\eqref{LSA:translation-dynamic}–\eqref{LSA:rotation-dynamic}.

For $m=\pm1$ (helical modes), the perturbation forces and torques are associated with the azimuthal dependence  $y \mp i z$ (up to complex conjugation). Hence, the projections of Eqs.~\eqref{LSA:translation-dynamic}–\eqref{LSA:kinematics}  along the $y$ and $z$ directions can be combined into a single set of equations in the bubble’s transverse plane \citep{tchoufag2014globaldisk}. 
Following \citet{jenny2004efficient}, this reduction is achieved by introducing the $U(1)$ variables:
\[
\widetilde{V}_\pm = \widetilde{V}_y \mp i\widetilde{V}_z, \qquad
\widetilde{X}_\pm = \widetilde{X}_y \mp i\widetilde{X}_z, \qquad
\widetilde{\Omega}_\pm = \widetilde{\Omega}_z \pm i\widetilde{\Omega}_y, \qquad
\widetilde{\Xi}_\pm = \widetilde{\Xi}_z \pm i\widetilde{\Xi}_y.
\]
Accordingly, for $|m|=1$, the bubble contribution to the eigenvector reduces to eight complex amplitudes:
\[
\widetilde{\boldsymbol{q}}^{b}
=
[\widetilde{V}_{mLB},\,\widetilde{X}_{mLB},\,
\widetilde{\Omega}_{mLB},\,\widetilde{\Xi}_{mLB},\,
\widetilde{V}_{mTB},\,\widetilde{X}_{mTB},\,
\widetilde{\Omega}_{mTB},\,\widetilde{\Xi}_{mTB} ].
\]
The complete coupled system
\cref{LSA-momentum,LSA-incompressible,LSA:extension-displacement,LSA:translation-dynamic,LSA:rotation-dynamic,LSA:kinematics},
supplemented by the corresponding boundary conditions,  
can then be recast as a generalised eigenvalue problem:
\begin{equation}
    \mathscr{A}_m \widetilde{\boldsymbol{q}}
    = \lambda\, \mathscr{B}_m \widetilde{\boldsymbol{q}},
\end{equation}
where $\widetilde{\boldsymbol{q}} = [\widetilde{\boldsymbol{q}}^{f},\,\widetilde{\boldsymbol{q}}^{b}]$ 
is the global eigenvector.  

In the present work, we restrict our analysis to the fundamental case $|m|=1$.  
Owing to the symmetry:
\[
(\widetilde{u}_r,\widetilde{u}_\theta,\widetilde{u}_x,p,
\widetilde{V}_{+},\widetilde{X}_{+},\widetilde{\Omega}_{+},\widetilde{\Xi}_{+},m)
\;\longrightarrow\;
(\widetilde{u}_r,-\widetilde{u}_\theta,\widetilde{u}_x,p,
\widetilde{V}_{-},\widetilde{X}_{-},\widetilde{\Omega}_{-},\widetilde{\Xi}_{-},-m),
\]
it is sufficient to consider only the $|m|=1$ mode without loss of generality.

\section{Numerical validations of the ALE--LSA method}
\label{app:ALE_validation}

\subsection{Grid and domain independence study} 

\begin{figure}
	\centering
	\setlength{\tabcolsep}{0pt} 
    \iflocalcompile
	\begin{tabular}{cc}
		\resizebox{0.5\textwidth}{!}{\input{fig/Appendix_B/verification-hmin-test/fig1.tex}} &
		\resizebox{0.5\textwidth}{!}{\input{fig/Appendix_B/verification-l1l2/l1l2-1.tex}}
	\end{tabular}
    \else
    \begin{tabular}{cc}
        \resizebox{0.5\textwidth}{!}{\includetikz{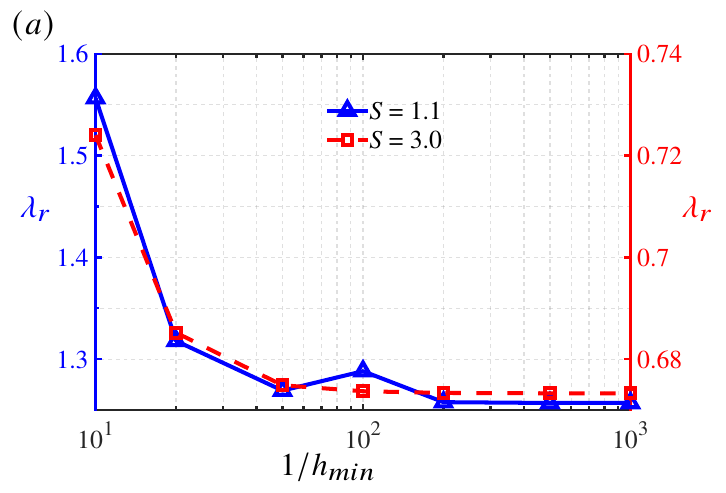}} &
        \resizebox{0.5\textwidth}{!}{\includetikz{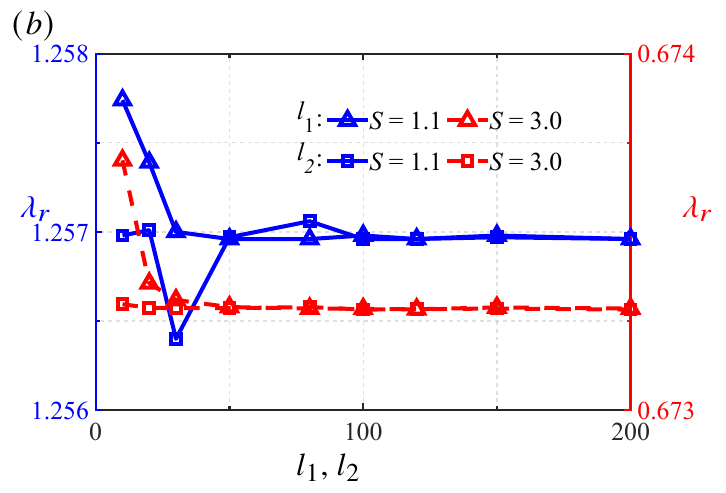}}
    \end{tabular}
    \fi
	\captionsetup{justification=justified, singlelinecheck=false, labelsep=period, width=\linewidth} 
	\caption{Dependence of the growth rate $\lambda_{r}$ on grid and domain parameters for two bubble separations, $S = 1.1$ and $S = 3.0$.  $(a)$ Variation of $\lambda_{r}$ with the smallest grid spacing $h_{\min}$ for a representative bubble pair at $(\Rey,\,\chi) = (220,\,1.0)$, with the domain dimensions fixed at $(l_{1},\,l_{2},\,h) = (120,\,120,\,60)$.  
$(b)$ Same as panel~$(a)$, but showing the influence of the inlet and outlet lengths $l_{1}$ and $l_{2}$, respectively, with the minimum grid spacing fixed at $h_{\min} = 10^{-3}$.}
	\label{validation-hmin}
\end{figure}

The unstructured triangular meshes used in the present study are generated by \textit{FreeFem++}, using a Delaunay--Voronoi algorithm, as already illustrated in \hyperref[flow-configuration]{figure~\ref{flow-configuration}$(c)$}.  
As discussed in \S~\ref{sec:LALE-numerical}, the smallest grid element is set to $h_{\min} = 10^{-3}$ in all computations, unless otherwise stated.  
To assess the convergence performance of this spatial resolution, we vary $h_{\min}$ within the range $[10^{-3},\,0.1]$ and compute the corresponding growth rate $\lambda_{r}$ of the stationary mode for a representative bubble pair at $(\Rey, \chi) = (220,\,1.0)$, considering two different separations, $S = 1.1$ and $S = 3.0$, which respectively correspond to small and large inter-bubble distances.  
All other domain parameters are fixed at $l_{1} = 120$, $l_{2} = 120$, $h = 60$, and $h_{\max} = 0.5$.
The variation of $\lambda_{r}$ with $1/h_{\min}$ is shown in \hyperref[validation-hmin]{figure~\ref{validation-hmin}$(a)$}. A clear convergence trend is observed for both separations: $\lambda_{r}$ approaches the asymptotic values $\lambda_{r}^{c} = 1.25696$ and $\lambda_{r}^{c} = 0.67328$ for $S = 1.1$ and $S = 3.0$, respectively, once the grid is refined below $h_{\min} < 5\times10^{-3}$.

As a next step, the influence of the inlet and outlet lengths, $l_{1}$ and $l_{2}$, is examined using the same bubble pair at $(\Rey, \chi) = (220,\,1.0)$ with two separations, $S = 1.1$ and $S = 3.0$.  
We first vary the inlet length over the range $l_{1} \in [10,\,200]$, while keeping the other grid parameters fixed at $l_{2} = 120$, $h = 60$, and $h_{\min} = 10^{-3}$.  
As shown in \hyperref[validation-hmin]{figure~\ref{validation-hmin}$(b)$}, the growth rate $\lambda_{r}$ exhibits negligible variation once $l_{1} > 100$ for both separations.  
A similar analysis is performed by varying the outlet length $l_{2} \in [10,\,200]$, with $l_{1} = 120$, $h = 60$, and $h_{\min} = 10^{-3}$ kept constant.  
Again, convergence is achieved for $l_{2} > 100$, confirming that the computational domain is sufficiently long to prevent artificial reflections or confinement effects.  
Consequently, a domain size of $(l_{1}, l_{2}) = (120,\,120)$ is adopted throughout the present study.  
Additionally, increasing the half-width of the domain from $h = 60$ to $h = 80$ results in a change in $\lambda_{r}$ smaller than $0.01\,\%$, further confirming the adequacy of the chosen computational domain, while these results are therefore omitted for brevity.

\subsection{Validation of the LSA results} 

The reliability of the numerical framework developed in this study is further assessed through a series of validation tests.  
These tests are designed to verify, respectively, the accuracy of the base-flow solver, the LSA solver for an isolated falling disk or bubble, and the full ALE--LSA solver for a tandem configuration.

\begin{figure}
    \centering
    \setlength{\tabcolsep}{-1.5pt}
    \iflocalcompile
    \begin{tabular}{cc}
    \resizebox{0.5\textwidth}{!}{\input{fig/Appendix_B/comparison-hallez2011/s_3R-reset-new.tex}} &
    \resizebox{0.5\textwidth}{!}{\input{fig/Appendix_B/comparison-hallez2011/equilibrium-distance-S_D-new.tex}}
    \end{tabular}
    \else
    \begin{tabular}{cc}
        \resizebox{0.5\textwidth}{!}{\includetikz{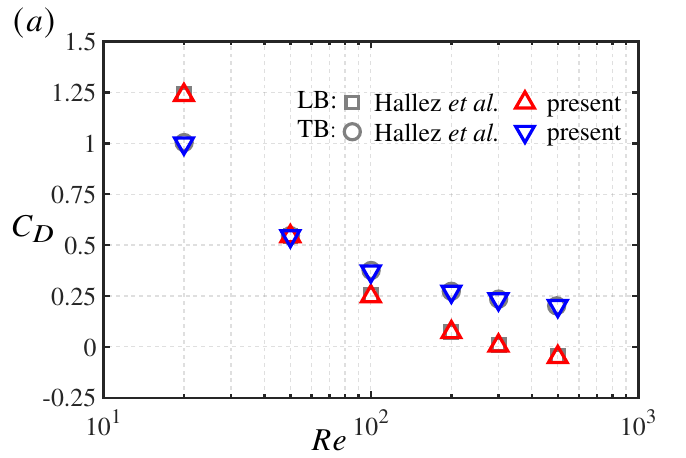}} &
        \resizebox{0.5\textwidth}{!}{\includetikz{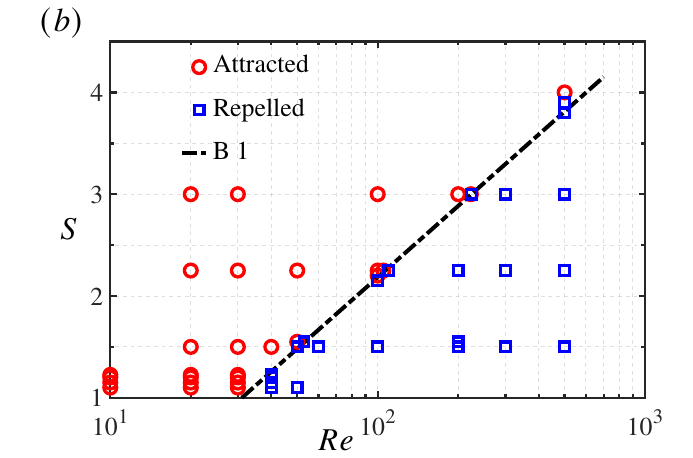}}
    \end{tabular}
    \fi
    \captionsetup{justification=justified, singlelinecheck=false, labelsep=period, width=\linewidth}
    \caption{Validation of the base-flow solver.  
$(a)$ Variation of the drag coefficients $C_{D}^{LB}$ and $C_{D}^{TB}$ with the Reynolds number for a spherical bubble pair at a fixed separation of $S = 1.5$.  
The numerical results of \citet{hallez2011interaction} are included for comparison.  
$(b)$ Equilibrium distance $S_e$ between two bubbles rising in-line.  
Symbols denote the present solutions: 
\protect\tikz\protect\draw[red, very thick] (0,0) circle (3pt);~attracted bubbles,~
\protect\tikz\protect\draw[draw={rgb,255:red,0;green,0;blue,255}, line width=1.2pt]
(0,0) -- (5pt,0) -- (5pt,5pt) -- (0,5pt) -- cycle;~repelled bubbles;~
\raisebox{0.5ex}{\protect\tikz\protect\draw[black, dash dot, thick] (0,0) -- ++(0.5,0);}~
dash-dotted line: empirical correlation of $S_e$ from \eqref{A.e1} proposed by \citet{zhang2021three}.}
    \label{validation-drag-sphercialbubble}
\end{figure}

\emph{\textbf{Test~1.}}  
We first solve the steady base flow past a fixed bubble pair and compare the resulting drag coefficients with the numerical data of \citet{hallez2011interaction}.  
In this test, both bubbles are spherical ($\chi = 1$) and separated by a fixed distance $S = 1.5$, so that the drag coefficients of the leading and trailing bubbles, $C_{D}^{LB}$ and $C_{D}^{TB}$, depend solely on the Reynolds number $\Rey$.  
As shown in \hyperref[validation-drag-sphercialbubble]{figure~\ref{validation-drag-sphercialbubble}$(a)$}, the present results are in excellent agreement with those reported by \citet{hallez2011interaction} over the entire range of $\Rey$ considered.
In addition, we examine the dependence of the equilibrium separation distance $\bar{S}_{e}$ on the Reynolds number for spherical bubble pairs.  
The present results are compared with the empirical correlation proposed by \citet{zhang2021three},
\begin{equation}
S_{e}(\Rey) = 2.3295 \log_{10}{\Rey} - 2.4752,
\label{A.e1}
\end{equation}
which provides improved accuracy in the regime $\Rey < 30$ compared with the earlier expression of \citet{hallez2011interaction}.  
As shown in \hyperref[validation-drag-sphercialbubble]{figure~\ref{validation-drag-sphercialbubble}$(b)$}, the computed equilibrium distances $\bar{S}_{e}$ agree perfectly with the empirical correlation, confirming the fidelity of the base-flow solver.

\begin{figure}
    \centerline{\includegraphics[width=1.1\textwidth]{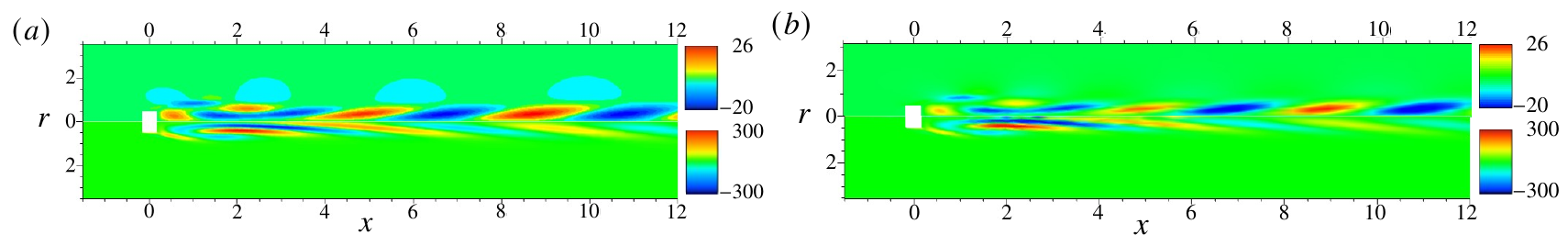}}
    \captionsetup{justification=justified, singlelinecheck=false, labelsep=period, width=\linewidth,font=small}
    \caption{Oscillatory mode associated with the path instability of a freely falling disk at $(\chi,\,I^*,\,\Rey) = (3,\,0.56,\,298)$.  The upper half of each panel shows contours of streamwise velocity, and the lower half shows contours of streamwise vorticity.  
$(a)$ Results of \citet{tchoufag2014globaldisk}, obtained using the generalized Kirchhoff formulation.  
$(b)$ Present results computed with the ALE method.  
All contours are displayed in the Eulerian reference frame.}
    \label{validation:freely-disk-mode}
\end{figure}

\emph{\textbf{Test~2.}}  
We next verify the accuracy of the present ALE--LSA framework by identifying the transition thresholds associated with buoyancy-driven path instabilities of freely moving bodies.  
To this end, the method is applied to two canonical problems: the freely falling circular disk \citep{tchoufag2014globaldisk} and the freely rising bubble \citep{tchoufag2014linearbubble}, with all translational and rotational degrees of freedom fully released.  
It is worth noting that \citet{tchoufag2014globaldisk,tchoufag2014linearbubble} solved the generalized Kirchhoff equations rather than the Navier--Stokes equations, by projecting the absolute flow variables onto a body-fixed coordinate system.  
Hence, their approach differs fundamentally from the present Arbitrary Lagrangian--Eulerian formulation, which directly accounts for the flow--structure coupling in an evolving physical domain.  
For the freely falling disk, we vary the dimensionless parameters $\chi$ and $I^{*}$, representing the aspect ratio and the dimensionless moment of inertia, respectively, in order to determine the critical Reynolds number $\Rey_{c}$ (or equivalently, the critical Archimedes number $Ar_{c}$, with $Ar^{2} = 3\Rey^{2}C_{D}/32$) at which the path loses stability. Note that $I^{*}$ is varied over a wide range here, representing ‘heavy’ disks falling in the fluid.  
The aspect ratio is fixed at $\chi = 10^4$, corresponding to an infinitesimally thin disk.  
The resulting critical Reynolds numbers $\Rey_{c}$ and oscillation frequencies $\lambda_{i}$ are listed in \hyperref[validation:freely-disk]{table~\ref{validation:freely-disk}}, together with the benchmark results from \citet{tchoufag2014globaldisk}.  
Despite the methodological differences, the present ALE--LSA predictions of both $\Rey_{c}$ and $\lambda_{i}$ agree remarkably well with the reference data across the entire range of $I^{*}$ considered.  
Furthermore, \hyperref[validation:freely-disk-mode]{figure~\ref{validation:freely-disk-mode}} compares the mode structures obtained for the disk at $(\chi,\,I^{*},\,\Rey) = (3,\,0.56,\,298)$.  
The iso-contours of the streamwise velocity perturbations (top half-panel) and the corresponding streamwise vorticity fields (bottom half-panel) exhibit identical spatial patterns to those reported by \citet{tchoufag2014globaldisk}, confirming that the present ALE--LSA framework accurately reproduces both the threshold and structure of the unstable modes.


\begin{table} 
    \begin{center}
        \def~{\hphantom{0}}
        \begin{tabular}{l@{\hspace{8pt}}c@{\hspace{12pt}}c@{\hspace{8pt}}c@{\hspace{8pt}}c@{\hspace{8pt}}c@{\hspace{8pt}}c} 
            \multicolumn{2}{c}{$I^*$}  											& 0.002			&   0.002		& 0.16		& 50 			& 50  			\\[3pt]
            \multirow{2}{*}{$Re_c$}				&\citet{tchoufag2014globaldisk}	& 113.0	 		& 	161.2   	& 33.0		& 59.5			& 273.0			\\
            &	present study    				& 113.0 		& 	158.5   	& 31.5		& 59.0			& 271.0			\\
            \midrule
            \multirow{2}{*}{$\lambda_i$}		&\citet{tchoufag2014globaldisk}	& 0.67273	 	& 1.6875	   	& 0.8239	& 0.04859		& 1.39327		\\
            &	present study    				& 0.67436		& 1.7045   		& 0.8291	& 0.04839		& 1.39324		\\
        \end{tabular}
        \captionsetup{justification=justified, singlelinecheck=false, labelsep=period, width=\linewidth,font=small}
        \caption{Comparison of the critical Reynolds numbers $\Rey_c$ and oscillation frequencies $\lambda_i$ for the path instability of a buoyancy-driven disk with aspect ratio $\chi = 10^4$.  
        Reference data from \citet{tchoufag2014globaldisk} are included for comparison.  
        Here, $I^*$ denotes the dimensionless moment of inertia of the disk, while all the values selected here correspond to  falling disks.}
        \label{validation:freely-disk}
    \end{center}
\end{table}

\begin{table} 
    \begin{center}
        \def~{\hphantom{0}}
        \setlength{\tabcolsep}{10pt}  
        {
            \begin{tabular}{l@{\hspace{6pt}}c@{\hspace{10pt}}c@{\hspace{10pt}}c@{\hspace{16pt}}c@{\hspace{10pt}}c} 
                \multicolumn{2}{c}{}  													&\multicolumn{2}{c}{oscillating mode 1} & \multicolumn{2}{c}{oscillating mode 2}  \\[3pt]
                                    				&									&$\Rey_{c, \min}$	&$\Rey_{c, \max}$			&$\Rey_{c, \min}$	&$\Rey_{c, \max}$		\\
                \multirow{2}{*}{$Re_c$}				&\citet{tchoufag2014linearbubble}	& 90.6	 		& 	132.0   			& 171.0			& 797.0					\\
                                                    &	present study    				& 93.5 			& 	128.0   			& 172.0			& 807.0					\\
                \midrule
                \multirow{2}{*}{$\lambda_i$}		&\citet{tchoufag2014linearbubble}	& 0.567	 		& 0.474	   				& 0.554			& 0.544				\\
                                                    &	present study    				& 0.567			& 0.474   				& 0.559			& 0.577				\\
            \end{tabular}
        }
        \captionsetup{justification=justified, singlelinecheck=false, labelsep=period, width=\linewidth,font=small}
        \caption{Critical Reynolds numbers ($\Rey_{c, \min}$, $\Rey_{c, \max}$) and associated oscillation frequencies $\lambda_i$ for the two oscillatory modes of a freely rising ellipsoidal bubble with aspect ratio $\chi = 2.5$.  
Reference results from \citet{tchoufag2014linearbubble}, obtained using the generalized Kirchhoff formulation, are included for comparison.}
        \label{validation:freely-bubble-single}
    \end{center}
\end{table}

We next extend the validation to the path instability of an oblate bubble rising freely under gravity.  
Compared with the disk cases discussed above, the key difference lies in the boundary condition imposed at the interface:  
the bubble surface satisfies a shear-free condition, rather than the no-slip constraint used for rigid bodies.  
In this configuration, the dynamics are governed solely by the aspect ratio $\chi$ and the Reynolds number $\Rey$, as the density ratio $\bar{\rho}$ is negligibly small.  
Fixing the aspect ratio at $\chi = 2.5$, corresponding to a highly deformed bubble, two distinct oscillatory modes are identified, each bounded by a lower and an upper critical Reynolds number, $\Rey_{c,\min}$ and $\Rey_{c,\max}$, respectively.  
The computed values of these critical thresholds, together with the associated oscillation frequency $\lambda_{i}$, are summarized in \hyperref[validation:freely-bubble-single]{table~\ref{validation:freely-bubble-single}},  
and compared with the benchmark results of \citet{tchoufag2014linearbubble}.  
The present ALE--LSA predictions exhibit excellent quantitative agreement with the reference data, confirming that the method accurately captures the onset and frequency of the path instability of a freely rising oblate bubble.  
This result further validates the robustness of the linear solver for stress-free and oblate interfaces.

\begin{figure}
    \centering
    \setlength{\tabcolsep}{0pt} 
    \iflocalcompile
    \begin{tabular}{ccc}
        \resizebox{!}{0.28\textwidth}{\input{fig/Appendix_B/Tirri/Tirri2023-sketch.tex}} &
        \resizebox{!}{0.28\textwidth}{\input{fig/Appendix_B/Tirri/Tirri2023-lambdar.tex}} &
        \resizebox{!}{0.28\textwidth}{\input{fig/Appendix_B/Tirri/Tirri2023-St.tex}}
    \end{tabular}
    \else
    \begin{tabular}{ccc}
        \resizebox{!}{0.28\textwidth}{\includetikz{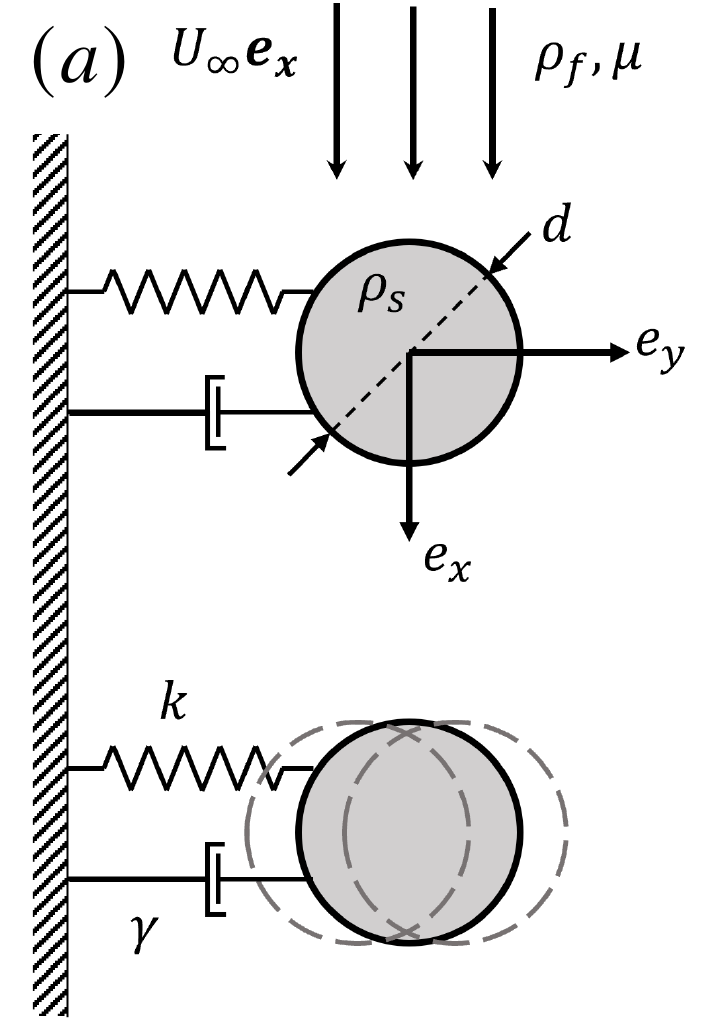}} &
        \resizebox{!}{0.28\textwidth}{\includetikz{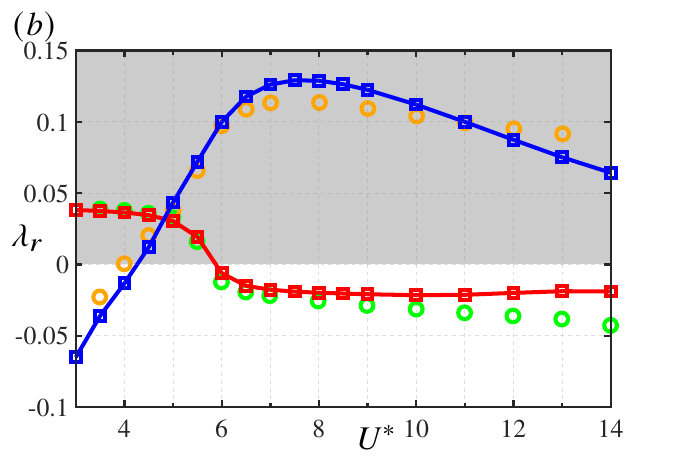}} &
        \resizebox{!}{0.28\textwidth}{\includetikz{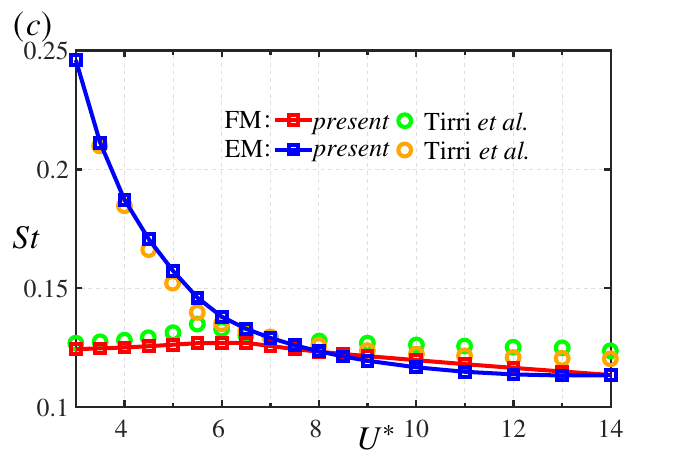}}
    \end{tabular}
    \fi
    \captionsetup{justification=justified, singlelinecheck=false, labelsep=period, width=\linewidth} 
    \caption{Flow past a tandem-cylinder configuration at $(\Rey,\,\rho^*) = (100,\,2.546)$, where $\rho^{*}$ denotes the ratio of solid to fluid density.  
$(a)$ Schematic of the problem setup, with two cylinders arranged in tandem and constrained to oscillate transversely to the incoming flow.  
$(b)$ Growth rate as a function of the reduced velocity $U^*$, defined as the ratio between the period of the natural structural oscillation and that of the convective flow motion.  
The unstable region is shaded in grey.  
$(c)$ Reduced frequency $St = \lambda_i d / (2\pi U_\infty)$ versus $U^*$.  
Reference data from \citet{tirri2023linear}, obtained using an immersed-boundary, matrix-free stability framework, are included for comparison.}
    \label{validation:Tirri:2023}
\end{figure}

\emph{\textbf{Test~3.}}  
As a final validation, we examine the flow past a tandem-cylinder configuration, previously investigated by \citet{tirri2023linear}.  
This test serves to verify the performance of the present ALE--LSA implementation for a twin-body system, since in \emph{\textbf{Tests~1}} and \emph{\textbf{2}} the fluid domain remained stationary in time.  
In this configuration, the relative motion between the two cylinders induces a time-dependent deformation of the computational domain, closely resembling the situation encountered in the bubble-pair problem.  
The setup of the problem is illustrated in \hyperref[validation:Tirri:2023]{figure~\ref{validation:Tirri:2023}$(a)$}.  
The two cylinders are arranged in tandem and constrained to oscillate transversely to the oncoming flow.  
The dynamics are governed by two non-dimensional parameters:  
the density ratio $\rho^{*} = \rho_{s}/\rho_{f}$, representing the ratio of solid to fluid density,  
and the reduced velocity $U^{*}$, which measures the ratio of two characteristic time scales, namely the period of the natural structural oscillation and that of the convective flow motion.  
In the present test, we set $\rho^{*} = 2.546$ and $\Rey = 100$.  
\hyperref[validation:Tirri:2023]{Figure~\ref{validation:Tirri:2023}$(b)$ and $(c)$} compare the growth rates $\lambda_{r}$ and the reduced oscillation frequencies $St$ obtained from the present ALE--LSA analysis with those reported by \citet{tirri2023linear}, as the reduced velocity $U^{*}$ varies from 3 to 14. Besides, the fluid mode (FM) and elastic mode (EM) are also distinguished following \citet{tirri2023linear}.  
Both quantities, in these two modes, exhibit excellent quantitative agreement with the benchmark data, confirming the accuracy and robustness of the present solver in capturing the coupled fluid–structure interaction and stability characteristics of twin-body configurations.  
Small discrepancies appear for $U^{*} > 10$, where the present ALE--LSA results deviate slightly from those of \citet{tirri2023linear}.  
These differences arise from methodological distinctions: the present formulation employs an explicit ALE framework that directly accounts for domain deformation \citep{bonnefis2024path}, whereas \citet{tirri2023linear} adopted an immersed-boundary, matrix-free stability approach.  
Nevertheless, these minor variations do not affect the identification of instability mechanisms or the prediction of the critical thresholds, thereby validating the reliability of the present ALE--LSA implementation for time-dependent multi-body systems.

\section{Exclusion of wake instability}
\label{app:wake}

This appendix confirms that the lateral motion of the TB identified in the present study does not originate from a local wake instability, but rather from a global hydrodynamic coupling between the two bubbles.  Before proceeding, it is instructive to recall the mechanisms underlying the wake instability of an isolated bubble.  
For a single bubble rising freely in a quiescent liquid, the steady axi-symmetric wake becomes unstable once the bubble exceeds a critical oblateness, $\chi_{c}^{iso}\!\simeq\!2.21$ \citep{magnaudet2007wake, ern2012wake}.  
This transition corresponds to the emergence of a pair of streamwise counter-rotating vortices in the near wake, forming the characteristic double-threaded structure identified by \citet{magnaudet2007wake}.  
The instability results from the convective amplification of surface-generated vorticity as it is advected downstream by the base flow \citep{tchoufag2013linearwake}.  

\begin{figure}
	\centering
	\setlength{\tabcolsep}{0pt} 
	\iflocalcompile
	\begin{tabular}{cc}
		\resizebox{0.5\textwidth}{!}{\input{\figfourone/lambdar_differentS.tex}} &
		\resizebox{0.5\textwidth}{!}{\input{\figfourone/fixed_s_Rec_2-new-deltas.tex}}
	\end{tabular}
	\else
	\begin{tabular}{cc}
		\resizebox{0.5\textwidth}{!}{\includetikz{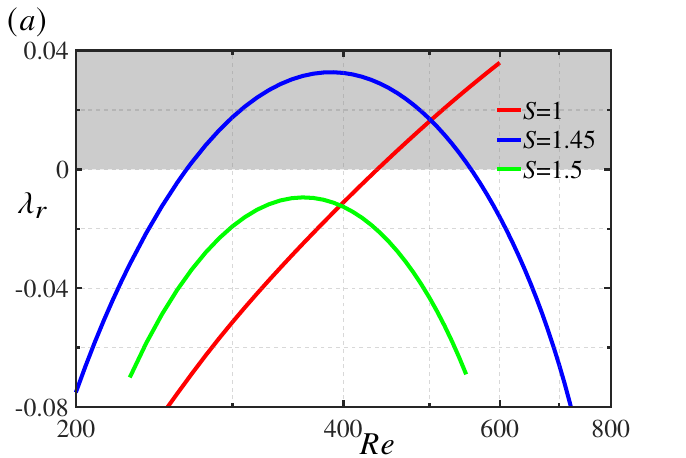}} &
		\resizebox{0.5\textwidth}{!}{\includetikz{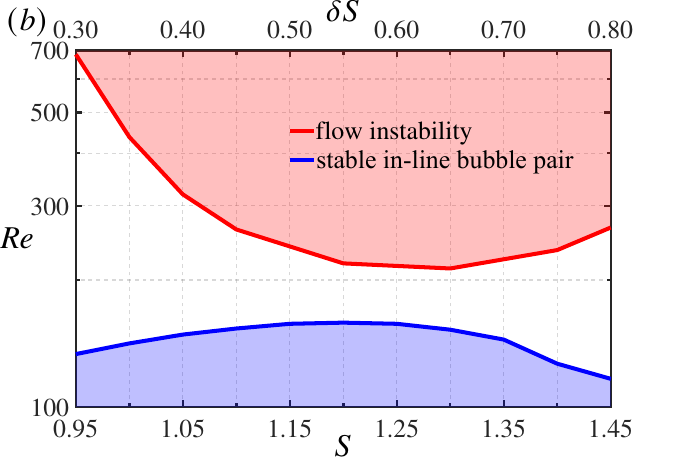}}
	\end{tabular}
	\fi	
	\captionsetup{justification=justified, singlelinecheck=false, labelsep=period, width=\linewidth} 
	\caption{Neutral stability curves for the wake instability of a frozen bubble pair with $\chi = 1.9$, corresponding to the most oblate configuration examined in the present ALE-LSA study.
$(a)$ Variation of the growth rate $\lambda_r$ of the stationary mode with $\Rey$ for different separations $S$.
$(b)$ Comparison between the neutral boundary for wake instability of a frozen pair (red region) and that for stable regime of a freely moving pair (blue region). }
	\label{wake:vs:path}
\end{figure}

In contrast, when the TB rises in the wake of another bubble, the surrounding flow is profoundly modified.  
The LB not only shields the TB from the ambient liquid but also imposes an almost axi-symmetric shear in the inter-bubble region.  
This shear alters both the incident velocity distribution and the vorticity flux at the TB surface, thereby reducing the effective critical aspect ratio for wake destabilisation, i.e. $\chi_{c}^{pair} < \chi_{c}^{iso}$ \citep{takagi1994drag}.  
To assess whether the TB wake could sustain a self-excited instability under such conditions, we performed a ``frozen-body'' linear stability analysis in which both bubbles were held fixed, thereby suppressing their translational and rotational degrees of freedom. In this auxiliary computation, the linearised momentum equations were solved for the perturbation field without invoking the ALE mapping, so as to isolate the intrinsic stability of the flow past a stationary bubble pair.  
The perturbations were decomposed into azimuthal Fourier modes of the form $\propto e^{im\theta+\lambda t}$, and the corresponding growth rates $\lambda_{r}$ were evaluated for $|m|=1$, known to yield the most amplified disturbances in axisymmetric wakes \citep{tchoufag2013linearwake,tchoufag2014globaldisk}.  
Calculations were carried out for the most critical shape considered here, $\chi = 1.9$, over the range $100 < \Rey < 800$ and for separations $S = 1.00$ ($\delta S = 0.35$), $1.45$ ($0.80$), and $1.50$ ($0.85$).  
The resulting variations of $\lambda_{r}$ with $\Rey$ are shown in \hyperref[wake:vs:path]{figure~\ref{wake:vs:path}$(a)$}.  

At the smallest separation ($S = 1.0$), increasing $\Rey$ enhances the wake sensitivity and eventually leads to instability ($\lambda_{r} > 0$), with the neutral threshold near $\Rey \simeq 420$.  
For $S = 1.45$, $\lambda_{r}$ exhibits a weakly non-monotonic dependence on $\Rey$, becoming positive within $270 \lesssim \Rey \lesssim 580$.  
The lower bound corresponds to viscous damping that limits vorticity amplification at the interface, whereas the upper bound reflects the $\Rey^{-1/2}$ scaling of interfacial vorticity flux, which progressively restores wake stability \citep{magnaudet2007wake}.  
Further increasing the separation to $S = 1.50$ preserves this trend but keeps $\lambda_{r}$ negative throughout the range considered, indicating that the wake remains linearly stable.  
These results demonstrate that, at sufficiently small separations, the presence of a neighbouring bubble can indeed promote wake sensitivity and lower the critical $(\chi, \Rey)$ thresholds relative to those of an isolated bubble.

However, as shown in \hyperref[wake:vs:path]{figure~\ref{wake:vs:path}$(b)$}, the neutral boundaries of the wake and path instabilities remain clearly distinct: the wake-unstable region (red shading) lies entirely at higher Reynolds numbers than the onset of path instability for the bubble pair (blue region).  
Hence, the lateral motion of the TB identified in the present study, occurring near the neutral curve of path instability, cannot be attributed to a local wake instability.  
Instead, it originates from a genuinely global hydrodynamic coupling between the two bubbles.  
The disturbed pressure and vorticity fields in the inter-bubble region, particularly those associated with the connected recirculation described in \S~\ref{sec:flow-structure}, provide the feedback necessary to sustain this collective instability.  
The following sections quantify this coupling through the neutral curves and eigenmode structures obtained from the full ALE--LSA analysis.

\section{Embedded–Boundary Method}
\label{app:EBM}

The Embedded–Boundary Method (EBM) employed in this study was originally developed in our previous work for flow past frozen droplets \citep{zhang2025lift,WEI2026114632}, by which $\mu^{*}$ characterizes the viscosity ratio of fluids interior and exterior of the interface. 
Although it was used for modeling oblate inviscid bubbles ($\mu^{*}=0$) here, the method is more general and can accommodate droplets with arbitrary viscosity ratios ($0<\mu^{*}<\infty$) and complex geometries. 
Here, we summarize the essential features of the method, while comprehensive algorithmic details and validation tests are provided by \citet{WEI2026114632}.

We consider a fixed droplet immersed in an external flow, with both the internal and external flow fields solved simultaneously. 
At the droplet interface $\mathcal{S}$, the velocity and stress fields must satisfy the interfacial boundary conditions:
\begin{eqnarray}
\left.
\begin{array}{ll}
u_{e,n} = u_{i,n} = 0, \\[0.15cm]
\mu_e \left(\dfrac{\partial u_{e,\tau_1}}{\partial n} - \kappa_1 u_{e,\tau_1}\right)
= \mu_i \left(\dfrac{\partial u_{i,\tau_1}}{\partial n} - \kappa_1 u_{i,\tau_1}\right), \\[0.15cm]
\mu_e \left(\dfrac{\partial u_{e,\tau_2}}{\partial n} - \kappa_2 u_{e,\tau_2}\right)
= \mu_i \left(\dfrac{\partial u_{i,\tau_2}}{\partial n} - \kappa_2 u_{i,\tau_2}\right),
\end{array}
\qquad
\right\}
\quad \mathrm{on}\ \Gamma ,
\label{eq:EBM_bc}
\end{eqnarray}
where $n$ denotes the unit normal to the interface, and $\tau_1$ and $\tau_2$ are the two principal tangential directions associated with the local curvatures $\kappa_1$ and $\kappa_2$. 
Superscripts $e$ and $i$ refer to quantities in the external and internal fluids, respectively. 
These relations express, respectively, the no-penetration condition and the continuity of tangential stresses across the interface. 
For inviscid bubbles ($\mu_i=0$), equations~(\ref{eq:EBM_bc}) reduce to the classical shear-free condition, corresponding to the flow past a fixed bubble.

While these interfacial conditions can be imposed exactly on body-fitted grids, such approaches become cumbersome and error-prone when applied to oblate or irregularly shaped droplets \citep{legendre2019basset,rachih2020numerical}. 
To overcome these limitations, we developed a Cartesian-grid EBM framework \citep{zhang2025lift,WEI2026114632} that avoids explicit surface meshing while retaining accurate enforcement of the jump conditions. 
In this method, the interface generally cuts through the Cartesian cells, dividing them into interior and exterior subregions. 
Accurate representation of the interfacial stress continuity therefore relies critically on determining the local principal curvatures $(\kappa_1,\kappa_2)$ and their orientations within these cut cells. 
The present implementation employs two main numerical strategies: 
(i) a height-function reconstruction of the interface to evaluate the local curvature tensor, and 
(ii) an embedded-boundary formulation that solves the internal and external velocity fields in a unified finite-volume framework, while consistently enforcing the interfacial jumps in (\ref{eq:EBM_bc}). 
Further algorithmic details and accuracy assessments are given by \citet{WEI2026114632}.

The solver is implemented within the open-source library \textit{Basilisk} \citep{popinet2015quadtree}, which employs a finite-volume discretisation on an adaptive quadtree/octree mesh, ensuring strict local conservation of mass and momentum. 
The original \textit{Basilisk} framework was primarily designed for rigid bodies with no-slip boundary conditions. 
The present EBM extension generalises this capability to interfaces with arbitrary viscosity contrasts, thereby enabling the unified simulation of inviscid bubbles and viscous droplets within the same computational framework. 
Validation tests reported by \citet{zhang2025lift} and \citet{WEI2026114632} demonstrated second-order accuracy in both velocity and pressure fields, and excellent agreement with analytical solutions for benchmark problems involving steady and unsteady flows past fixed bubbles and droplets. 


\begin{figure}
    \centering
    \setlength{\tabcolsep}{10pt} 
    \iflocalcompile
    \begin{tabular}{cc}
        \resizebox{!}{0.26\textwidth}{\input{fig/Appendix_C/EBM-mesh.tex}} &
        \resizebox{!}{0.26\textwidth}{\input{fig/Appendix_C/EBM-mesh-detail.tex}}
    \end{tabular}
    \else
    \begin{tabular}{cc}
        \resizebox{!}{0.26\textwidth}{\includetikz{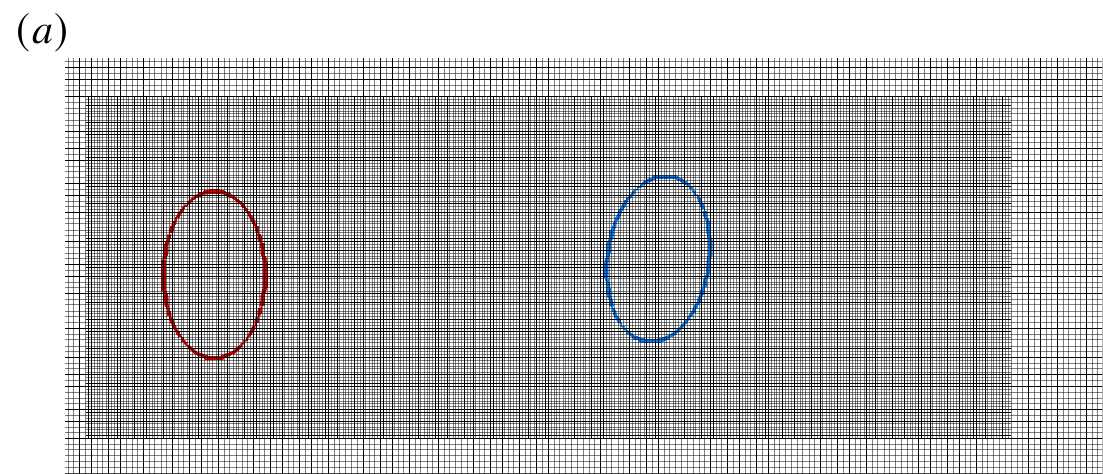}} &
        \resizebox{!}{0.26\textwidth}{\includetikz{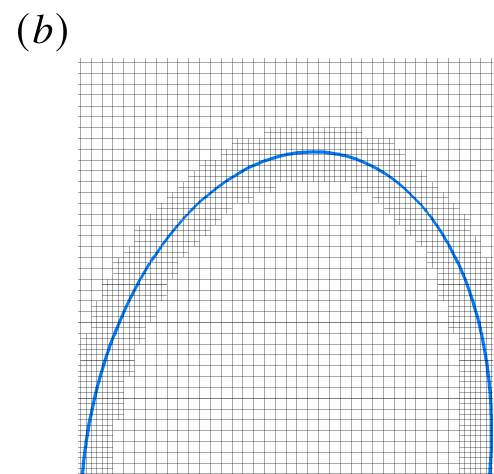}}
    \end{tabular}
    \fi
    \captionsetup{justification=justified, singlelinecheck=false, labelsep=period, width=\linewidth} 
    \caption{Overview of the grid distribution used in the EBM–DNS framework.  
$(a)$ Computational setup corresponding to the bubble pair investigated in figure~7, with the leading bubble aligned horizontally as $\Theta_{LB} = 0^{\circ}$ (red line) and the trailing bubble inclined by $\Theta_{TB} = 8^{\circ}$ (blue line).  
The geometrical parameters are $(\chi,\,S,\,S_r) = (1.6,\,3,\,0.1)$.  
For clarity, the finest refinement level of the actual grid has been omitted.  
$(b)$ Close-up view of the mesh refinement in the vicinity of the trailing bubble.}
\label{appc-f-1}
\end{figure}

In the present study, the EBM--DNS framework was adapted to compute the hydrodynamic forces and torques acting on the bubble pair, serving as an auxiliary numerical approach to corroborate the findings obtained from the ALE--LSA analysis. 
\hyperref[appc-f-1]{Figure~\ref{appc-f-1}$(a)$} shows the Cartesian grid used in the simulations, corresponding to the configuration analysed in figure~7, i.e. $(\chi, S) = (1.6, 3)$, where the TB is inclined by $\Theta_{TB} = 8^{\circ}$.  
A close-up view of the locally refined mesh in the vicinity of the TB is presented in \hyperref[appc-f-1]{figure~\ref{appc-f-1}$(b)$}.  
The smallest grid spacing near the interface was set to $\Delta/2R = 1/160$, with approximately ten grid layers resolved on either side of each interface.  
This resolution ensures at least eleven grid points across the viscous boundary layer, providing sufficient accuracy for evaluating the hydrodynamic forces and torques acting on the bubbles.  
Systematic grid-refinement tests reported by \citet{zhang2025lift} and \citet{WEI2026114632} confirmed that this level of resolution yields grid-independent predictions for both the lift and torque coefficients, validating the numerical accuracy of the present EBM--DNS computations.

\section{More influential factors on the ocillatory mode}
\label{app:oscillatory}
\subsection{Effect of mutual interaction between the two bubbles}
\label{sec:oscillatory_1}

This section quantifies the influence of the two-way hydrodynamic coupling between the bubbles on the oscillatory mode.  
We focus on the bubble pair with $\chi = 1.9$, using the same setup as that described in \hyperref[LB:fixed:chi1.6and2.0:neutralcurve]{figure~\ref{LB:fixed:chi1.6and2.0:neutralcurve}$(a)$}.  
A complementary series of ALE--LSA computations was performed in which the LB was artificially frozen, so that it still modified the flow field experienced by the TB but no longer responded dynamically to its perturbations.  

\begin{figure}
	\centering
	\setlength{\tabcolsep}{0pt}
	\iflocalcompile
		\resizebox{0.5\textwidth}{!}{\input{\figfourfive/chi1.9-LBfixed-osc.tex}}
	\else
		\resizebox{0.5\textwidth}{!}{\includetikz{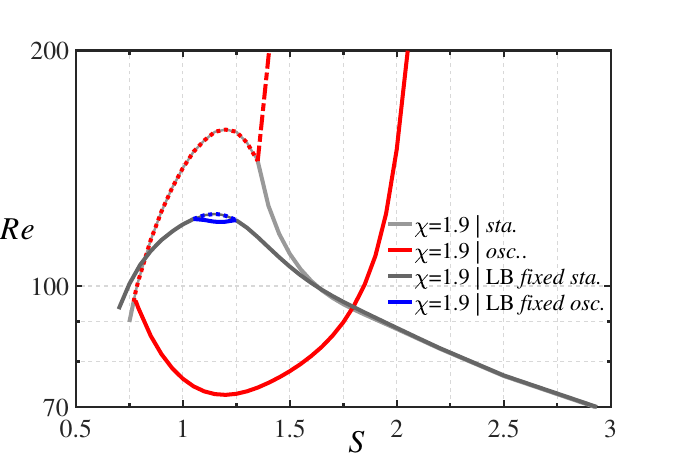}}
	\fi
	\captionsetup{justification=justified, singlelinecheck=false, labelsep=period, width=\linewidth}
	\caption{Neutral curves of the oscillatory mode in the $(\Rey, S)$ plane for $\chi = 1.9$, comparing configurations where the LB is either fixed or free to move.  Coloured lines indicate the oscillatory branches, and grey lines reproduce the stationary modes. Dotted and dashed coloured lines correspond to the reference neutral curves shown in figure~\ref{neutral:curve:osci}$(a)$.
}
\label{LB:fixed:chi1.9:oscillatory}
\end{figure}

The corresponding neutral curves for $\chi = 1.9$ are presented in \hyperref[LB:fixed:chi1.9:oscillatory]{figure~\ref{LB:fixed:chi1.9:oscillatory}}.  
The oscillatory branch is found to be markedly altered when the LB motion is suppressed, and the response is even more sensitive than that of the stationary branch discussed previously.  
As shown in the figure, the parameter domain supporting oscillatory instability shrinks drastically once the LB is held fixed, and the associated reduced frequency $St$ decreases significantly (not shown here).  
The transition scenario also changes qualitatively: for $\chi = 1.9$, the oscillatory branch no longer coexists with the stationary mode (Cases~ii and~iii), but instead transforms into two distinct stationary branches through a codimension-two exceptional point (Case~i).  
This behaviour confirms that the oscillatory instability is not an intrinsic property of the TB wake, but rather a manifestation of the dynamic interaction between the two bubbles.  

Within the hydrodynamic--spring framework 
(\hyperref[sec:LSA_oscillatory]{\S~\ref{sec:LSA_oscillatory}}), 
the LB and TB act as two coupled masses linked by an elastic, vortical ``spring'' formed by the inter-bubble recirculation.  
Allowing the LB to move introduces a reactive component that phase-lags the TB motion, thereby maintaining the energy exchange required to sustain oscillation.  
When the LB is fixed, one end of the hydrodynamic spring becomes rigidly anchored, suppressing this reactive feedback and extinguishing the oscillatory mode.  
Physically, the lateral oscillation of the TB induces a counter-rotation and lateral displacement of the LB, which modulate the local shear and vorticity within the gap.  
This phase-lagged coupling between pressure and torque forms the feedback loop that sustains the global oscillation.  
Once the LB is immobilised, the feedback collapses, leaving only a damped response and a purely stationary instability.  
Hence, the mutual hydrodynamic interaction is not a secondary correction, but a fundamental mechanism governing the onset, frequency selection, and persistence of the oscillatory instability in bubble pairs.

\subsection{Effect of distinct aspect ratios of the two bubbles}
\label{sec:oscillatory_2}
\begin{figure}
	\centering
	\setlength{\tabcolsep}{0pt}
	\iflocalcompile
	\begin{tabular}{cc}
		\resizebox{0.5\textwidth}{!}{\input{fig/4.5differentchi/chi1.9-1.7-different-chi-osci.tex}} &
		\resizebox{0.5\textwidth}{!}{\input{fig/4.5differentchi/chi1.9-1.8-different-chi-osci.tex}}
	\end{tabular}
	\else
	\begin{tabular}{cc}
		\resizebox{0.5\textwidth}{!}{\includetikz{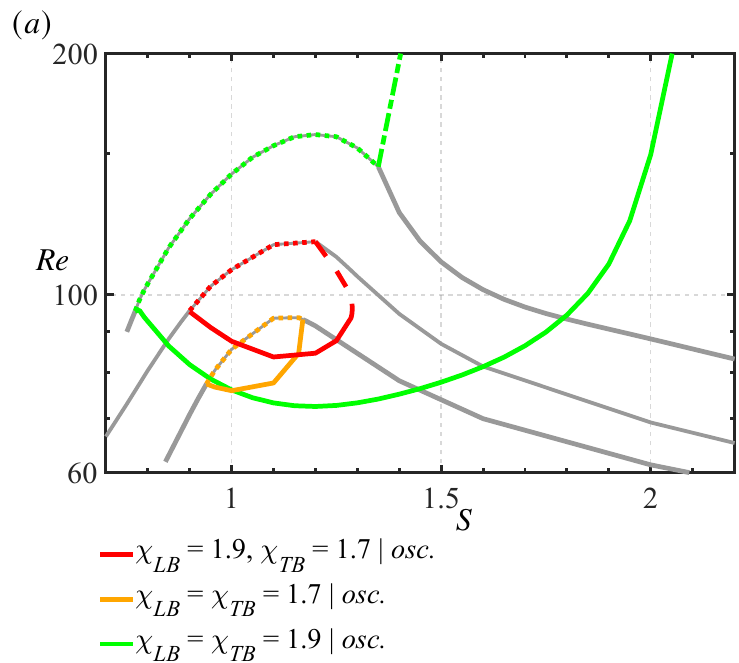}} &
		\resizebox{0.5\textwidth}{!}{\includetikz{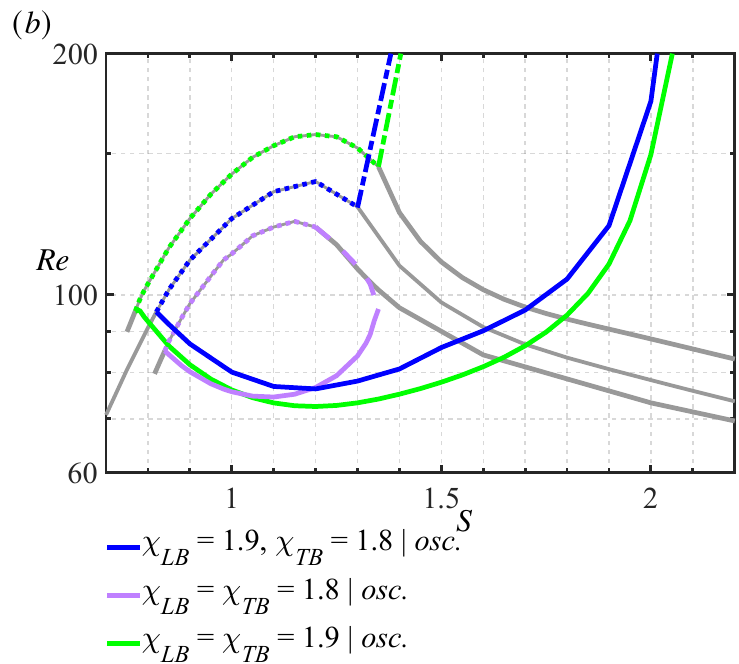}}
	\end{tabular}
	\fi
	\captionsetup{justification=justified, singlelinecheck=false, labelsep=period, width=\linewidth}
	\caption{
Neutral stability curves of the oscillatory mode for unequal-$\chi$ bubble pairs with a fixed LB aspect ratio $\chi_{LB} = 1.9$.  
$(a)$ Case $(\chi_{LB}, \chi_{TB}) = (1.9, 1.7)$ and $(b)$ case $(\chi_{LB}, \chi_{TB}) = (1.9, 1.8)$.  
Coloured solid lines denote the oscillatory branches, while grey lines correspond to reference results for equal-$\chi$ configurations with $\chi_{LB} = \chi_{TB} = 1.7$, $1.8$, and $1.9$.  
In both panels, the oscillatory regions of the unequal-$\chi$ pairs lie between those of the two symmetric limits, with narrower unstable domains and modified transition scenarios.  
Specifically, the $(1.9, 1.7)$ pair exhibits a hybrid Case~(i)–(ii) transition, whereas the $(1.9, 1.8)$ pair follows a combined Case~(i)–(iii) transition, indicating that the oscillatory instability is more sensitive to the deformation of the leading bubble than to that of the trailing one.
}
\label{chi1.9:different:chi:osci}
\end{figure}

We now investigate a more realistic in-line bubble pair in which the TB is more spherical than the LB, and quantify how such shape asymmetry modifies the oscillatory stability of the system.  
Three oblate configurations are examined with $\chi_{LB} = 1.9$ and $\chi_{TB} = 1.9$, $1.8$, and $1.7$.  
The corresponding oscillatory neutral curves are displayed in 
\hyperref[chi1.9:different:chi:osci]{figures~\ref{chi1.9:different:chi:osci}$(a,b)$}, 
which illustrate the unequal-$\chi$ cases $(\chi_{LB}, \chi_{TB}) = (1.9, 1.7)$ and $(1.9, 1.8)$, respectively.  
In both configurations, the oscillatory branches depart significantly from their equal-$\chi$ counterparts.  
The unstable domains become narrower than those of the more oblate pair, yet broader than those corresponding to the less oblate one, and the transition scenarios are altered accordingly.  
For instance, in \hyperref[chi1.9:different:chi:osci]{figure~\ref{chi1.9:different:chi:osci}$(a)$}, the equal-$\chi$ pair with $\chi_{LB} = \chi_{TB} = 1.7$ exhibits a Case~(i) transition, whereas $\chi_{LB} = \chi_{TB} = 1.9$ displays a combined Case~(i)–(iii) behaviour.  
In contrast, the asymmetric configuration $(\chi_{LB}, \chi_{TB}) = (1.9, 1.7)$ produces a hybrid Case~(i)–(ii) transition.  
Similarly, in \hyperref[chi1.9:different:chi:osci]{figure~\ref{chi1.9:different:chi:osci}$(b)$}, the pair $(\chi_{LB}, \chi_{TB}) = (1.9, 1.8)$ behaves more like the equal-$\chi$ configuration $(1.9, 1.9)$, following a combined Case~(i)–(iii) scenario.  
These comparisons indicate that the oscillatory instability is more sensitive to the deformation of the LB than to that of the TB.  
Hence, the oscillatory mode is primarily governed by the flow topology established by the coupled interaction between the two bubbles.  

Physically, the recirculating structure in the inter-bubble gap originates from vorticity shed by the LB, which accumulates between the two interfaces and interacts with the TB through blockage and shear coupling.  
The strength and extent of this recirculation depend on the deformation of both bubbles; however, the LB largely determines the topology of the vortical field, whereas the TB mainly modulates the amplitude of the coupling.  
Thus, shape asymmetry acts primarily as a secondary modulation of the global hydrodynamic feedback, rather than altering the fundamental nature of the instability.  
From the viewpoint of the hydrodynamic–spring analogy introduced in 
\hyperref[sec:LSA_oscillatory]{\S~\ref{sec:LSA_oscillatory}}, 
the deformation of each bubble governs distinct but complementary aspects of the coupled dynamics.  
The LB, which sheds and sustains the vortical recirculation in the inter-bubble gap, primarily controls the effective stiffness, $K_{h}$, of the hydrodynamic spring.  
A more oblate LB enhances vorticity generation and strengthens the pressure gradients along the wake axis, thereby reinforcing the elastic coupling between the two bubbles.  
In contrast, the TB sets the effective damping and rotational compliance of the system, as its shape determines how efficiently the incident vortical flux is converted into inclination and lateral motion, which in turn controls the phase lag between torque and displacement.  
Hence, while the LB dictates how strongly the system stores hydrodynamic energy through vortex stretching and pressure coupling, the TB regulates how this energy is released or dissipated through viscous damping and rotational response.  

When the two bubbles have different deformations, the symmetry of the spring–mass–damper system is broken.  
The resulting imbalance between stiffness and damping alters not only the extent of the oscillatory domain but also the transition scenario between stationary and oscillatory modes.  
Specifically, an oblate LB coupled with a less deformed TB weakens the damping relative to the stiffness, promoting mixed Case~(i)–(ii) transitions characterised by underdamped oscillations that evolve toward steady displacement.  
Conversely, when both bubbles are highly oblate, the strong vortical coupling sustains a stiffer spring with lower damping, leading to Case~(i)–(iii) coexistence, where oscillatory and stationary modes overlap.  
This rationalises the widening and bifurcation of the neutral lobes observed in 
\hyperref[chi1.9:different:chi:osci]{figure~\ref{chi1.9:different:chi:osci}}, 
showing that the relative magnitudes of $K_{h}$ and $\varsigma$ determine whether the pair exhibits stationary, oscillatory, or mixed-mode instability, 
thereby explaining the diversity of transition scenarios identified in unequal-$\chi$ configurations.

\bibliographystyle{jfm}
\bibliography{jfm}

\end{document}